\documentclass[twocolumn,floatfix]{aastex631}
\usepackage{amsmath}
\usepackage{rotating}
\usepackage{multirow}

\begin{document}

\title{Polarized Light from Massive Protoclusters (POLIMAP). I.\\Dissecting the role of magnetic fields in the massive infrared dark cloud G28.37+0.07}

\author[0000-0003-1964-970X]{Chi-Yan Law}
\email{cylaw.astro@gmail.com}
\affiliation{Department of Space, Earth \& Environment, Chalmers University of Technology, SE-412 96 Gothenburg, Sweden}
\affiliation{European Southern Observatory, Karl-Schwarzschild-Strasse 2, D-85748 Garching, Germany}
\author[0000-0002-3389-9142]{Jonathan C. Tan}
\affiliation{Department of Space, Earth \& Environment, Chalmers University of Technology, SE-412 96 Gothenburg, Sweden}
\affiliation{Department of Astronomy, University of Virginia, Charlottesville, VA 22904-4235, USA}
\author[0000-0003-2337-0277]{Raphael Skalidis}
\affiliation{Division Office Physics, Math and Astronomy, California Institute of Technology,  Bridge Annex, Pasadena, CA 91125, USA}
\author[0000-0001-5551-9502]{Larry Morgan}
\affiliation{Green Bank Observatory, 155 Observatory Rd, Green Bank, West Virginia, WV 24944, USA}
\author[0000-0001-6216-8931]{Duo Xu}
\affiliation{Department of Astronomy, University of Virginia, Charlottesville, VA 22904-4235, USA}
\author[0000-0003-4040-4934]{Felipe de Oliveira Alves}
\affiliation{Institut de Ciencies de l\'{E}spai (IEEC-CSIC), Carrer de Can Magrans, 08193 Cerdanyola del Vall\'{e}s, Barcelona, Spain.}
\author[0000-0003-0410-4504]{Ashley T. Barnes}
\affiliation{European Southern Observatory, Karl-Schwarzschild-Strasse 2, D-85748 Garching,Germany}
\author[0000-0002-4013-6469]{Natalie Butterfield}
\affiliation{Green Bank Observatory, 155 Observatory Rd, Green Bank, West Virginia, WV 24944, USA}
\author[0000-0003-1481-7911]{Paola Caselli}
\affiliation{Max-Planck-Institute for Extraterrestrial Physics: Garching, Bayern, DE}
\author[0000-0001-5551-9502]{Giuliana Cosentino}
\affiliation{Department of Space, Earth \& Environment, Chalmers University of Technology, SE-412 96 Gothenburg, Sweden}
\author[0000-0001-5551-9502]{Francesco Fontani}
\affiliation{Osservatorio Astrofisico di Arcetri, Largo Enrico Fermi, 5, 50125 Firenze FI, Italy}
\author[0000-0001-9656-7682]{Jonathan D. Henshaw}
\affiliation{Astrophysics Research Institute, Liverpool John Moores University, Liverpool, UK}
\author[0000-0003-4493-8714]{Izaskun Jimenez-Serra}
\affiliation{Centro de Astrobiologia, Instituto Nacional de Tecnica Aeroespacial Ctra de Torrejon a((INTA/CSIC)), km 4, 28850 Ajalvir, Madrid, Spain}
\author[0000-0001-6216-8931]{Wanggi	Lim}
\affiliation{Caltech/IPAC,MC 314-6 (Keith Spalding Building) 1200 E California Blvd Pasadena, CA 91125}




\begin{abstract}
Magnetic fields may play a crucial role in setting the initial conditions of massive star and star cluster formation. To investigate this, we report SOFIA-HAWC+ $214\:\mu$m observations of polarized thermal dust emission and high-resolution GBT-Argus C$^{18}$O(1-0) observations toward the massive Infrared Dark Cloud (IRDC) G28.37+0.07. Considering the local dispersion of $B$-field orientations, we produce a map of $B$-field strength of the IRDC, which exhibits values between $\sim0.03 - 1\:$mG based on a refined Davis–Chandrasekhar–Fermi (r-DCF) method proposed by Skalidis \& Tassis. Comparing to a map of inferred density, the IRDC exhibits a $B-n$ relation with a power law index of $0.51\pm0.02$, which is consistent with a scenario of magnetically-regulated anisotropic collapse. Consideration of the mass-to-flux ratio map indicates that magnetic fields are dynamically important in most regions of the IRDC. A virial analysis of a sample of massive, dense cores in the IRDC, including evaluation of magnetic and kinetic internal and surface terms, indicates consistency with virial equilibrium, sub-Alfv\'enic conditions and a dominant role for $B-$fields in regulating collapse. A clear alignment of magnetic field morphology with direction of steepest column density gradient is also detected. However, there is no preferred orientation of protostellar outflow directions with the $B-$field. Overall, these results indicate that magnetic fields play a crucial role in regulating massive star and star cluster formation and so need to be accounted for in theoretical models of these processes.
\end{abstract}



\section{Introduction}\label{sec:intro}
Magnetic ($B$) fields have been suggested to play a significant role in the birth of massive stars and star clusters \citep[e.g., see reviews by][]{2014prpl.conf..149T,2022arXiv220311179P}. However, many aspects remain uncertain, including the relative importance of $B$-fields compared to turbulence, converging flows, feedback and self-gravity in setting the initial molecular cloud conditions and fragmentation properties that ultimately lead to massive star and star cluster formation. Depending on the morphology and strength of $B$-fields, various models of magnetically regulated star formation have been proposed, including: fragmentation of magnetized filamentary clouds \citep[e.g.,][]{2017NatAs...1E.158L}; collisions of magnetised clouds \citep[e.g.,][]{2015ApJ...811...56W}; and collapse after compression in feedback bubbles \citep[e.g.,][]{2015A&A...580A..49I}. 
For massive star formation/accretion, two main scenarios are (i) Core Accretion \citep[e.g., the Turbulent Core Accretion (TCA) model of][]{2003ApJ...585..850M}
and (ii) Competitive Accretion \citep[e.g.,][]{2001MNRAS.324..573B,2019MNRAS.490.3061V,2020ApJ...900...82P,2022MNRAS.512..216G}. The former is a scaled-up version of the standard model of low-mass star formation \citep{1987ARA&A..25...23S}, although with the internal pressure of the massive prestellar core dominated by turbulence and/or magnetic fields rather than thermal pressure. The fiducial TCA model involves near equipartition between turbulence and magnetic fields, resulting in typical field strengths in massive pre-stellar cores of $\sim 1\:$mG. In Competitive Accretion, stars chaotically gain their mass in the crowded centres of protoclusters, typically via the global collapse of a cluster-forming clump, without passing through the massive prestellar core phase. This mode has generally been seen in simulations that are either unmagnetized or have relatively weak $B$-fields.


To uncover the connections between larger-scale environments and smaller-scale star formation properties and to test theoretical star formation models, it is crucial to obtain observational constraints on the magnetic field morphologies, strengths, and energy balances of molecular clouds that are progenitors of massive  stars and star clusters. Infrared dark clouds (IRDCs), owing to their dense ($n_{\rm H}\geq10^{5}$cm$^{-3}$) and cold ($T\leq$ 20~K) conditions, are thought to be examples of such clouds \citep[e.g.,][]{2014prpl.conf..149T}.
However, measuring the magnetic field strength and morphology in IRDCs is challenging due to their relatively far distances ($d\gtrsim 3\:$kpc).
Current and future advances in the polarimetric capabilities of various single dish telescopes (e.g., SOFIA-HAWC+, JCMT-POL2, LMT-TolTEC, and IRAM-30m-NIKA2) and the full polarization capabilities of interferometric facilities (e.g., ALMA, SMA, and VLA) are enabling studies of $B$-fields in IRDCs \citep[e.g.,][]{2015ApJ...799...74P,2023ApJ...942...95X}.
Such studies typically rely on mapping polarized thermal dust emission to infer the plane of sky component magnetic field morphology and utilizing the \citet{1951PhRv...81..890D} and \citet{1953ApJ...118..113C} (DCF) method to infer the magnetic field strength. Other techniques to infer magnetic fields include via the Zeeman effect \citep[e.g.,][]{2019FrASS...6...66C,2022Natur.601...49C}, employing velocity anisotropy observed in the spectral line cubes \citep{2023ApJ...942...95X}, or via the Goldreich-Kylafis effect \citep[e.g.,][]{1981ApJ...243L..75G,1982ApJ...253..606G,2008A&A...492..757F}. However, these non-DCF methods have so far been challenging to apply to IRDCs. 

Observations of the magnetic field properties toward IRDCs have revealed ordered magnetic field morphology and relatively strong $B$-fields of up to $\sim$mG strength, which would be strong enough to influence the cloud dynamics, including fragmentation. A correlation between magnetic field morphology and filament density structure indicating perpendicular alignment with filament orientations has been reported \citep[e.g.,][]{2018ApJ...859..151L,2019ApJ...883...95S}. However, these studies have concerned only a handful of IRDCs, e.g., G11.11-0.12 (the ``Snake''), G34.34+0.24, G35.39-0.33, and the ``Brick'' \citep{2015ApJ...799...74P,2017ApJ...836..199H,2018ApJ...859..151L,2019ApJ...883...95S,2019ApJ...878...10T,2022MNRAS.512.4272V,2022arXiv220703695C,2023arXiv230210543B,2023ApJ...942...95X}. Hence, a systematic study of IRDCs that span a range of masses, sizes, and environments 
is an important next step further to understanding the links between the magnetic field and IRDC properties and, thus, better constrain the initial conditions of massive star and star cluster formation. 

To more systemically map and constrain the magnetic fields properties in IRDCs, we have initiated the Polarized Light from Massive Protoclusters (POLIMAP) (PI: J. C. Tan) survey with an original goal to map the polarized dust continuum emission of a sample of 10 IRDCs (Clouds A to J) from the sample of \citet{2009ApJ...696..484B,2012ApJ...754....5B}. POLIMAP utilizes SOFIA-HAWC+ observations at $214\:{\rm \mu m}$ (Project ID: 09\_0104)\footnote{We note that the observation toward G28.37+0.07 is shared with the large program (SIMPLIFI) with proposal ID 09$\_$0215 (PI: T. Pillai)}, complemented by joint Green Bank Telescope (GBT)-Argus $^{13}$CO(1-0) and C$^{18}$O(1-0) observations to probe molecular gas kinematics at high angular resolution of $\sim 7^{\prime\prime}$. Before the shutdown of SOFIA, POLIMAP HAWC+ observations were completed for IRDCs B, C, F, H, I, J, while GBT-Argus observations have been completed for IRDCs B, C, F, G, H, I, J. In this paper we present the first results from POLIMAP for the massive IRDC G28.37+0.07 (Cloud C).



\begin{figure*}[h!]
\centering
\includegraphics[width=0.9\textwidth]{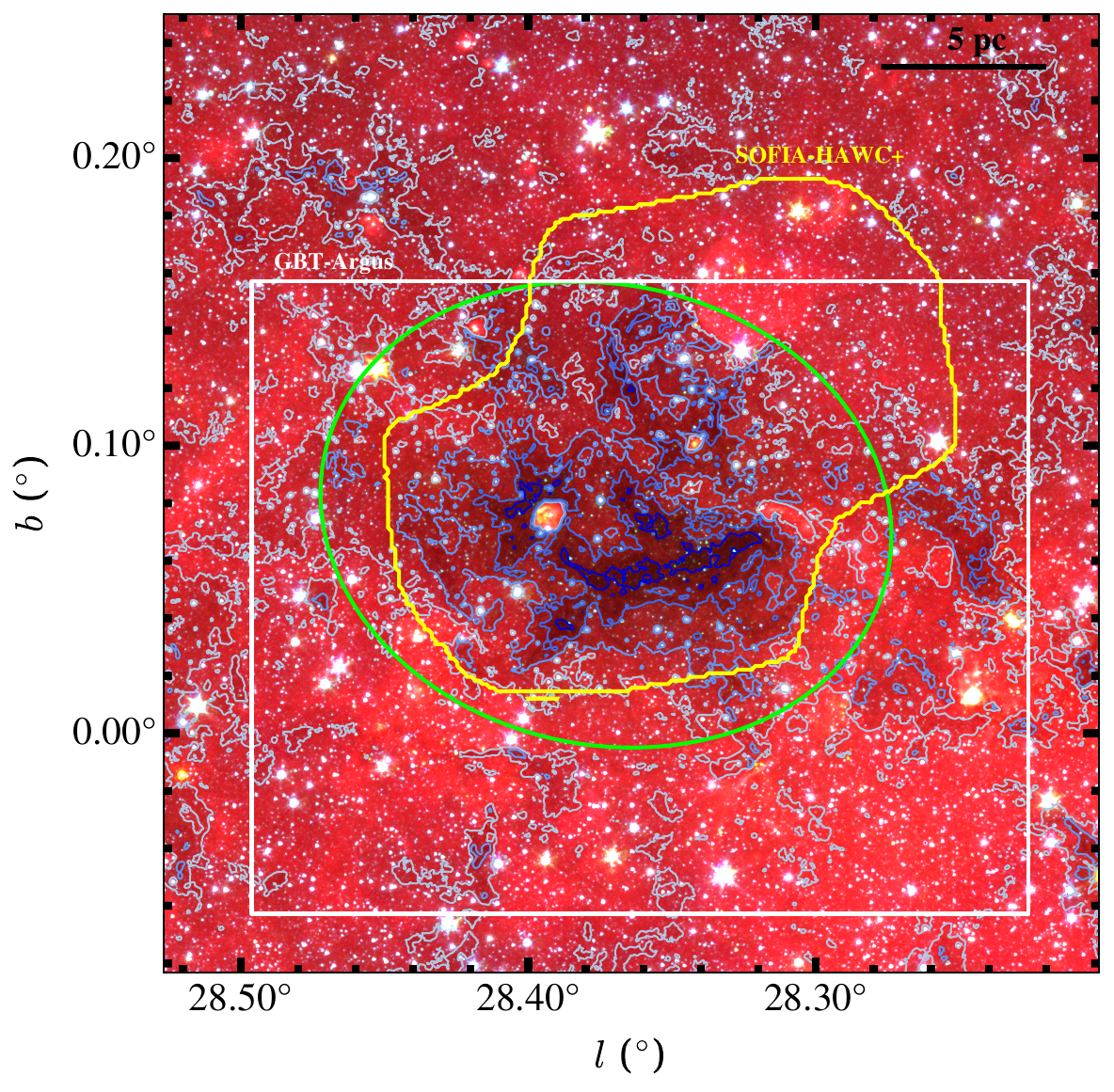}
\caption{Global MIR view of IRDC G28.37+0.07. MIR extinction (including NIR correction) contours ($A_V=7, 15, 30, 60\:$mag) \citep{2013A&A...549A..53K} are overlaid on a 3-colour Spitzer GLIMPSE image (RGB: 8, 5.8 and 4.5 $\rm \mu m$) \citep{2009PASP..121..213C}. The yellow contour shows the region mapped by SOFIA-HAWC+ at 214~$\rm \mu m$. The white rectangle is the region mapped by GBT-Argus to trace $^{13}$CO and $\rm C^{18}O$(1-0) emission. The green ellipse shows the IRDC region defined by \citet{2006ApJ...639..227S}.} 
\label{fig:Figure1_paper}
\end{figure*}

Located at a distance of $d=5.0$~kpc from the Sun \citep{2006ApJ...639..227S},
G28.37+0.07 is one of the most massive IRDCs known, with a mass estimated to be about $70,000\:M_\odot$ (Butler et al. 2014; Lim \& Tan 2014)
and hosts multiple massive dense cores \citep[][]{2013ApJ...779...96T,2023arXiv230315499B}. 
Figure~\ref{fig:Figure1_paper} shows a {\it Spitzer} mid-infrared (MIR) view of the IRDC, along with MIR extinction (MIREX) contours that help illustrate the cloud's global structure \citep{2012ApJ...754....5B,2013A&A...549A..53K}.
Overall physical conditions of the cloud, including the mass surface density distributions, gas kinematics, excitation conditions, core mass function, high-mass protostellar populations and their outflows, have been explored in a variety of studies \citep[e.g.,][]{2009ApJ...696..484B,2012ApJ...754....5B,2013A&A...549A..53K,2015ApJ...809..154H,2018MNRAS.474.3760C,2019ApJ...874..104K,2020ApJ...897..136M,2022A&A...662A..39E}. However, there has only been limited study of magnetic fields in this IRDC \citep[e.g.,][]{2020ApJ...895..142L}, especially on global scales.


This paper is structured as follows. In \S\ref{sec:observation} we describe the observations and data reduction procedures used for the SOFIA and GBT observations, as well as various ancillary datasets. 
In \S\ref{sec:results} we present our main observational results for the $214\:{\rm \mu m}$ polarimetric imaging (tracing $B$-field orientation), $\rm C^{18}O$(1-0) emission, derived $B$-field strength map, and associated $B-N$ and $B-n$ relations.
In \S\ref{sec:dynamics} we discuss the dynamical implications of the magnetic field, including presentation of the mass-to-flux ratio map and a virial analysis of dense cores in the IRDC. 
In \S\ref{sec:fragmentation} we explore the connections between magnetic field morphology and gradients in the mass surface density map and orientations of protostellar outflows.
We present a discussion and summary of our results in \S\ref{sec:conclusions}.


\section{Observations}\label{sec:observation}

\subsection{SOFIA-HAWC+}\label{subsec:SOFIA}

We observed the POLIMAP IRDC sample with the Stratospheric Observatory for Infrared Astronomy (SOFIA) using the HAWC+ instrument \citep{2018JAI.....740008H} observing in band E ($\rm 214\:\mu m$). HAWC+ is a Half-Wave Plate (HWP) and wire-grid polarimeter operating from $\sim 50$ to $214\:{\rm \mu m}$. Rotating the HWP allows the four required angles to be probed at $\theta$ = (0, 22.5, 45, and 67.5~degrees). Note, HAWC+ is not sensitive to circular polarization and thus only measures the linearly polarized emissions. 

The SOFIA-HAWC+ observations of G28.37+0.07 were carried out on 21, 27 \& 28 Sept. 2022 (Proposal ID: 09\_0104) as part of SOFIA Cycle 9. The full width at half maximum (FWHM) of the beam is $18.^{\prime\prime}2$, and the surveyed map size is about $12.^{\prime}0 \times 11.^{\prime}4$ (see Fig.~\ref{fig:Figure1_paper}). The observations were performed using the Nod-Match chop mode with a Lissajous scan pattern to have fast enough measurements such that the variations in the background do not overwhelm the polarization signal. 

The raw data were processed by the HAWC+ data reduction pipeline version 3.2.0\footnote{https://www.sofia.usra.edu/data/data-pipelines\#faq}. This pipeline includes different data processing steps, such as corrections for dead pixels and the intrinsic polarization of the instrument and telescope. The final data output of the pipeline is `Level 4', flux-calibrated, and the resulting pixel size of these data products are $4.^{\prime\prime}55$. 

\subsection{GBT-Argus}\label{subsec:GBT}

We mapped $^{13}$CO(1-0) and C$^{18}$O(1-0) emissions of the IRDC sample with Argus instrument of the GBT between Jan to April 2022. Argus \citep{2014SPIE.9153E..0PS} is a 16-element heterodyne `radio camera' operating in the 85-116 GHz range. Argus has 16 heterodyne pixel receivers mounted in a 4$\times$4 layout and with a beam separation of 30.\arcsec4 in both the elevation and cross-elevation directions. 

The observations utilized a `fast-mapping' method in which the heterodyne pixel receivers are scanned across the sky in rows of Galactic longitude, separated by $5.^{\prime\prime}58$ (i.e., $0.8\times$ the beam size). This might be expected to undersample the plane of the sky in comparison with Nyquist sampling, in which a separation of less than 0.5$~\times$ of the beam size is required. However, the effects of sky rotation in relation to the scan direction and the fact that multiple beams observe any given point in the sky allow for a relaxation of this requirement without a significant effect on the data quality. The map coverage toward G28.37+0.07 has an area of $16.^{\prime}2 \times 13.^{\prime}2$ (see Fig.~\ref{fig:Figure1_paper}).

The data were reduced using the GBTIDL package, with calibration achieved through observations of the Argus `Vane', essentially a warm load placed in the beam of the receiver. Atmospheric opacity estimates were taken from the forecast values obtained from the CLEO\footnote{https://www.gb.nrao.edu/rmaddale/GBT/CLEOManual/index.html} application, which are the average of the values from three surrounding local sites. After baseline fitting and removal were performed on the spectra of each Argus feed, the data were then gridded using the GBO tool {\it gbtgridder} with a Gaussian beam assumption and a pixel size of $3.^{\prime\prime}0$. The resulting cube was then subjected to further iterative baseline fitting and removal to equalize background noise levels. Both the final $^{13}$CO(1-0) and C$^{18}$O(1-0) position-position-velocity (PPV) cubes have a pixel scale of $2.^{\prime\prime}57$ and a spectral resolution of $0.182\,{\rm km\,s^{-1}}$. 
The root-mean-square (RMS) noise level is estimated by taking the average value of RMS noise levels measured from spectra obtained within beam-sized regions over the velocity range of 50-65 ${\rm km\,s^{-1}}$, i.e., away from the emission of the IRDC or other clouds. We find a noise level of 0.55~K per channel (of $0.182\,{\rm km\,s^{-1}}$ velocity range).


\subsection{Ancillary Data}


We utilize previously published maps of the mass surface density, $\Sigma$, of the IRDC. First, \citet{2012ApJ...754....5B} constructed a mid-infrared (MIR) extinction (MIREX) map based on $8\,\mu m$ Spitzer-IRAC imaging data. A near-infrared (NIR) extinction corrected version of this map was published by \citet{2013A&A...549A..53K}, which improves the accuracy of the large-scale, lower column density base level of the map. These MIREX maps have a relatively high angular resolution of $2^{\prime\prime}$, set by the Spitzer-IRAC resolution. They also have the advantage of not requiring knowledge of the temperature of the IRDC material, with the main uncertainty arising from the choice of the opacity per unit mass at MIR wavelengths, estimated to be $\sim30\%$ from comparison of different dust models \citep{2012ApJ...754....5B}. However, the MIREX maps have a disadvantage in that they cannot measure $\Sigma$ in MIR-bright regions, e.g., MIR-bright sources, which appear as ``holes'' in the map.
In addition, the MIREX method has a maximum level of $\Sigma$ that it is able to probe, which for IRDC C is $\sim 0.5\:{\rm g\:cm}^{-2}$, so in high-$\Sigma$ regions it will tend to underestimate the true mass surface density.

An alternative method of estimating $\Sigma$ is from modelling the far-infrared (FIR) emission. \citet{2016ApJ...829L..19L} derived a $\Sigma$ map of IRDC C by fitting a graybody emission model to Herschel-PACS and SPIRE fluxes from 160 to 500~$\rm \mu m$. This analysis also yields a map of the dust temperature of the IRDC. Since the IRDC is close to the Galactic plane, there is significant FIR emission from material in the diffuse ISM along the line of sight. \citet{2016ApJ...829L..19L} explored various methods for correcting for this background and foreground emission. Our analysis uses the map derived from their Galactic Gaussian (GG) foreground-background subtraction method. The FIR emission derived $\Sigma$ map has a resolution of $18^{\prime\prime}$ (i.e., the ``hi-res'' version), while the temperature map has a resolution of $36^{\prime\prime}$ \citep[see][for details]{2016ApJ...829L..19L}.

\section{Results}\label{sec:results}

\subsection{214~$\mu m$ Polarimetric Imaging}

\begin{figure*}[h!]
\centering
\includegraphics[width=\textwidth]{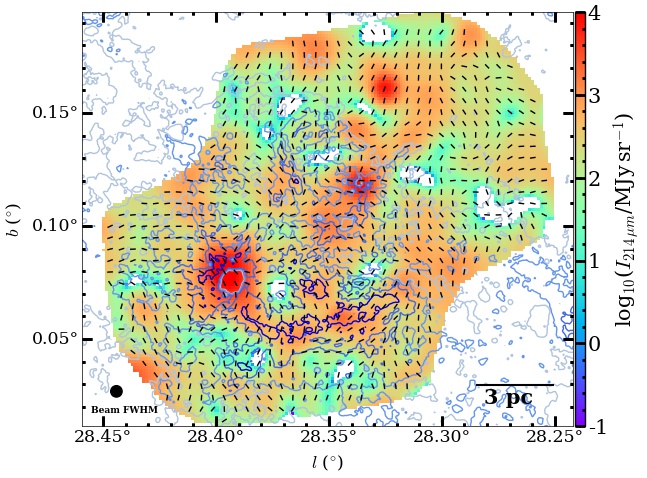}
\caption{The SOFIA-HAWC+ Stokes $I$ intensity map of G28.37+0.07 at $214\:\mu m$. Inferred magnetic field position angles are shown with black line segments, separated from each other by a beam size of 18$^{\prime\prime}.2$, i.e., $0.441$~pc. The size of the beam is shown by the black circle in the lower left corner. The contours show constant levels of extinction (same as Fig.~\ref{fig:Figure1_paper}).
}
\label{fig:SOFIA_cloudC}
\end{figure*}


\begin{figure*}[]
\centering
\includegraphics[width=\textwidth]{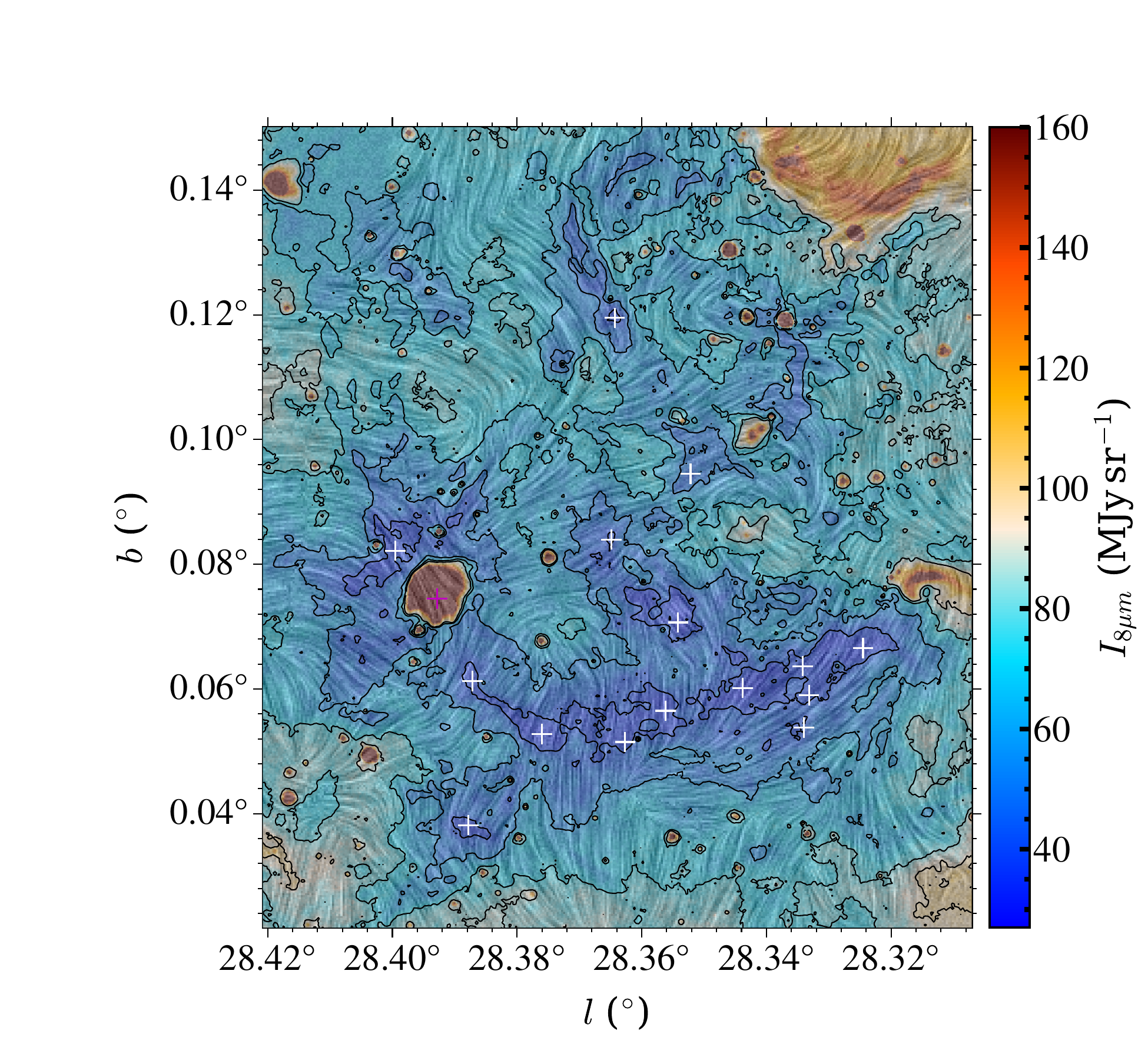}
\caption{$8\,\mu m$ intensity map of G28.37+0.07 with the inferred plane of sky magnetic field component overlaid as a ``drapery'' image. The drapery pattern is produced using the line integral convolution method \citep{Cabral:1993:IVF:166117.166151} after regridding the original polarisation angle data to a finer spatial scale. The extinction contours of Figure~\ref{fig:Figure1_paper} are shown in grey. The magenta $+$ marks the position of massive protostar Cp23 \citep{2020ApJ...897..136M}, while the white $+$ symbols mark the dense starless/early-stage core/clump sample of \citet{2014ApJ...782L..30B}. Note, each symbol size is equivalent to one SOFIA-HAWC+ beam of $18.^{\prime\prime}2$.} 
\label{fig:cloudC-fragmentation}
\end{figure*}

The $214\:{\rm \mu m}$ intensity (Stokes $I$) map of the IRDC is shown in Figure~\ref{fig:SOFIA_cloudC}. This map also shows 90-degree-rotated polarization vectors, which are expected to trace the material's emission-weighted plane-of-sky magnetic field orientations along the line of sight. The spacing of these vectors is set to one SOFIA-HAWC+ beam at $214\:{\rm \mu m}$, i.e., $18.^{\prime\prime}2$. The inferred magnetic field position angle orientations follow the IAU convention, measured counterclockwise from Galactic north. In Figure~\ref{fig:cloudC-fragmentation} we also show the expected $B$-field orientation in the form of a ``drapery'' pattern overlaid on the Spitzer IRAC $8\:{\rm \mu m}$ map. Appendix~\ref{sec:AppA} presents maps of Stokes $Q$, $U$, and the debiased polarization fraction ($P^\prime$),
and signal-to-noise ratio (SNR) of Stokes $I$. Since SOFIA-HAWC+ does not measure circular polarization, the Stokes parameter $V$ is assumed to be 0.
%

The Stokes $I$ map shows strong, concentrated emissions at several locations. In particular, the brightest region in Figure~\ref{fig:Figure1_paper} corresponds to a MIR-bright region at the eastern side of the IRDC, with $I_{\rm 214 \mu m} \simeq 10^4\:{\rm MJy\:sr}^{-1}$. The main spine of the IRDC extends SW and then W from this location and is seen as an emission feature at $214\:{\rm \mu m}$ with intensity values of $\sim 10^3\:{\rm MJy\:sr}^{-1}$. Two other FIR-bright regions with $I_{\rm 214 \mu m} \simeq 10^4\:{\rm MJy\:sr}^{-1}$ are present to the north of the main IRDC spine, with corresponding MIR-bright sources visible in Fig.~\ref{fig:Figure1_paper}.

The typical uncertainties of Stokes $I$ in the main part of the map are at a near constant level of $\sim9\:{\rm MJy\:sr}^{-1}$, but with local peaks of $\sim16\:{\rm MJy\:sr}^{-1}$ at the locations of the brightest FIR regions. Thus the SNR map of Stokes $I$ (see Appendix~\ref{sec:AppA})
shows typical values of $\sim 150$ in the main IRDC spine, rising to $\sim 500$ in the FIR-bright sources.


\begin{table*}[]
\footnotesize
\centering
\caption{The Stokes parameters, rotated polarization angles, polarization fraction, and the corresponding uncertainties of selected clumps from \citet{2014ApJ...782L..30B}.}
\label{tab:summary_data_table}
\begin{tabular}{cccccccccccccc}
\hline
\hline
Name &\multicolumn{2}{c}{Coordinates} & $I$ & $\Delta I$ & $Q$ & $\Delta Q$ & $U$ & $\Delta U$ & $\theta$ & $\Delta \theta$ & $P$ & $\Delta P$ & $P^{\prime}$ \\
\hline
 & l (deg) & b (deg) & (MJy/sr) &(MJy/sr) & (MJy/sr) & (MJy/sr) & (MJy/sr) & (MJy/sr) & (deg) & (deg)& (\%)&(\%)&(\%) \\
\hline
Cp23$^{\dagger}$&28.3969  & 0.0793& $2.3\times10^{4}$ & 15.7 & -99 & 11.4 & 63.6 & 11.5 & 28.7 & 2.76 & 0.520 & 0.0512 & 0.518 \\
C1&28.3245	 & 0.0665 & $3.98\times10^{2}$ & 12.4 & 39.3 & 16.8 & 17.1 & 17.0 & -33.2 & 11.5 & 10.8 & 4.20 & 10.0\\
C2&28.3438	 & 0.0602 & $1.60\times10^{3}$ & 11.6 & 24.0 & 13.6 & 12.8 &13.9 & -31.2 & 14.6 & 1.76 & 0.859 & 1.56 \\
C3&28.3522	 & 0.0945 & $1.02\times10^{3}$ &8.04 & 32.4 & 9.96 & -25.9 & 10.1 & -64.0 & 7.14 & 4.10 & 1.02 &4.06 \\
C4&28.3542 & 0.0707 & $8.25\times10^{2}$ & 9.15 & -18.6 & 11.1 & 1.36 & 11.2 &42.7 & 16.8 & 2.29 & 1.35& 1.86 \\
C5&28.3562 & 0.0565 & $4.97\times10^{2}$ & 10.6 & 3.76 & 13.1 & 23.4 & 13.3 & -4.70 & 15.9 & 4.83 & 2.69 & 4.04 \\
C6&28.3627 & 0.0515 & $9.32\times10^{2}$ & 11.0 & 7.39 & 14.1 &  22.2& 14.2 & -9.28 & 17.4 & 2.50 & 1.52 &2.00 \\
C7&28.3643 & 0.1195& $4.53\times10^{2}$ & 8.40 & -11.64 & 10.8 & 12.6 & 10.7 & 20.7 & 18.5 & 3.79 & 2.43 & 2.93 \\
C8&28.3878 &0.0382 & $0.437\times10^{2}$ & 11.6 & 6.35 & 16.2 & 11.0 & 16.2 & -15.0 & 34.4 & 46.1 & 70.3 & 0$^{*}$ \\
C9&28.3995 & 0.0822 & $1.29\times10^{4}$ & 15.4 & -84.1 & 12.0 & 47.6 & 12.1 & 30.4 & 3.71 & 0.831 & 0.111 & 0.847\\
C10&28.3649 & 0.0840 & $2.74\times10^{2}$ & 7.92 & 21.2 & 10.0 & -3.49 & 10.2 & -49.5 & 13.2 & 8.08 & 3.77 & 7.24\\
C11&28.3760 & 0.0528 & $4.80\times10^{2}$ & 9.89 & 15.0 & 12.9 & 22.0 & 13.0 & -17.4 & 13.9 & 5.67 & 2.70 & 5.02\\
C12&28.3872 & 0.0613 & $1.10\times10^{2}$ & 9.36 & 42.6 & 11.0 & 17.3 & 11.0 & -33.8 & 6.68 & 4.30 & 1.03 & 4.23\\
C13&28.3332 & 0.0590 & $6.58\times10^{2}$ & 12.2 & 40.8 & 15.9 & -1.45 & 16.3 & -46.2 & 11.3 & 6.28 & 2.43 & 5.84\\
C14&28.3340& 0.0538 & $4.33\times10^{2}$ & 11.5 & 36.5 & 14.9 & 4.24 & 15.2 & -41.9 & 12.5 & 8.66 & 3.57 & 7.93\\
C15&28.3342& 0.0637 & $4.34\times10^{2}$ & 11.5 & 35.7 & 14.9 & 3.13 & 15.1 & -42.8 & 12.2 & 8.50 & 3.60 & 7.76\\
C16&28.3299& 0.0672 & $3.69\times10^{2}$ & 11.4 & 30.8 & 15.2 & 13.7 & 15.4 & -32.7 & 13.7 & 9.10 & 4.15 & 8.12\\
\hline
\end{tabular}%
\tablecomments{$^{\dagger}$\citet{2020ApJ...897..136M}. $^{*}$If the error in the polarization fraction is larger than the polarization fraction, then the debiasing fail, and the debiased polarization will be flagged as $0\%$.}
\end{table*}


In this paper, we focus on analysis of the orientation of the polarization vectors. We defer analysis of the degree of polarization, i.e., polarization fraction, and the implications for dust grain alignment mechanisms and size distributions for a future paper in this series. Nevertheless, we note that the map of $P^\prime$ (see Appendix~\ref{sec:AppA}) shows typical values from a few to $\sim 10\%$ in the main spine of the IRDC, but with the values dropping to $\sim 0.5\%$ toward the brightest MIR region. Furthermore, in Table~\ref{tab:summary_data_table} we list the values of the derived Stokes parameters $I$, $Q$, $U$, magnetic field position angles ($\theta$), and degree of polarization ($P$ and $P^\prime$) for selected regions in the IRDC, i.e., the massive protostar \citep[Cp23][]{2020ApJ...897..136M}, which corresponds to the brightest FIR source, and 16 massive starless/early-stage core/clumps from the sample of \citet{2014ApJ...782L..30B} (the locations of these sources are shown in Fig.~\ref{fig:cloudC-fragmentation}). For each of these sources, the quantities have been evaluated within a circular aperture of radius of $9^{\prime\prime}$, i.e., at the resolution of the 214~$\rm \mu m$ HAWC+ image. The polarization percentages shown in Table~\ref{tab:summary_data_table} follow what was noted above, i.e., the bright protostellar source has $P^\prime\simeq0.5\%$, while the starless/early-stage cores/clumps generally have larger values, typically from a few to $\sim 10\%$. We return to the discussion of the dynamics of these sources in \S\ref{sec:energy_cal}.

\begin{figure}
\includegraphics[width=\columnwidth]{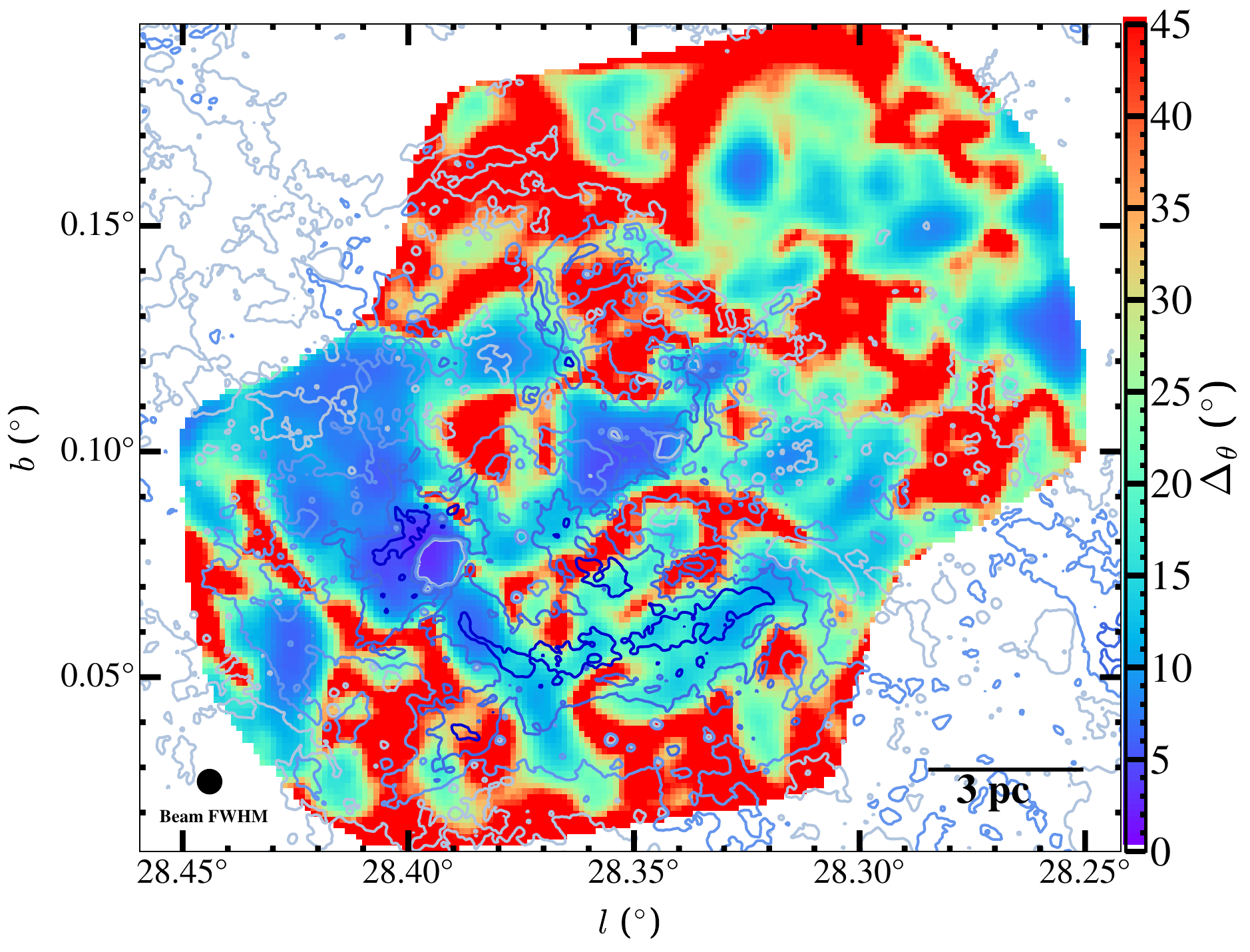}
\caption{Map of polarization angle uncertainties toward G28.37+0.07 (note scale is truncated at 45$^\circ$). Generally, the typical uncertainties in the main spine of the IRDC range from a few to 30$^\circ$.}
\label{fig:angle_error}
\end{figure}

When considering the derived orientations of the polarization vectors, it is important to note the uncertainties in $\theta$. Figure~\ref{fig:angle_error} presents a map of these uncertainties. These are also presented in Table~\ref{tab:summary_data_table} for the example sources, which range from a few to $\sim30^\circ$. Such values are also typical in the main spine of the IRDC, but larger uncertainties are found in regions away from the cloud that are relatively faint in their FIR emission.

\subsection{GBT C$^{18}$O(1-0) spectrum and moment maps}

\begin{figure}
\includegraphics[width=\columnwidth]{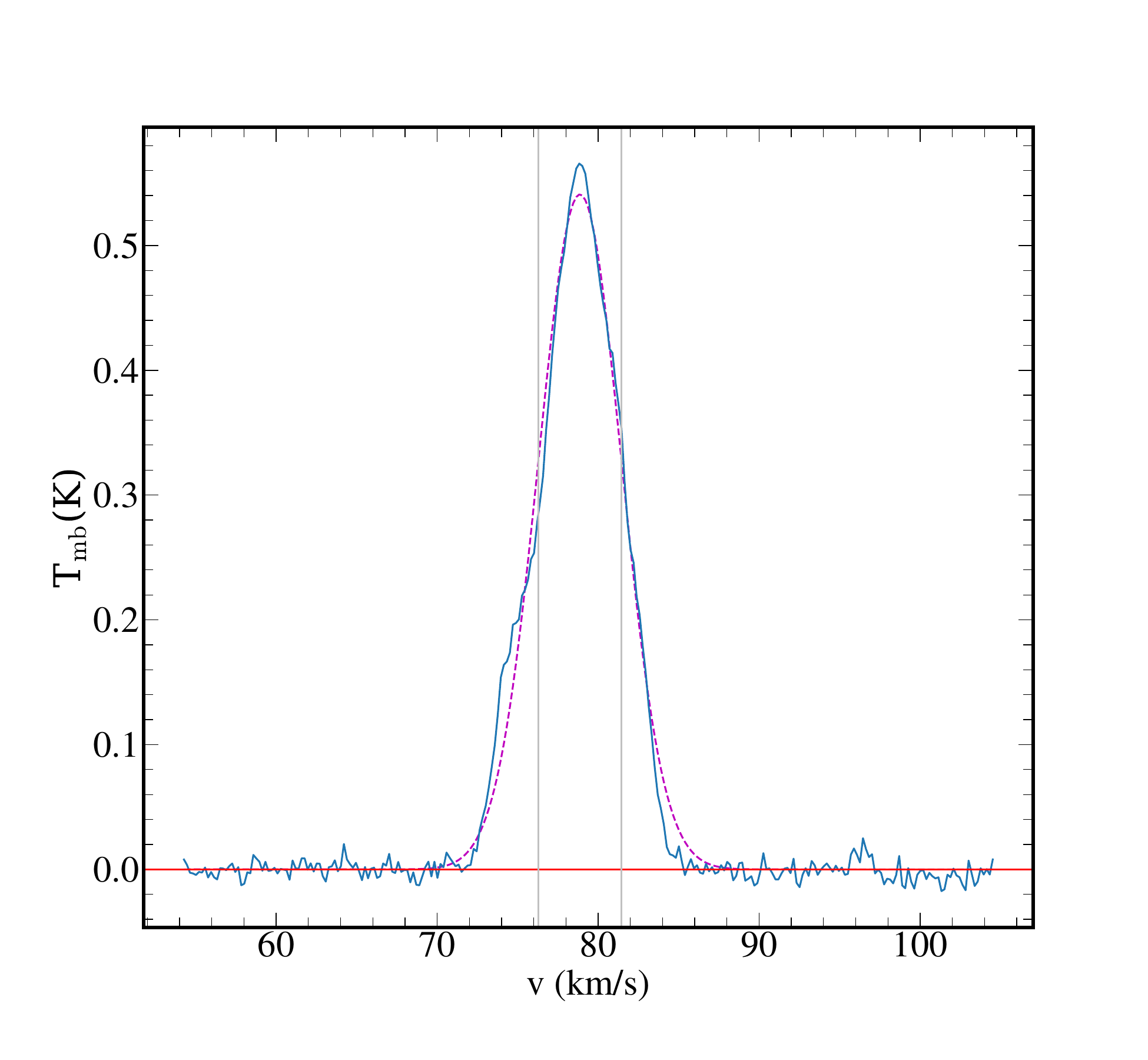}
\caption{GBT C$^{18}$O(1-0) spectrum averaged over the ellipse in Figure~\ref{fig:Figure1_paper}. Based on this spectrum, we set the velocity range of 70 to 86~km/s to construct the moment maps (see text). The average velocity centroid of 78.9~km/s and the total 1D velocity dispersion, i.e., $\pm\sigma_{v}$ with $\sigma_{v}=2.58\:$km/s, are shown by the vertical grey lines.}
\label{fig:summary:C18O_spectrum}
\end{figure}

In Figure~\ref{fig:summary:C18O_spectrum}, we present the average spectrum of C$^{18}$O(1-0) emission observed by the GBT extracted from the IRDC ellipse region (see Fig.~\ref{fig:Figure1_paper}). This spectrum shows a single main emission feature extending from about 70 to 86~km/s and with a central peak near 79~km/s, which is the well-established value for dense gas in G28.37+0.07 \citep[e.g.,][]{2013ApJ...779...96T,2015ApJ...809..154H}. 
A simple Gaussian fit to this feature results in an average velocity centroid of $78.9\,$km\,s$^{-1}$ and a total 1-D velocity dispersion of $\sigma_{v}=2.58\,$km\,s$^{-1}$.

\begin{figure*}
\centering
\begin{minipage}{0.49\textwidth}
\includegraphics[width=\textwidth]{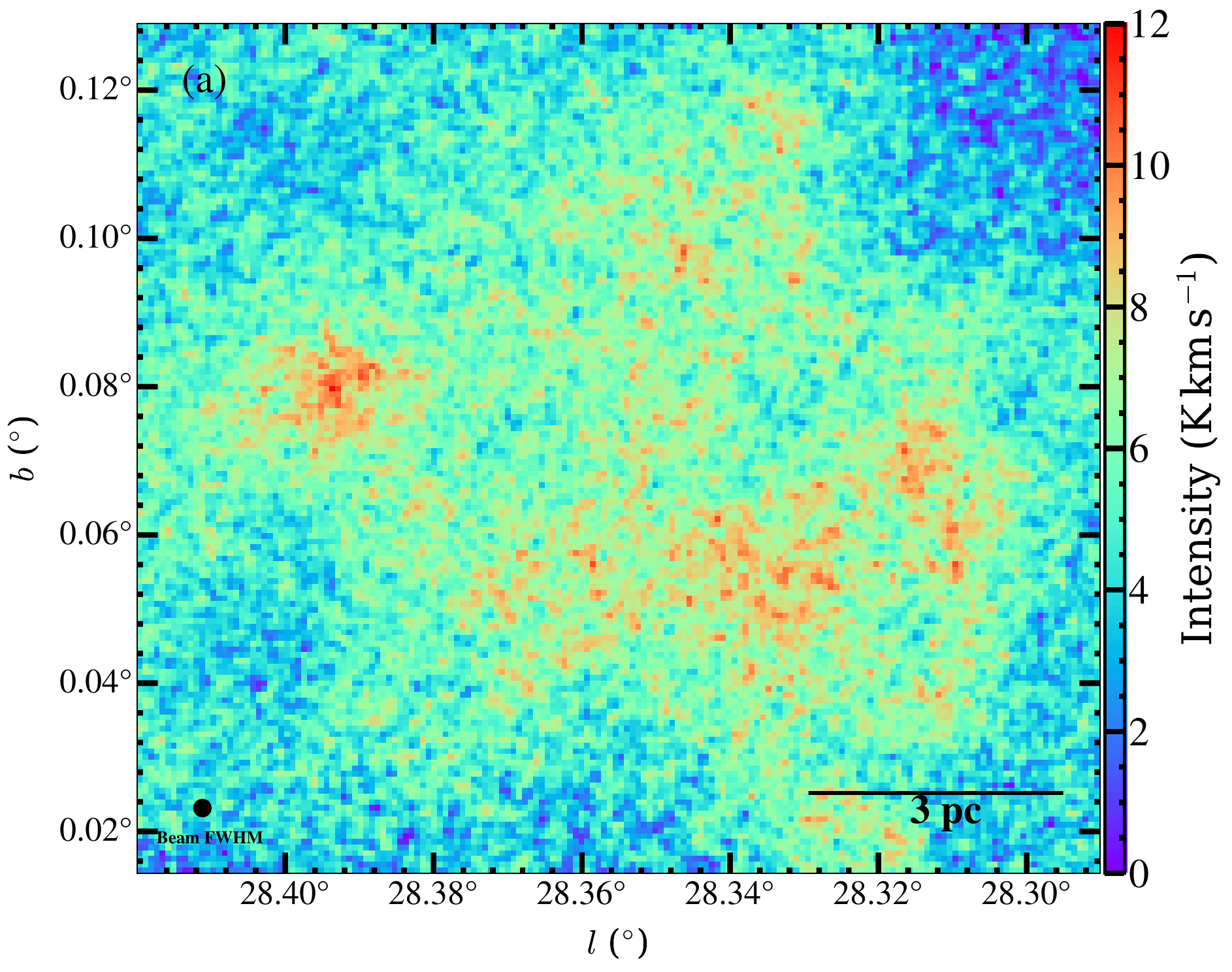}
\end{minipage}
\begin{minipage}{0.49\textwidth}
\includegraphics[width=\textwidth]{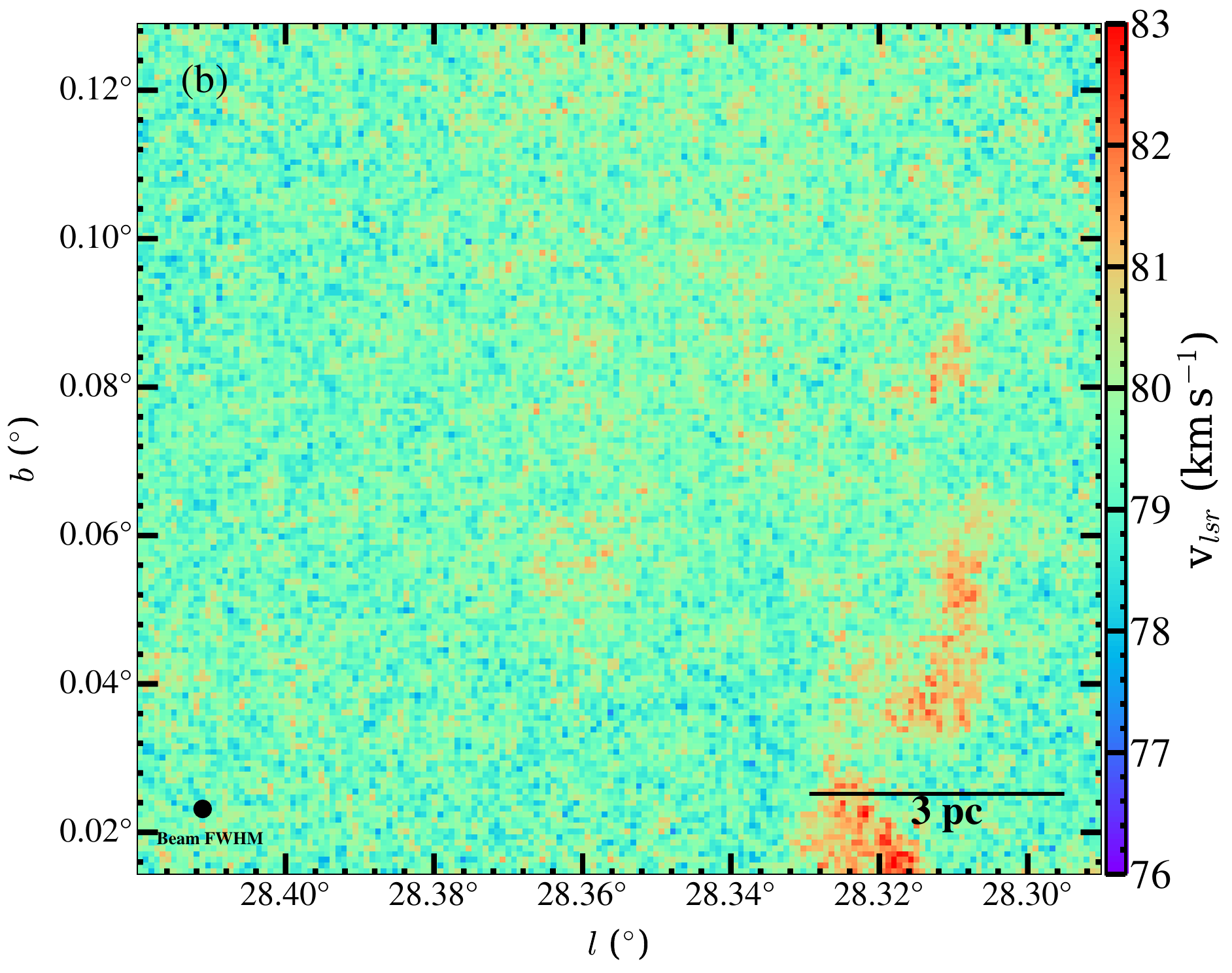}
\end{minipage}
\begin{minipage}{0.49\textwidth}
\includegraphics[width=\textwidth]{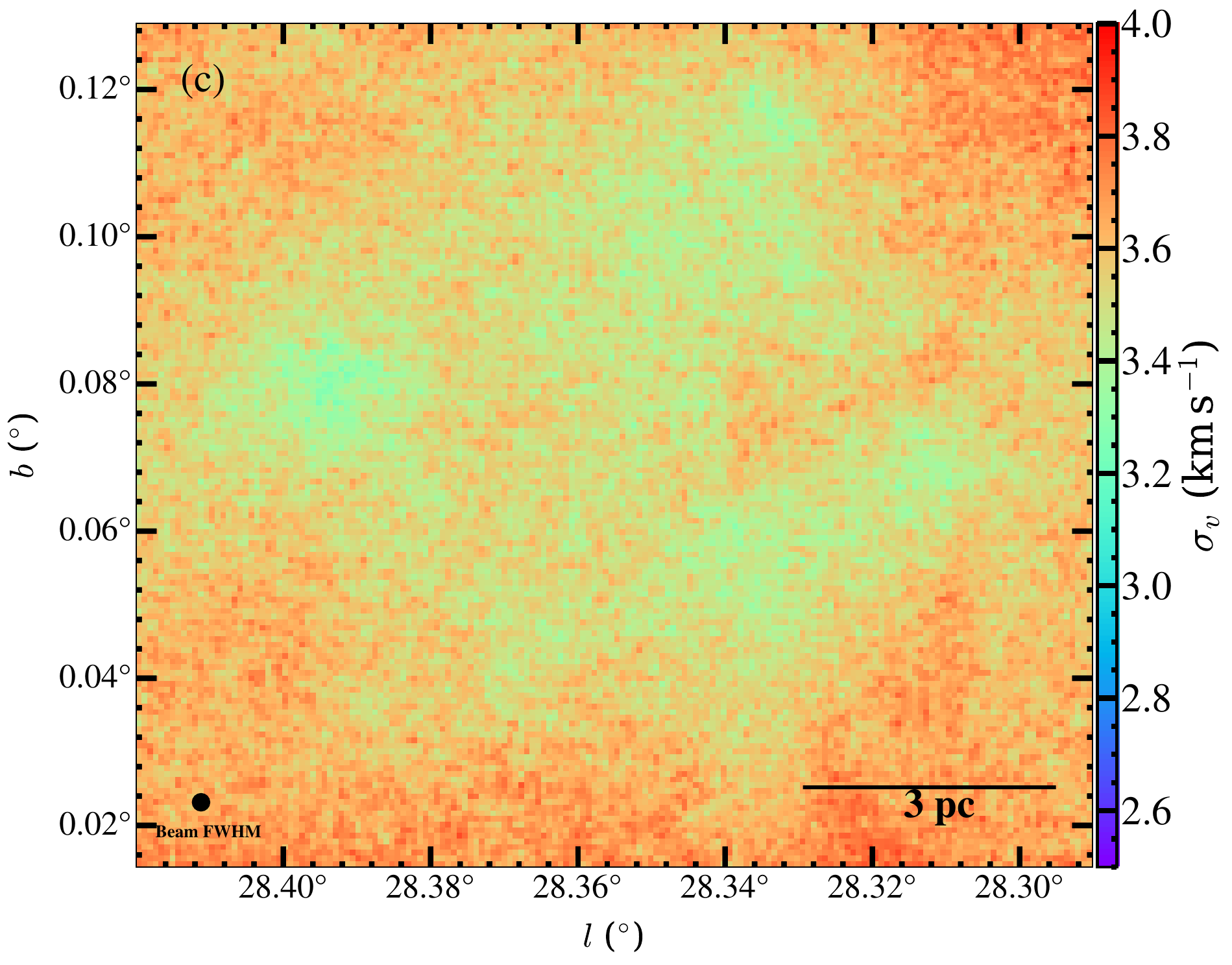}
\end{minipage}
\begin{minipage}{0.49\textwidth}
\includegraphics[width=\textwidth]{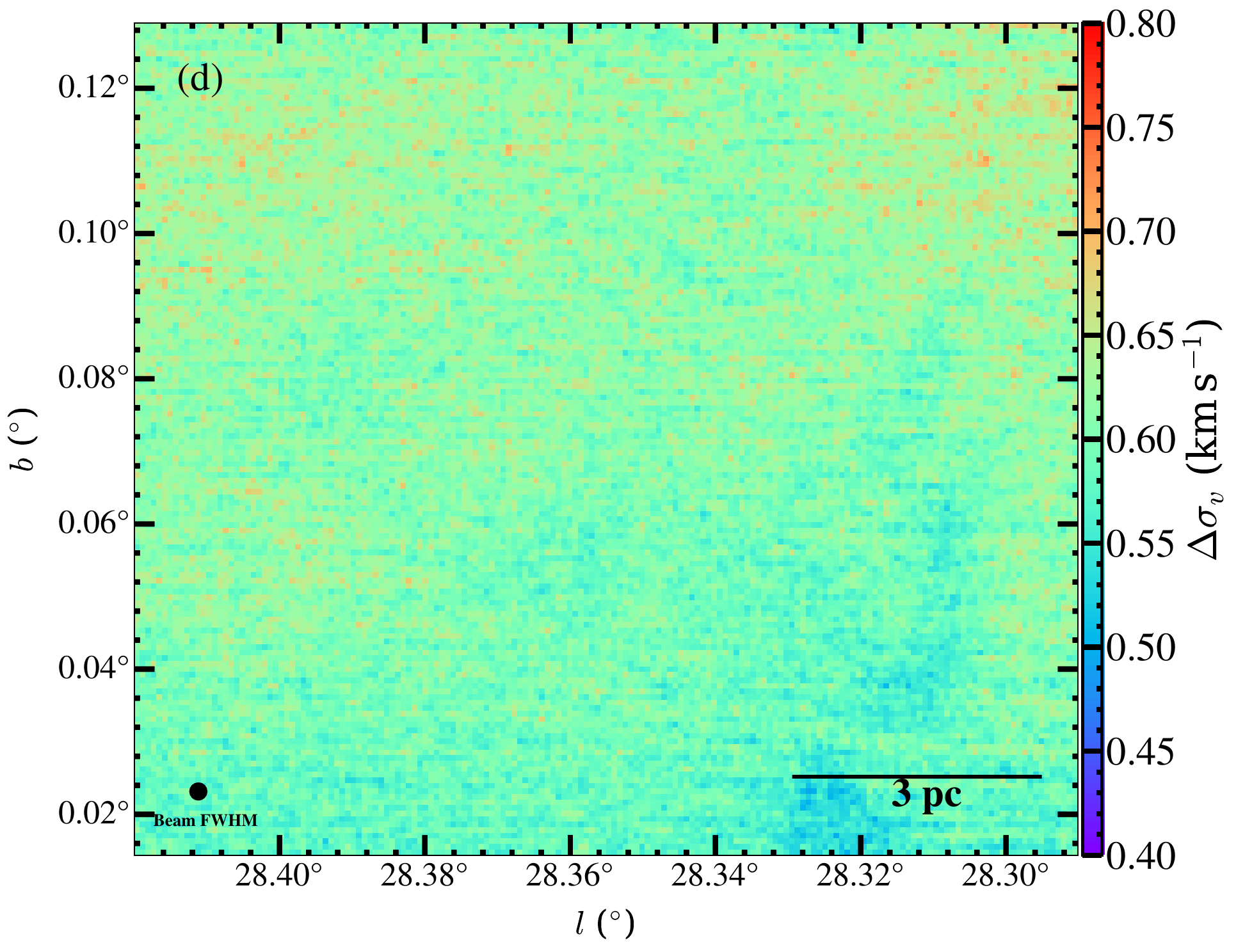}
\end{minipage}
\caption{{\it (a) Top left:} Integrated intensity map of C$^{18}$O(1-0) emission of IRDC G28.37+0.07, evaluated from 70 to 86 km$\,$s$^{-1}$. 
{\it (b) Top right:} As (a), but now showing the first moment map.
{\it (c) Bottom left}: As (a), but now showing the second moment map.
{\it (d) Bottom right:} As (a), but now showing the uncertainty in the second moment map.
}
\label{fig:GBT_cloudC}
\end{figure*}

In Figure~\ref{fig:GBT_cloudC} we present the zero, first and second moment maps of C$^{18}$O(1-0) emission of G28.37+0.07, evaluated over the velocity range from 70 to 86~km/s. Peak values of integrated intensity reach values of about 12~K~km/s. However, there is significant emission at levels of $\sim 6\:$K~km/s over most of the mapped region. Over most of the field of view, the first moment map shows quite constant values between about 78 and 80~km/s. However, some localized, higher-velocity features are seen in the SW region of the map. The second moment map shows values of just over 3~km/s in the dense regions of the IRDC, with moderately larger values in the surroundings. Finally, Figure~\ref{fig:GBT_cloudC}d presents a map of the uncertainty of the velocity dispersion, estimated utilizing the {\it bettermoments} python package \citep{2019RNAAS...3...74T}.
We see it takes typical values of about 0.5~km/s, i.e., about a 15\% uncertainty.

\subsection{Mapping Magnetic Field Strength}\label{sec:field_strength}



We construct maps of magnetic field strength of G28.37+0.07 using the data presented in the previous sections. We first use the Davis–Chandrasekhar–Fermi (DCF) method \citep{1951PhRv...81..890D,1951ApJ...114..206D,1953ApJ...118..113C} to estimate the plane of sky component of the magnetic field strength. The DCF method estimates magnetic field strength via:
\begin{equation}
\label{eqn:classic_DCF}
B_{\rm 2D}=Q \sqrt{4 \pi \rho} \frac{\sigma_{v}}{\sigma_\theta},
\end{equation}
where $Q$ is a factor related to the geometry \citep[usually set equal to 0.5]{2001ApJ...561..800H,2001ApJ...559.1005P,2001ApJ...546..980O}, $\rho$ is the gas density, $\sigma_{v}$ is the 1-D velocity dispersion, and $\sigma_\theta$ is the dispersion in polarization position angles in radians.


A modified version of the DCF method has been proposed by \citet{2021A&A...647A.186S} (ST), which we will refer to as ``refined'' DCF (r-DCF). Here the magnetic field strength is estimated via:
\begin{equation}
\label{eqn:refined_DCF}
B_{\rm 2D}=\sqrt{2 \pi \rho} \frac{\sigma_{v}}{\sqrt{\sigma_\theta}}.
\end{equation}
In general the r-DCF method is expected to be more accurate in regions sub- to trans-Alfvénic turbulence, where it derives lower magnetic field strengths than the DCF method. We also note that from the above equations, the two methods will give the same estimate of field strength (for the same input physical conditions and with $Q=0.5$) when $\sigma_\theta=0.5$~radians, i.e., $28.6^\circ$.

The procedure to construct these magnetic field strength maps is as follows. We first re-grid the original Stokes parameter maps into a coarser pixel scale with a size that corresponds to the SOFIA band-E beam FWHM $=18.2^{\prime\prime}$ so that the measurements in each pixel are spatially independent of each other. We then compute the polarization angle quantities, including the polarization position angles and the rotated angles (i.e., magnetic field directions) from these re-gridded maps. We then measure $\sigma_\theta$ from these maps via
\begin{equation}
\label{eqn:ang_disp_pol}
\sigma_\theta=\sqrt{\left[\frac{1}{N} \sum_{i=1}^N\left(90-\mid(90-\mid(\theta_i-\langle\theta\rangle\right)\mid)\mid)^2\right]},
\end{equation}
where the summation is done over a set of measured polarization angles $\theta_i$ 
and $\langle\theta\rangle$ is the mean polarization orientation within the region of consideration. The expression inside involves the absolute values of angle difference to ensure the angles remain in the range -90$^\circ$ to +90$^{\circ}$ \citep{2019A&A...622A.145P}.  

To construct the highest possible resolution $B$-field strength map we work with a $2\times2$ pixel grid (i.e., $36.^{\prime\prime}4 \times 36.^{\prime\prime}4$) as an input window in which to evaluate $\sigma_\theta$. This window is then translated one pixel at a time for the next evaluation. Thus most pixels in the image, except for those at the edge, receive four measurements of $\sigma_\theta$, with the final reported value being the average of these measurements. We also carry out a version of our analysis with a $3\times3$ pixel grid window, with these results presented in Appendix~\ref{sec:AppC}.



\begin{figure*}[h!]
\begin{minipage}{0.49\textwidth}
\includegraphics[width=\textwidth]{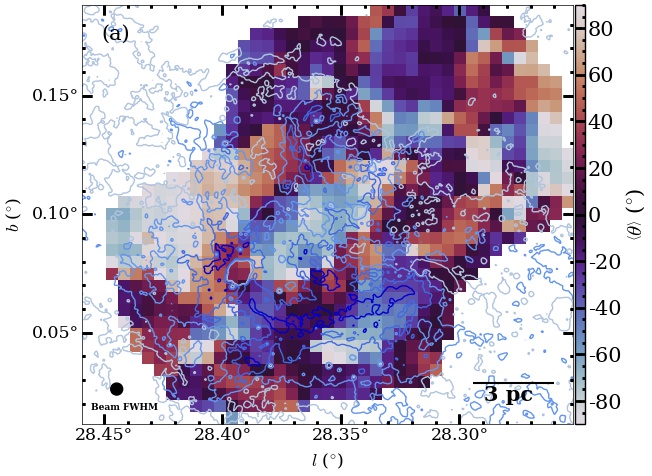}
\end{minipage}
\begin{minipage}{0.49\textwidth}
\includegraphics[width=\textwidth]{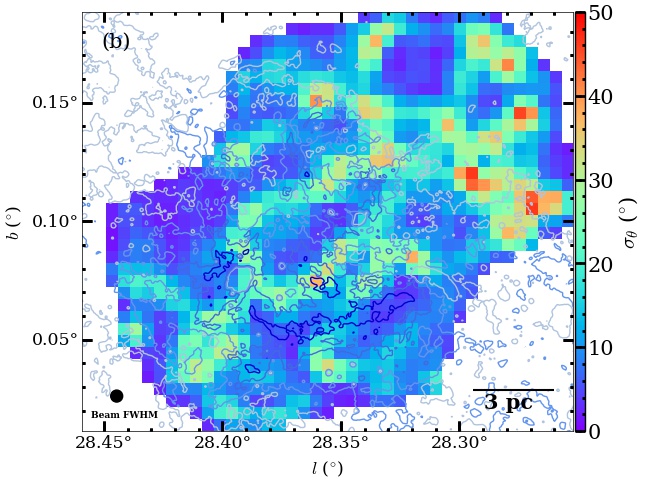}
\end{minipage}
\begin{minipage}{0.49\textwidth}
\includegraphics[width=\textwidth]{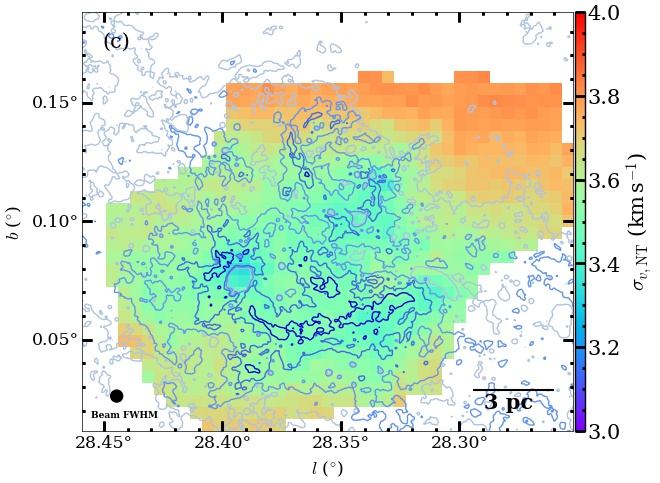}
\end{minipage}
\begin{minipage}{0.49\textwidth}
\includegraphics[width=\textwidth]{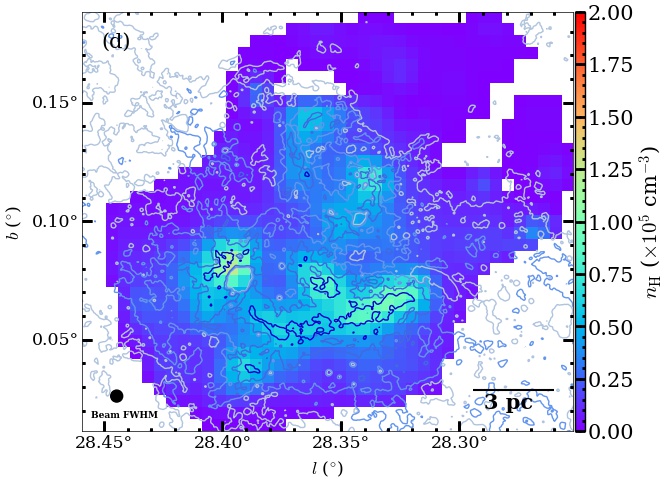}
\end{minipage}
\caption{{\it (a) Top Left:} Map of mean magnetic field position angle, $\langle\theta\rangle$.
{\it (b) Top Right:} Map of dispersion in magnetic field position angle, $\sigma_\theta$. 
{\it (c) Bottom Left:} Map of 1-D non-thermal velocity dispersion, $\sigma_v$.
{\it (d) Bottom Right:} Map of mass-weighted H nuclei number density, $n_{\rm H}$.
}
\label{figure:summary_B_2b2}
\end{figure*}

The map of the mean polarization angle, $\langle\theta\rangle$, is presented in Figure~\ref{figure:summary_B_2b2}a and the map of polarization angle dispersion in Figure~\ref{figure:summary_B_2b2}b. We notice a typical angle dispersion toward the main spine of the IRDC of a few degrees to $\sim 15^\circ$. We also note an increase in the angle dispersion in the midpoint of the main spine, which is where a global change of field orientation from mainly perpendicular to mainly parallel to the filament is seen.
A few thin ``lines'' of higher dispersion of $\sim25^\circ$ are consistent with boundaries between regions that have particular orientations of field directions.




To measure $\sigma_v$ we utilize the C$^{18}$O(1-0) data shown in Figure~\ref{fig:GBT_cloudC}c, but with a small correction to subtract off, in quadrature, the thermal broadening of the line. For this we utilize the dust temperature map of \citet{2016ApJ...829L..19L}, assuming gas and dust are at similar temperatures. However, it should be noted that this correction is a very minor effect, given the observed values of velocity dispersion of $\gtrsim 3\:$km/s and the cold temperatures ($\lesssim 15\:$K) of the IRDC.



To estimate the density of the region we utilize the mass-weighted density map of \citep{2023arXiv230401670X}, which is derived from the mass surface density map via machine learning methods, with the algorithm trained on a set of numerical simulations of magnetized GMCs. The density map is shown in Figure~\ref{figure:summary_B_2b2}d.



\begin{figure*}[]
\begin{minipage}{0.49\textwidth}
\includegraphics[width=\textwidth]{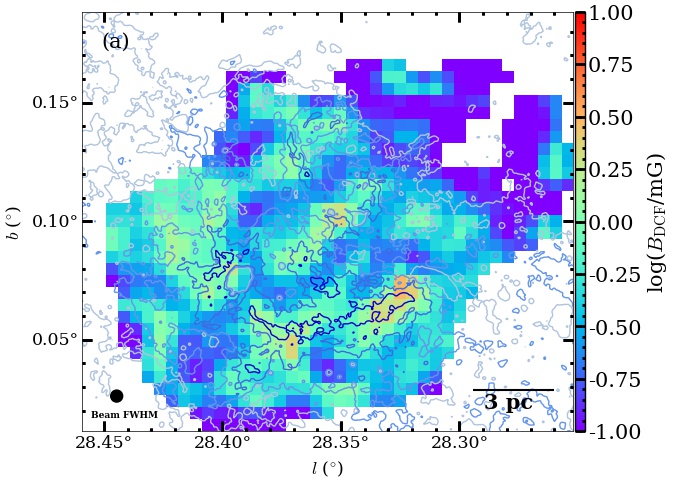}
\end{minipage}
\begin{minipage}{0.49\textwidth}
\includegraphics[width=\textwidth]{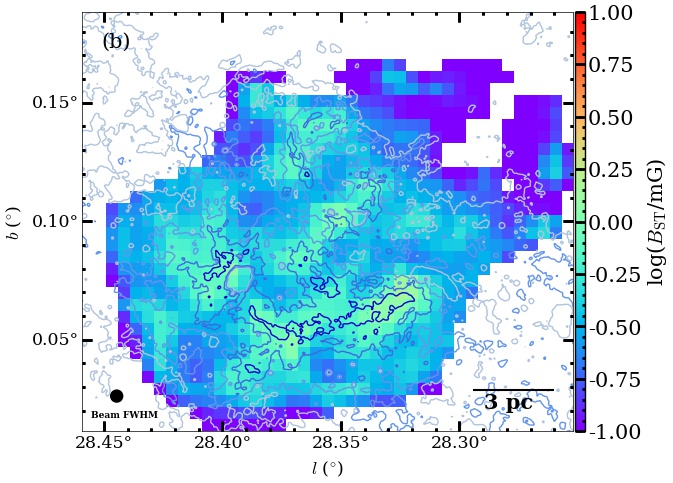}
\end{minipage}
\begin{minipage}{0.49\textwidth}
\includegraphics[width=\textwidth]{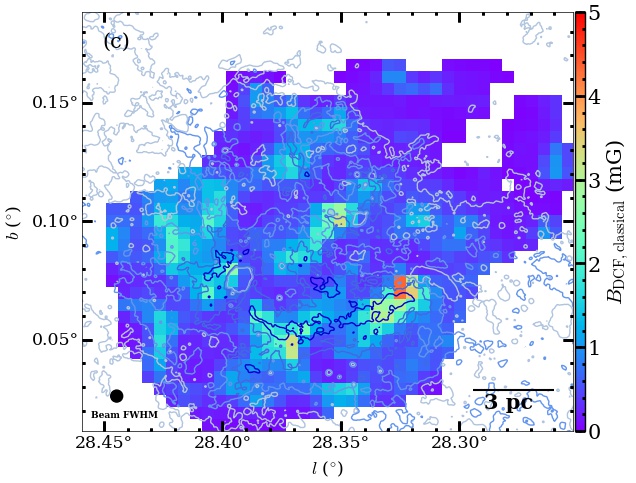}
\end{minipage}
\begin{minipage}{0.49\textwidth}
\includegraphics[width=\textwidth]{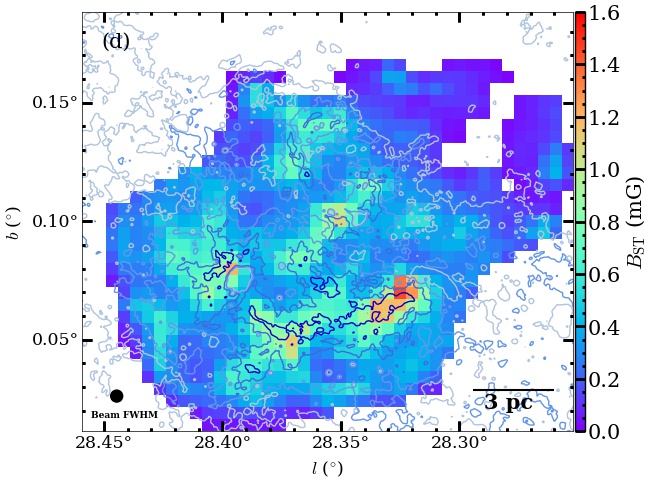}
\end{minipage}
\caption{Plane of sky magnetic field strength maps of G28.37+0.07 using the DCF and r-DCF (or ST) methods (Left column: DCF; right column: r-DCF). The top row shows the field strength maps with a logarithmic scale, while the bottom row shows the same maps but in linear scale. 
}
\label{figure:strength}
\end{figure*}

Combining the above information, we then create two versions of the magnetic field strength map, i.e., one based on the DCF (Eq.~\ref{eqn:classic_DCF}) and one based on the r-DCF (Eq.~\ref{eqn:refined_DCF}). Logarithmically and linearly scaled versions of these maps are shown in Figure~\ref{figure:strength}. We note that the resolution of the maps is $36.4^{\prime\prime}$, i.e., $0.85\:$pc, set by the 2 x 2 pixel grid size used for the $\sigma_\theta$ measurement. As expected, the DCF method returns higher $B$-field strength estimates than the r-DCF method. In the DCF $B$-field strengths map we see $B$-field strengths up to about 3~mG in localized regions, especially at the western end of the main filament of the IRDC. However, we regard the r-DCF map as being more accurate. Here the peak $B$-field strengths are about 1.5~mG at the western end of the IRDC. Going along the main spine to the east, these values decrease to be $\sim 0.5\:$mG.

\begin{figure}[]
\includegraphics[width=\columnwidth]{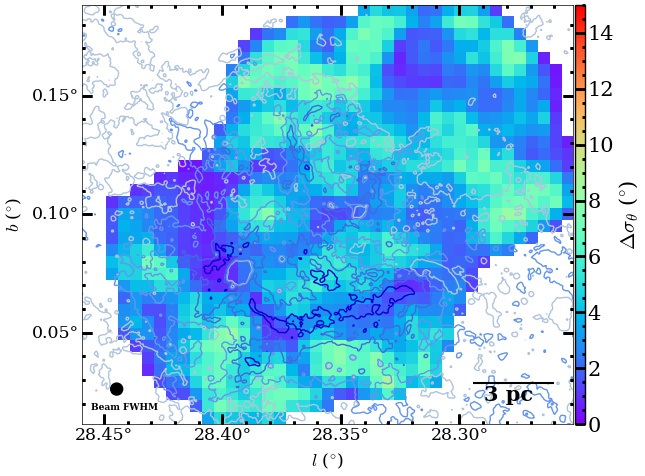}
\caption{Map of uncertainty in $\sigma_{\theta}$, i.e., $\Delta \sigma_{\theta}$ (see text).
}
\label{figure:ang_dispersion_error}
\end{figure}

\begin{figure*}[]
\begin{minipage}{0.49\textwidth}
\includegraphics[width=\textwidth]{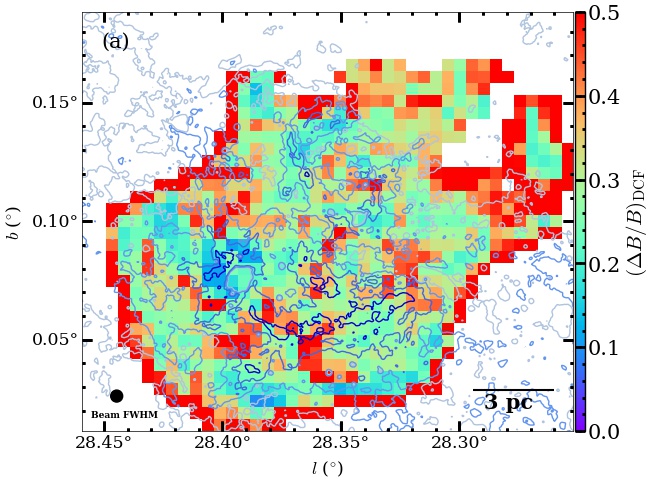}
\end{minipage}
\begin{minipage}{0.49\textwidth}
\includegraphics[width=\textwidth]{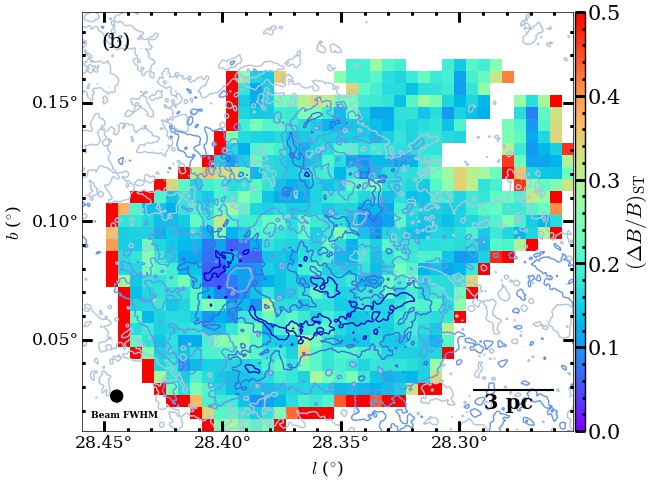}
\end{minipage}
\caption{{\it (a) Left:} Map of fractional uncertainty in plane of sky magnetic field strength as estimated via the DCF method. {\it (b) Right:} As (a), but now for the r-DCF method.}
\label{figure:summary_B_error}
\end{figure*}

The uncertainties in magnetic field strength are computed given estimated errors for the density, velocity dispersion, and polarization angle dispersion. Uncertainties in velocity dispersion over a 2 x 2 pixel grid are relatively small, i.e., about 5-10\%, and so are negligible in the final uncertainty of field strength. For the uncertainty in density, based on the results of \citet{2023arXiv230401670X}, we adopt $\Delta n_{\rm H}/n_{\rm H} = 0.5$. We estimate the uncertainty in $\sigma_{\theta}$ by sampling the individual polarization orientations within their estimated errors, running our algorithms on these maps, and then examining the dispersion in the resulting derived values of $\sigma_{\theta}$. A map of the uncertainty in $\sigma_{\theta}$, i.e., $\Delta \sigma_{\theta}$, is shown in Figure~\ref{figure:ang_dispersion_error}. Maps of the overall uncertainties of $B$-field strength via the DCF and r-DCF methods are shown in Figure~\ref{figure:summary_B_error}. These have typical values of 20\% and 15\% in the IRDC for the DCF and r-DCF maps, respectively.

We also note that observational uncertainties in polarization position angle tend to inflate the measured dispersion in these angles, thus leading to a systematic bias of underestimation of $B$-field strengths. These uncertainties then also set a minimum value for $\sigma_{\theta}$, and thus a maximum value for $B$-field strength that can be measured in a given region \citep[see also][]{2017ApJ...846..122P,2023A&A...673A..76S}.

\subsection{The $B-N$ and $B-n$ relations}

The dependencies of magnetic field strength on mass surface density, $\Sigma$ (or equivalently $N_{\rm H}$), and volume density, $\rho$ (or equivalently $n_{\rm H}$), including both mean and dispersion, are constraints on theoretical models and simulations of molecular clouds as they undergo collapse to form stars. The magnetic field strength estimated by the DCF and r-DCF methods is the plane of sky component, i.e., $B_{\rm 2D}$. To make comparison with Zeeman-based relations that measure 1D line of sight magnetic field components \citep[e.g.,][]{2004ApJ...600..279C}, we estimate the 1D field strength to be $B_{\rm 1D}= B_{\rm 2D}/\sqrt{2}$.



\begin{figure*}[]
\begin{minipage}{0.49\textwidth}
\includegraphics[width=\textwidth]{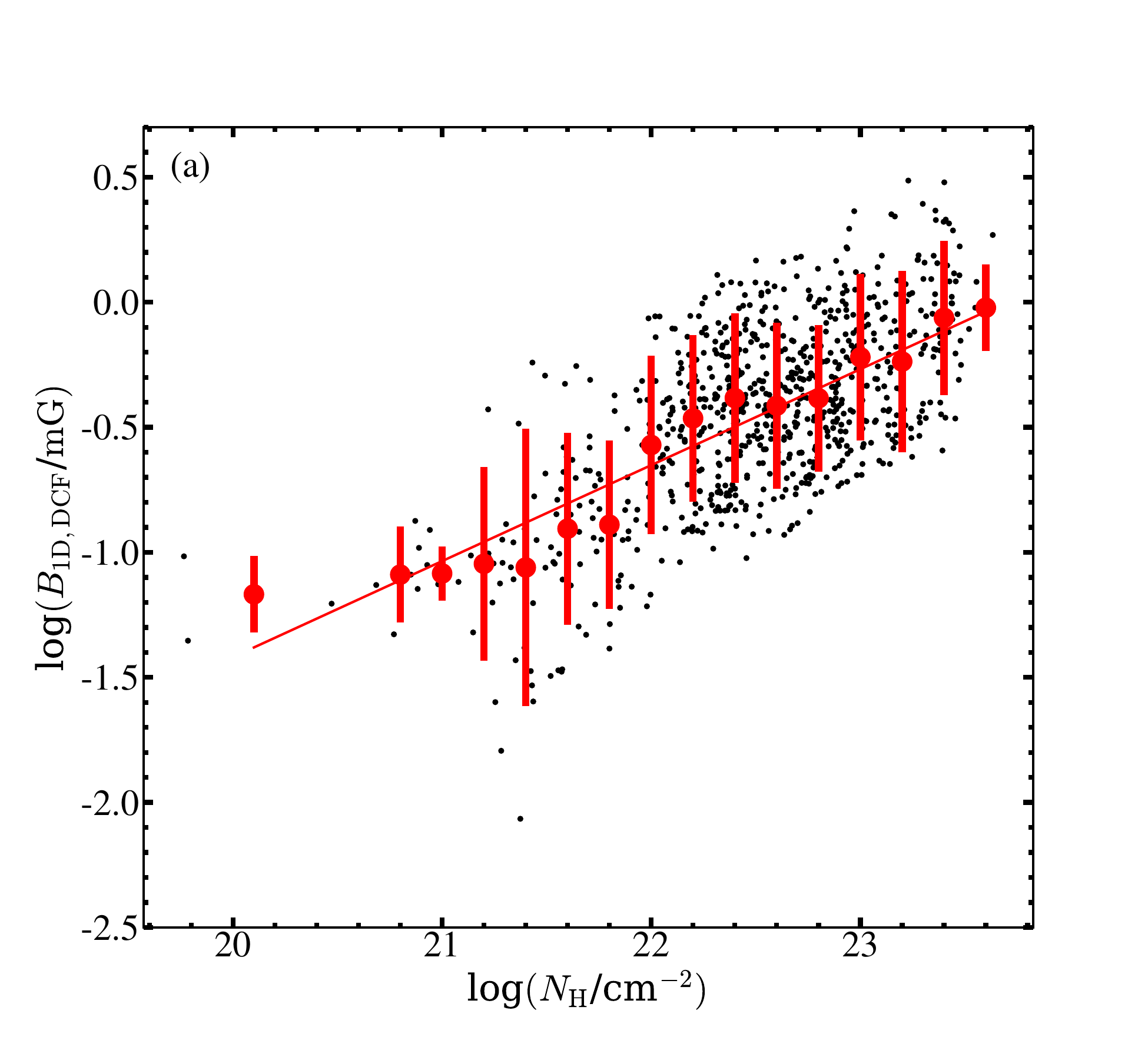}
\end{minipage}
\begin{minipage}{0.49\textwidth}
\includegraphics[width=\textwidth]{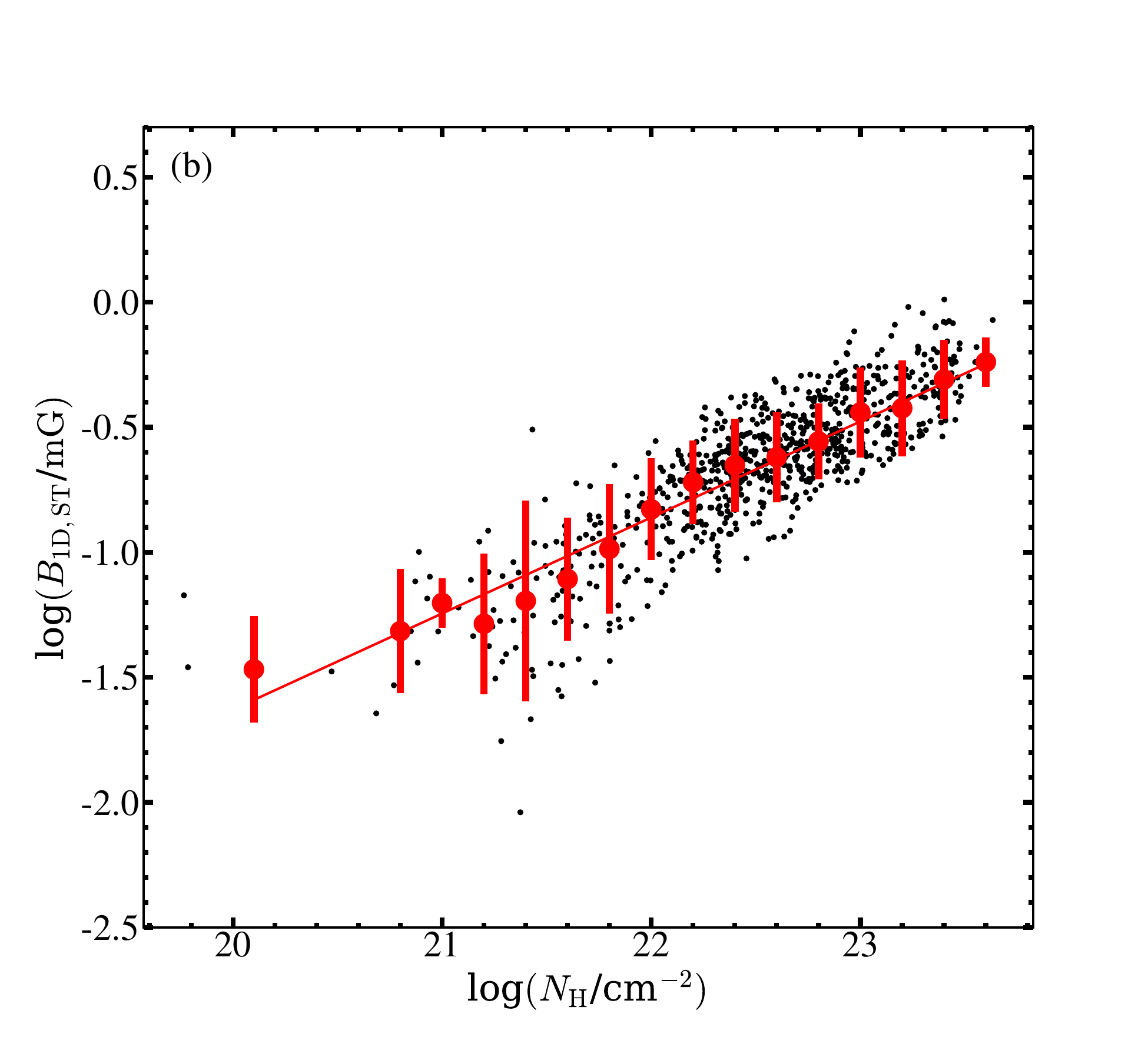}
\end{minipage}
\caption{{\it (a) Left:} $B_{\rm 1D}$ versus $N_{\rm H}$ relation for IRDC G28.37+0.07, with magnetic field estimated via the DCF method and $N_{\rm H}$ derived from the Herschel FIR $\Sigma$ map. Black points show data from all pixels in the maps, while red points show median values of binned data, with error bars showing the dispersion about the average. A power law fit to the data (see text) is shown by the solid line. {\it (b) Right:} As (a), but now for the r-DCF method for magnetic field strength.
}
\label{figure:summary_B_N}
\end{figure*}

Figure~\ref{figure:summary_B_N} presents the DCF and r-DCF $B_{\rm 1D}$ versus $N_{\rm H}$ relations for IRDC G28.37+0.07. Note that here we have measured $N_{\rm H}$ from the Herschel, i.e., FIR, derived $\Sigma$ map. In addition to the full data set shown by the black dots, we also display the median values of binned data, using bin widths of 0.2~dex in $N_{\rm H}$ unless such a bin would contain fewer than five data points (relevant only for the lowest column density bin). We fit a power law relation of the form
\begin{equation}
B_{\rm 1D} = B_{N22} N_{\rm H,22}^{\alpha_{B-N}},
\end{equation}
where $N_{\rm H,22}\equiv N_{\rm H}/10^{22}\:{\rm cm}^{-2}$. For the DCF method we derive $\alpha_{B-N}=0.384\pm0.028$ and $B_{N22} = 0.269 \pm 0.151\:$mG. For the r-DCF method we find $\alpha_{B-N}=0.384\pm0.017$ and $B_{N22} = 0.149 \pm 0.089\:$mG. We note that the values of $\alpha_{B-N}$ derived from the DCF and r-DCF methods are nearly identical.

\begin{figure*}[]
\begin{minipage}{0.49\textwidth}
\includegraphics[width=\textwidth]{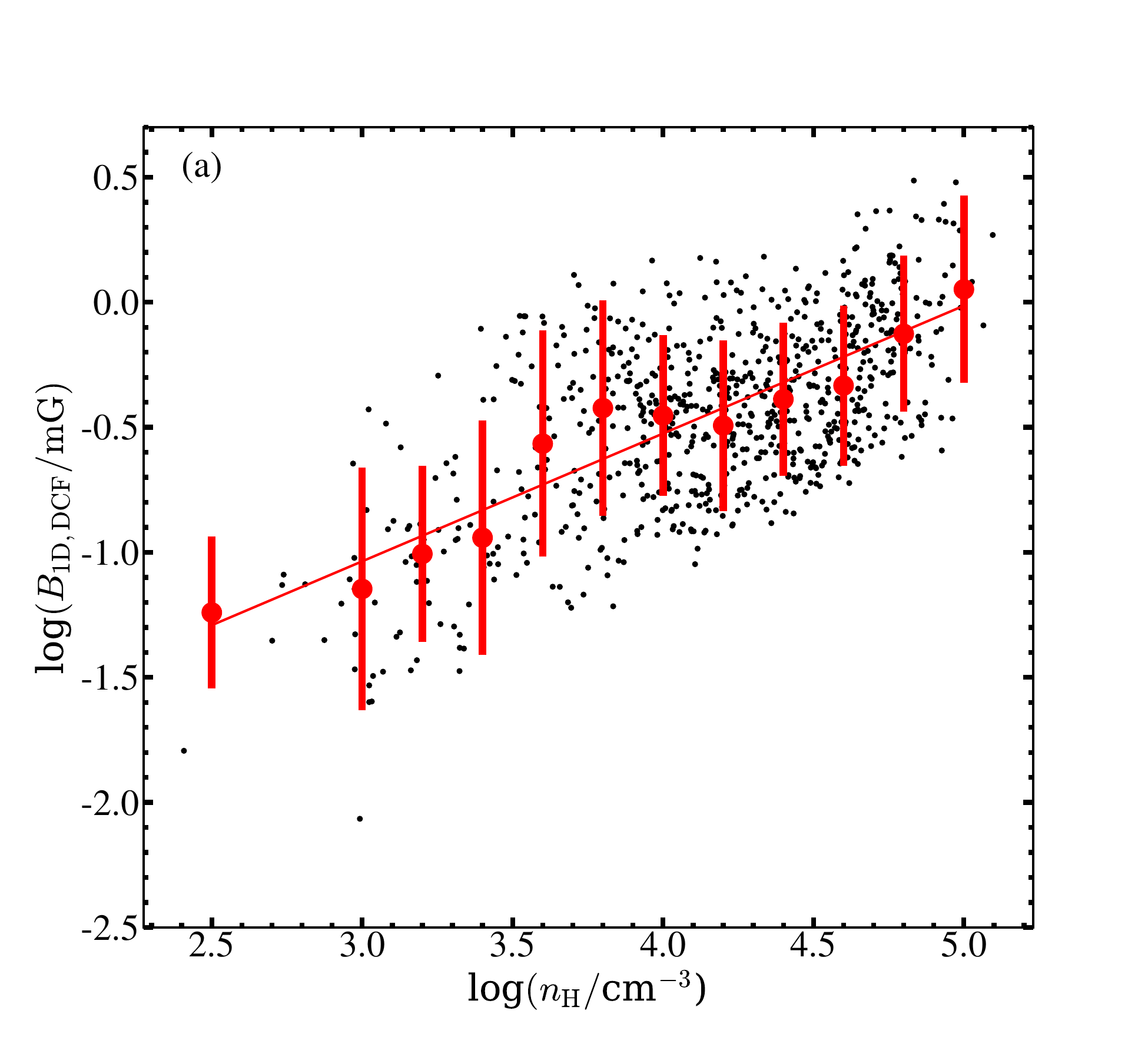}
\end{minipage}
\begin{minipage}{0.49\textwidth}
\includegraphics[width=\textwidth]{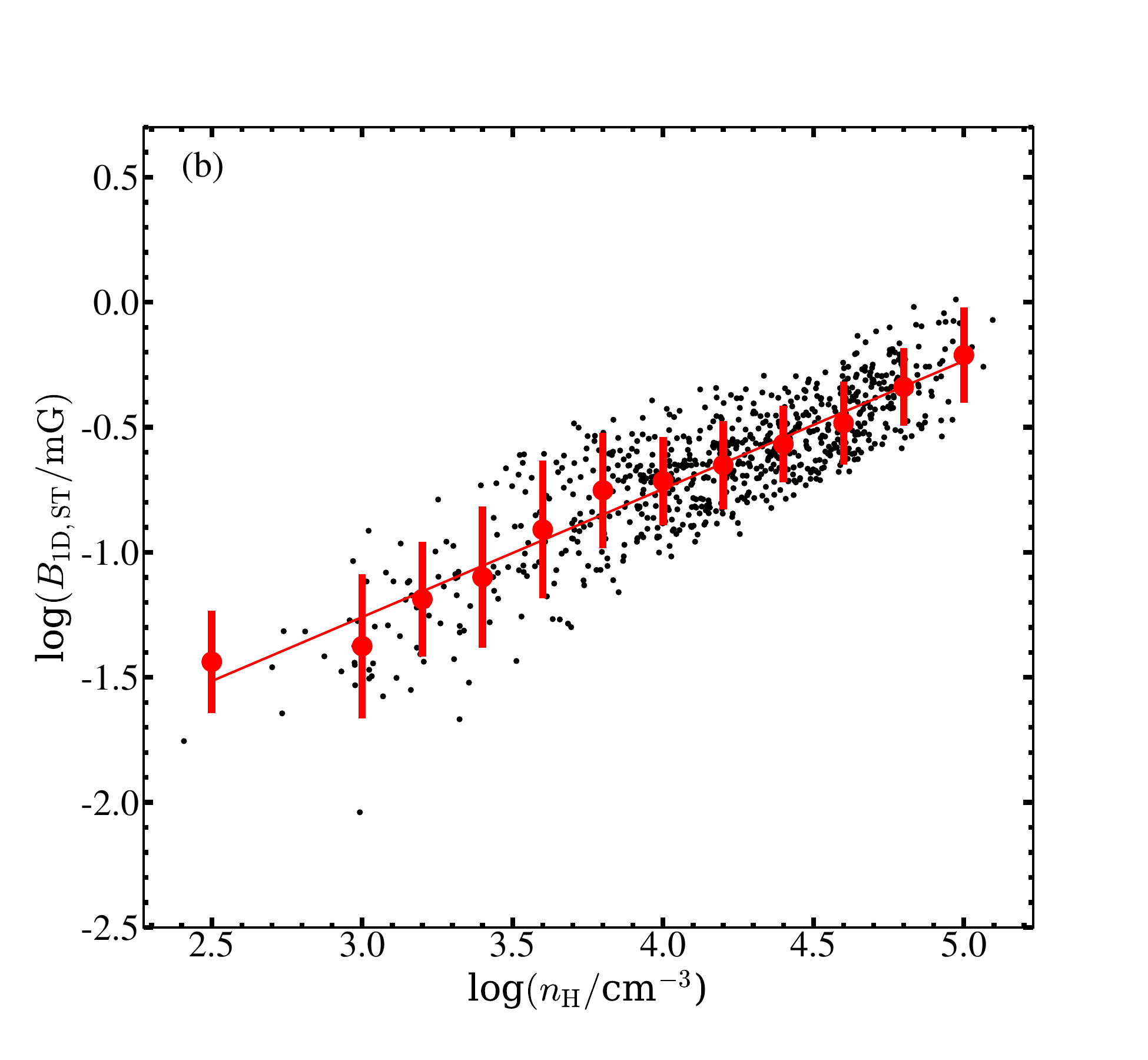}
\end{minipage}
\caption{{\it (a) Left:} $B_{\rm 1D}$ versus $n_{\rm H}$ relation for IRDC G28.37+0.07, with magnetic field estimated via the DCF method and $n_{\rm H}$ estimated as the machine learning inferred value derived from the Herschel, i.e., FIR-based, $\Sigma$ map. Black points show data from all pixels in the maps, while red points show median values of binned data, with error bars showing the dispersion about the average. A power law fit to the data (see text) is shown by the solid line. {\it (b) Right:} As (a), but now for the r-DCF method for magnetic field strength.}
\label{figure:summary_B_n}
\end{figure*}

Figure~\ref{figure:summary_B_n} presents the DCF and r-DCF $B_{\rm 1D}$ versus $n_{\rm H}$ relations for IRDC G28.37+0.07. Note that here we have estimated $n_{\rm H}$ as the machine learning inferred value derived from the Herschel, i.e., FIR-based, $\Sigma$ map. We fit a power law to the binned median values of the form:
\begin{equation}
B_{\rm 1D} = B_{n4} n_{\rm H,4}^{\alpha_{B-n}},
\end{equation}
where $n_{\rm H,4}\equiv n_{\rm H}/10^{4}\:{\rm cm}^{-3}$. For the DCF method we derive $\alpha_{B-n}=0.511\pm0.045$ and $B_{n4} = 0.352 \pm 0.182\:$mG. For the r-DCF method we find $\alpha_{B-n}=0.512\pm0.024$ and $B_{n4} = 0.193 \pm 0.092\:$mG. Again, the DCF and r-DCF methods return very similar values of the power law index, with $\alpha_{B-n}\simeq 0.5$.



The indices of the $B-N$ and $B-n$ relations provide constraints on the relative importance of magnetic fields in cloud contraction and dense core collapse \citep{2010ApJ...725..466C,2015Natur.520..518L,2022A&A...665A..77S}. In the paper by \citet{2010ApJ...725..466C}, the $B-n$ relation was characterized by a piece-wise function in which the magnetic field remains uniform until a threshold density of $n_{\rm H}\sim 300\:$cm$^{-3}$). At higher densities, the field strength increases with an index $\alpha_{B-n}\simeq 2/3$.
Such an index is expected for the case of isotropic collapse where the magnetic field has only a minor role in the cloud dynamics \citep[][]{1966MNRAS.133..265M,2010ApJ...725..466C}. On the other hand, for magnetic field regulated contraction  \citep[e.g., via ambipolar diffusion;][]{1976ApJ...206..753M,1976ApJ...207..141M} a scaling index shallower than 0.67 is expected, as suggested by multiple studies \citep[e.g.,][]{2015MNRAS.451.4384T,2015Natur.520..518L,2023ApJ...946L..46C}. Thus, our results appear to favor a field regulated contraction and collapse.

\section{Dynamical Implications of the Magnetic Field}\label{sec:dynamics}

We have constructed magnetic field strength maps of IRDC G28.37+0.07 based the SOFIA-HAWC+ polarized dust emission data via the DCF and r-DCF (ST) methods combining the GBT-Argus 1-D velocity dispersion map of C$^{18}$O(1-0) and the number density map derived from a machine learning method that is inferred from a Herschel FIR derived $\Sigma$ map\citep{2023arXiv230401670X}. In this section we characterize the dynamical state of IRDC G28.37+0.07, its dense starless/early-stage core/clump population \citep{2014ApJ...782L..30B} and the massive protostellar core Cp23 residing at the MIR/FIR peak position \citep{2020ApJ...897..136M}.

\subsection{Mass-to-flux ratio map}\label{subsec:mass_flux}

We first consider the mass-to-flux ratio ($M/\Phi$) map of IRDC G28.37+0.07 to evaluate the relative importance of the magnetic field to gravity. The equation that computes the corresponding mass-to-flux ratio maps in units of the critical value is \citep{1978PASJ...30..671N}:
\begin{equation}
\label{eqn:mass_to_flux}
\mu_{\Phi}=\frac{(M / \Phi)}{(M / \Phi)_\mathrm{ crit}}=2 \pi \sqrt{G} \left(\frac{\Sigma}{B_{\rm 1D}}\right),
\end{equation}
where $G$ is the gravitational constant, $\Sigma$ is the mass surface density 
and $B_{\rm 1D}$ is the 1-D magnetic field strength (obtained by scaling the plane of sky magnetic field by $1/\sqrt{2}$). $(M / \Phi)_{\rm crit}$ is the critical mass-to-flux ratio, where defined as 
\begin{equation}
\label{eqn:mass_to_flux_crit}
\left(\frac{M}{\Phi}\right)_\mathrm{crit} \equiv \frac{1}{2 \pi \sqrt{G}}.
\end{equation}
If $\mu_{\Phi} > 1$, magnetic fields are not able to support the cloud against gravity. Conversely, if $\mu_{\Phi}<1$ then magnetic fields are strong enough to counter gravitational collapse.

\begin{figure*}[]
\begin{minipage}{0.49\textwidth}
\includegraphics[width=\textwidth]{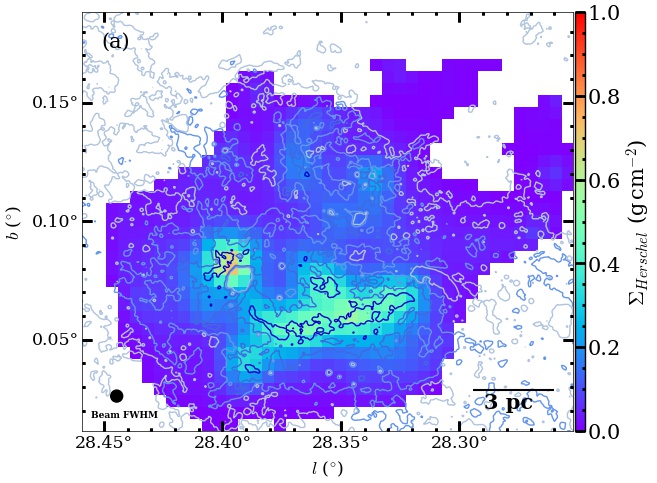}
\end{minipage}
\begin{minipage}{0.49\textwidth}
\includegraphics[width=\columnwidth]{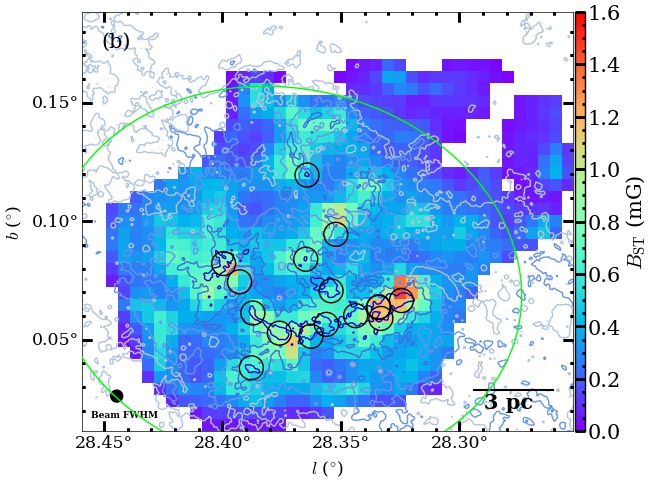}
\end{minipage}
\begin{minipage}{0.49\textwidth}
\includegraphics[width=\textwidth]{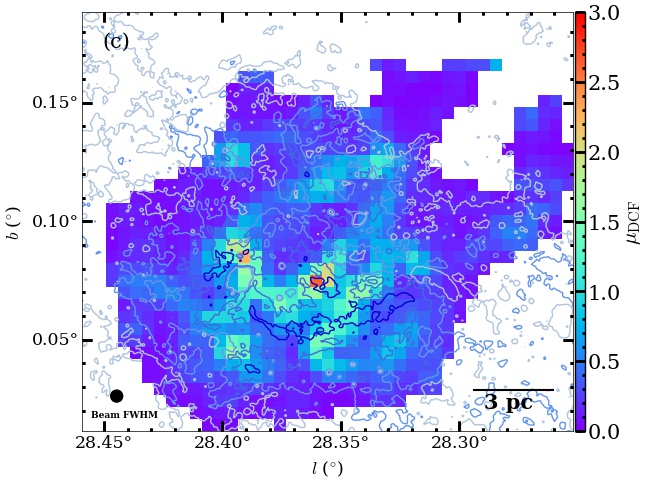}
\end{minipage}
\begin{minipage}{0.49\textwidth}
\includegraphics[width=\textwidth]{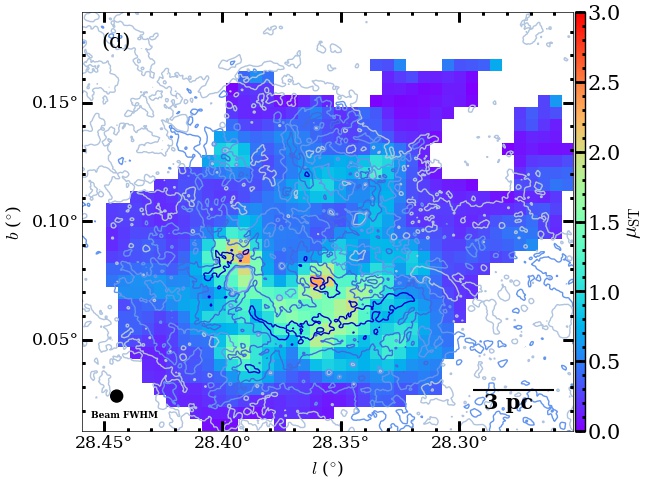}
\end{minipage}

\caption{{\it (a):} Mass surface density in pixel scale equivalent to SOFIA beam FWHM. 
{\it (b):} Dense cores from \citet{2014ApJ...782L..30B} overlaid on the magnetic field strength G28.37+0.07 measured via the ST method. Each circle is centered at the coordinate of each dense core and with a fix aperture radius equivalent to 1 SOFIA-HAWC+ beam FWHM. The ellipse shows the IRDC region defined by \citet{2006ApJ...639..227S}
{\it (c):} Mass to flux ratio ($\mu$) map toward G28.37+0.07 based on magnetic field strength estimated from classical DCF method. 
{\it (d):} As (c), but now showing mass to flux ratio ($\mu$) based on magnetic field strength estimated from refined DCF method \citep{2021A&A...647A.186S}. 
}
\label{figure:mass_flux_ratio}

\end{figure*}

To evaluate the mass-to-flux ratio ($M/\Phi$) conditions in the IRDC, we use our derived magnetic field strength maps (i.e., DCF and r-DCF cases, scaled by $1/\sqrt{2}$) and the FIR-derived mass surface density map. These maps are shown in Figure~\ref{figure:mass_flux_ratio}. As expected, the DCF-based map shows smaller values of $\mu_\Phi$ than the r-DCF-based map. Considering the cloud defined by the ellipse from \citet{2006ApJ...639..227S}, we find an average mass-to-flux ratio of $\sim0.38\pm 0.42$ for DCF and $0.54\pm 0.36$ for r-DCF. Considering the main spine of the IRDC, we notice that whether or not this structure is magnetically supercritical ($\mu_\Phi>1$) or not, depends on whether the DCF or r-DCF method is used. To the north of the main IRDC spine, we see two localized spots and a connecting arc region where there are high, supercritical mass-to-flux ratios.



\subsection{Energetics}\label{sec:energy_cal}


Here, we study the energy budget of the global IRDC and some of its dense substructures to characterise the relative importance of gravity, turbulence, and magnetic fields.

The dynamical state of a spherical cloud is described by the virial theorem. Following \citet{1992ApJ...395..140B} (hereafter BM92), a cloud in virial equilibrium satisfies the following condition:
\begin{equation}
0=2 E_{{K}}+E_{{B}}+E_{{G}},
\label{eq:virialeq}
\end{equation}
where $E_K$, $E_B$, and $E_G$ are the total kinetic, magnetic, and gravitational energies, respectively.
The gravitational energy of an ellipsoidal cloud is given by 
\begin{equation}
E_G=-\frac{3}{5} a_1 a_2 \frac{G M^2}{R},
\end{equation}
where $G$ is the gravitational constant, $M$ is the cloud mass, $R$ is the cloud radius, $a_{1}$ is a parameter that describes the effect of the internal density distribution of the cloud, and $a_{2}$ is a parameter that describes the effect of cloud ellipticity. 
For a power-law density distribution, $\rho \propto r^{-k_{\rho}}$, $a_1=\left(1-k_\rho / 3\right) /\left(1-2 k_\rho / 5\right)$. Following the study of \citet{2012ApJ...754....5B}, we adopt $k_{\rho} =1$ so that $a_{1} = 10/9$.
We will only consider the effects of ellipticity for the case of the global IRDC, which has a ratio of major to minor axes of 1.25. For this case, the value of $a_2=1.05$ (BM92). 


Following BM92, the kinetic energy is defined by
\begin{equation}
E_K \equiv \frac{3}{2} (\rho \sigma_{v}^2 -\rho_0 \sigma_{v,0}^2) V,
\end{equation}
where $\sigma_{v}$ is the total 1-D velocity dispersion, $V$ is the volume of the cloud, and subscript ``0'' refers to ambient conditions around the cloud.
Finally, the magnetic energy is
\begin{equation}
E_{{B}} =\frac{1}{8 \pi} \left(B_{\rm 3D}^2-B_{\rm 3D,0}^2\right) V,
\end{equation}
where $B_{\rm 3D}=\sqrt{3}B_{\rm 1D}$ is the total magnetic field in the cloud.


With our mass surface density, velocity dispersion and magnetic field strength maps we are able to estimate $E_K$, $E_B$, and $E_G$, and thus test equation~\ref{eq:virialeq}. For a cloud defined by a given aperture on the sky, the mass is most simply estimated from the total mass surface density times the projected area of the aperture. However, we expect that some of the material along the line of sight should be considered to be part of the surrounding ambient cloud, so for the dense core sample within the IRDC we also evaluate ``background-subtracted'' masses in which the background $\Sigma$ is evaluated in an annulus from $R$ to $2R$ around a given aperture. 

Our estimates of the IRDC and core masses are listed in Table~\ref{tab:core_summary}. The core masses span about an order of magnitude, ranging from $\sim 200\,M_{\odot}$ to $\sim 2000\,M_{\odot}$. We also notice that the FIR-based masses are generally larger than the MIREX-based masses by about a factor of 1.5. This level of difference is within the systematic uncertainties of the two methods, e.g., involving assumptions about dust opacities. We also notice that in some cases (i.e., C3, C9-MIR, C12, C14) the background subtracted mass becomes very small, i.e., $<20\%$ of the non-background-subtracted mass. Especially in these instances, we consider that the background subtracted mass is not reliably measured and we will set the level of the background as a measure of the uncertainty. Table~\ref{tab:core_summary} also reports an estimate of the volume density (in terms of $n_{\rm H}$) of the sources, with this based on a simple assumption of spherical geometry of the structure. The density of the background, which is used for the surface terms around the core, is estimated in a similar way.

Table~\ref{tab:core_summary} next reports the 1D velocity dispersion of the sources, based on the C$^{18}$O(1-0) emission. 
The velocity dispersion of the background region is also estimated from the spectrum of its C$^{18}$O(1-0) emission. Note that we do not attempt to estimate a background subtracted velocity dispersion.
The final primary measurements reported in Table~\ref{tab:core_summary} are the total magnetic field strengths, $B_{\rm 3D}$, as measured from the r-DCF (ST) map (i.e., scaling $B_{\rm 2D}$ by $\sqrt{3/2}$) of the source and background. We note that the magnetic field in the annular region is often at a similar level to that in the core. For this reason, we also do not attempt to estimate a background subtracted magnetic field strength for the core.

\begin{figure*}[h!]
\begin{minipage}{0.49\textwidth}
\includegraphics[width=\textwidth]{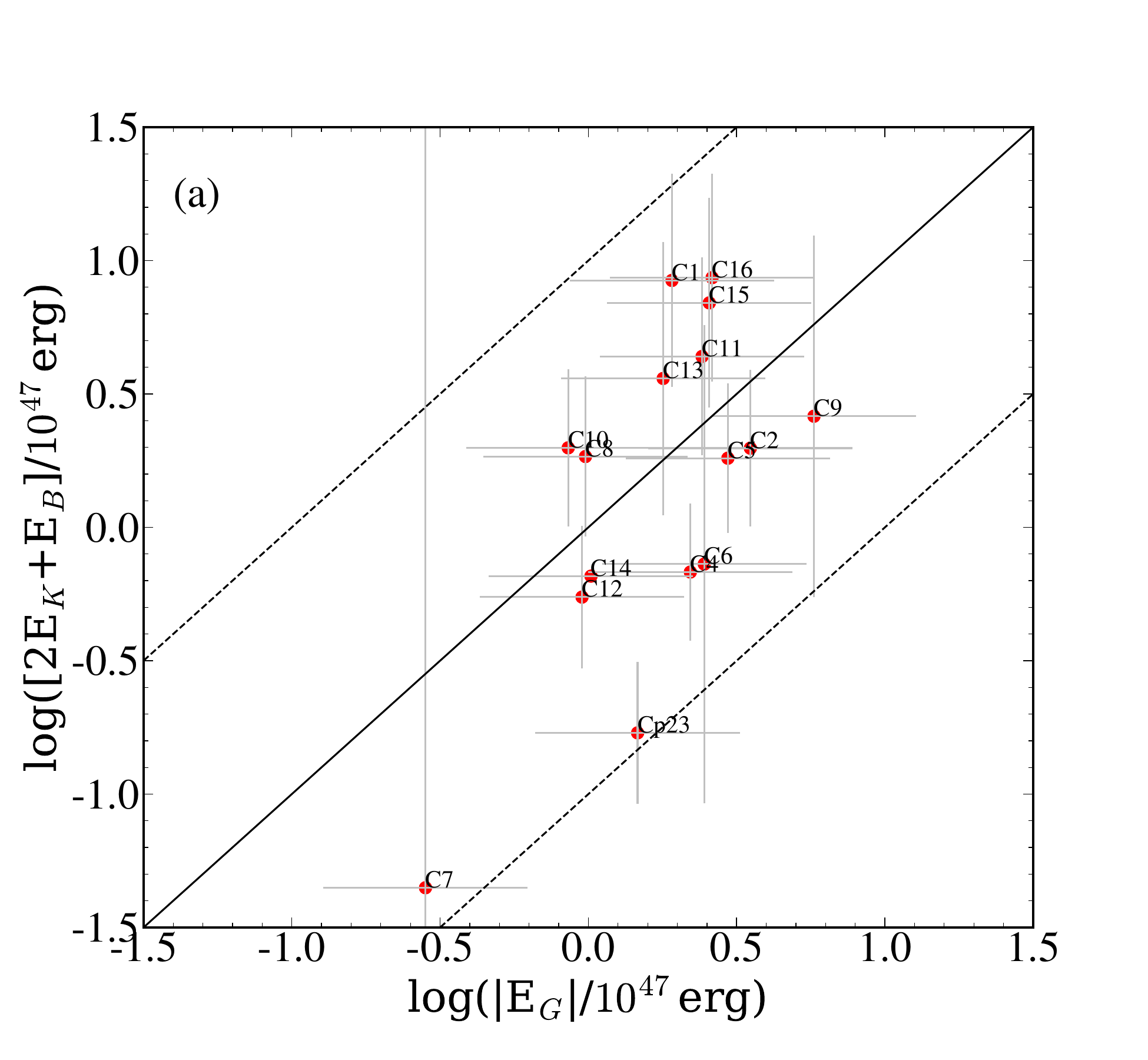}
\end{minipage}
\begin{minipage}{0.49\textwidth}
\includegraphics[width=\textwidth]{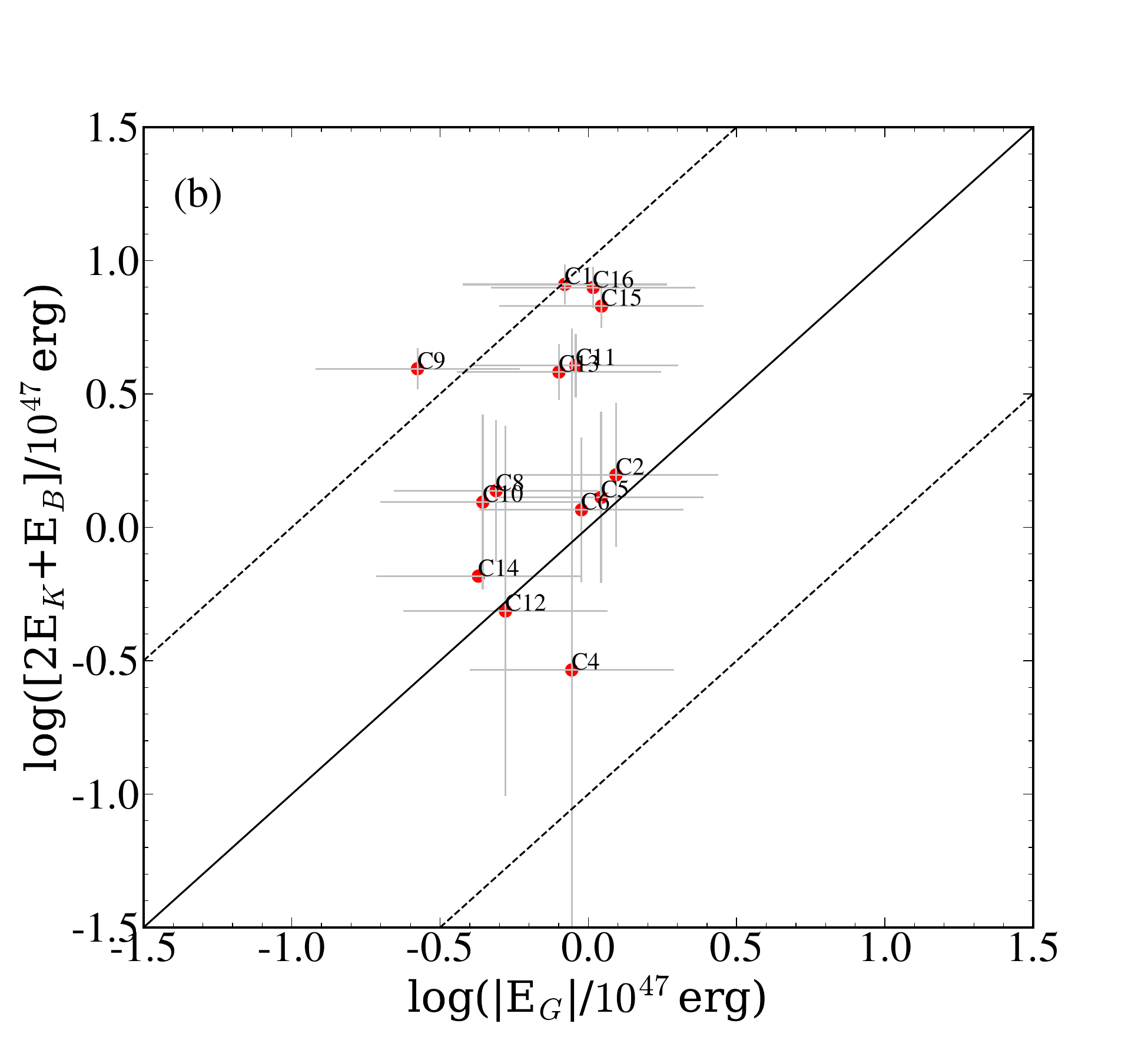}
\end{minipage}
\begin{minipage}{0.49\textwidth}
\includegraphics[width=\textwidth]{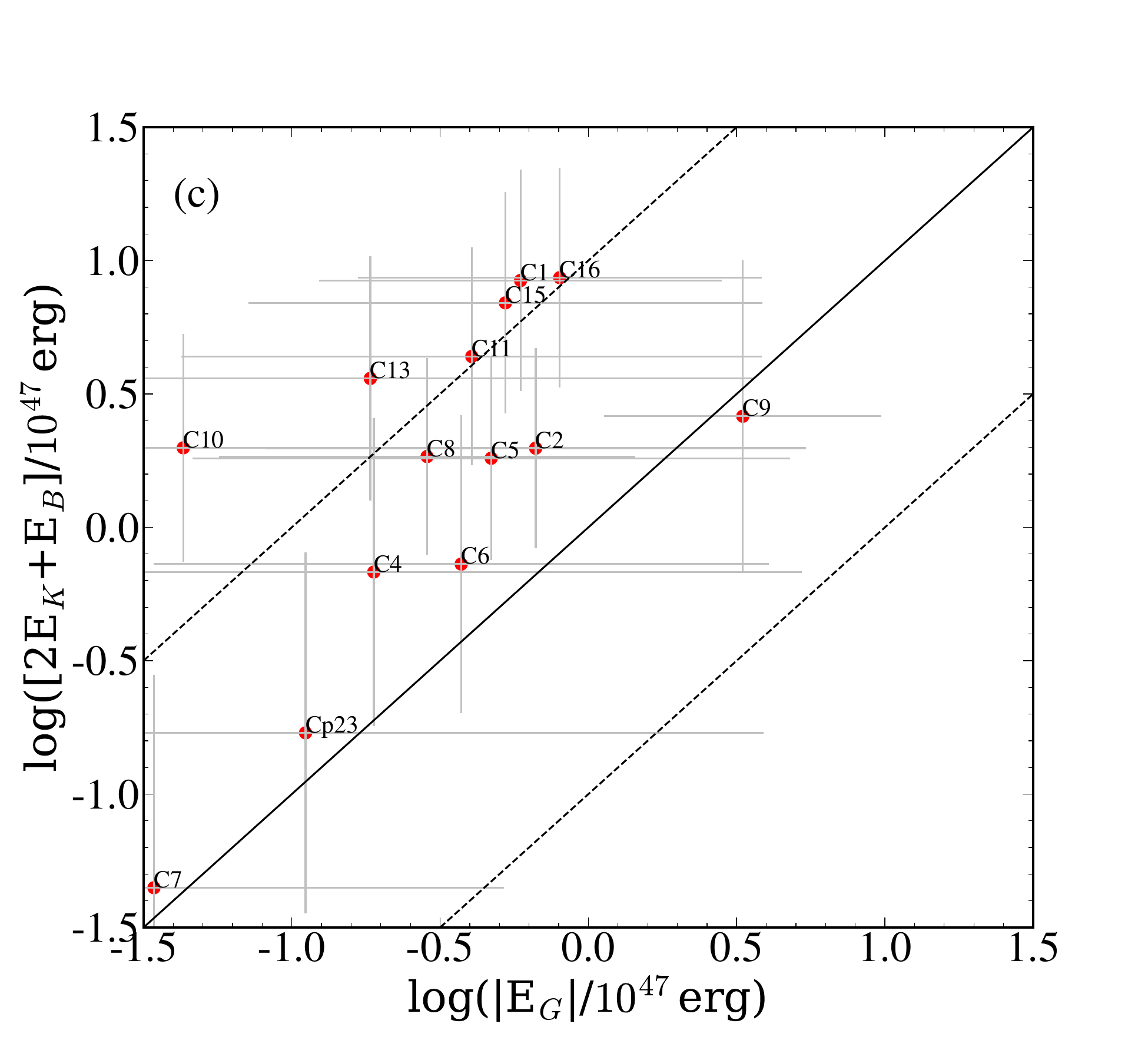}
\end{minipage}
\begin{minipage}{0.49\textwidth}
\includegraphics[width=\textwidth]{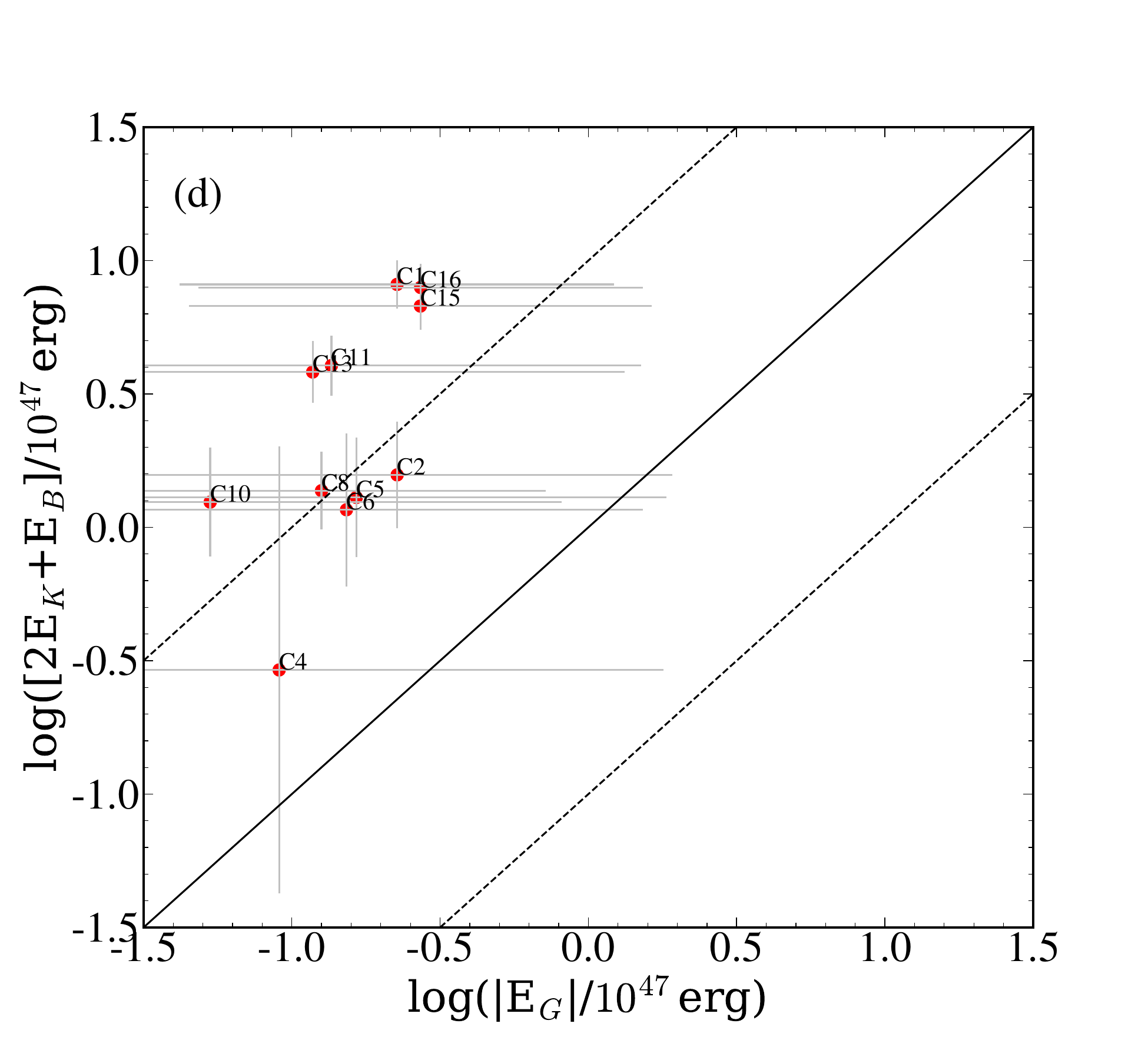}
\end{minipage}
\caption{{\it (a) Top Left:} Test of virial equilibrium of IRDC dense cores. Here masses are estimated from the FIR $\Sigma$ map and no background subtraction is done when evaluating source mass, velocity dispersion and $B$-field strength.
%
{\it (b) Top Right:} As (a), but now with masses estimated from the MIREX $\Sigma$ map.
{\it (c) Bottom Left:} As (a), but now with background subtraction when evaluating source mass.
%
{\it (d) Bottom Right:} As (c), but now with masses estimated from the MIREX $\Sigma$ map.
%
%
}
\label{figure:Kvir_vs_M}
\end{figure*}

With the above measurements of primary quantities, we then estimate $E_G$, $E_K$, and $E_B$. Note that $E_K$ and $E_B$ include a subtraction of surface terms and in certain cases where the background is relatively strong, these can then take negative values. To test the condition of virial equilibrium, in Figure~\ref{figure:Kvir_vs_M} we plot $2E_K + E_B$ versus $|E_G|$. Here we examine four cases resulting from whether we measure mass and density via the FIR or MIREX $\Sigma$ maps and whether or not we carry out background subtraction to measure the core properties of mass and density.

From Figure~\ref{figure:Kvir_vs_M} we see that for the FIR-derived core properties and the case before background subtraction, the core population scatters quite closely about the condition of virial equilibrium. The most extreme outlier to the gravity dominated regime is the massive protostellar source Cp23. The most extreme outlier to the $2E_K+E_B$ dominated regime is the C1 core/clump that is a prime candidate for hosting massive pre-stellar and earl-stage cores \citep{2013ApJ...779...96T,2016ApJ...821L...3T,2018ApJ...867...94K}. Indeed, here we see that $E_B$ dominates the energetics. The effect of using MIREX based masses and densities is a modest shift toward supervirial conditions, expected given the systematically lower masses found by the MIREX method. However, as discussed by \citet{2012ApJ...754....5B}, the MIREX masses may be somewhat underestimated in the densest regions given the effects of ``saturation'' limiting the maximum $\Sigma$ that can be probed by the method.

The effect of background subtraction on the mass and density estimates is more significant. As noted above, there are cases where the mass of a core is reduced by a large factor ($>5$) after attempting background subtraction. Such estimates thus become highly uncertain and care should be taken with the interpretation of the results.

\begin{figure*}[h!]
\begin{minipage}{0.49\textwidth}
\includegraphics[width=\textwidth]{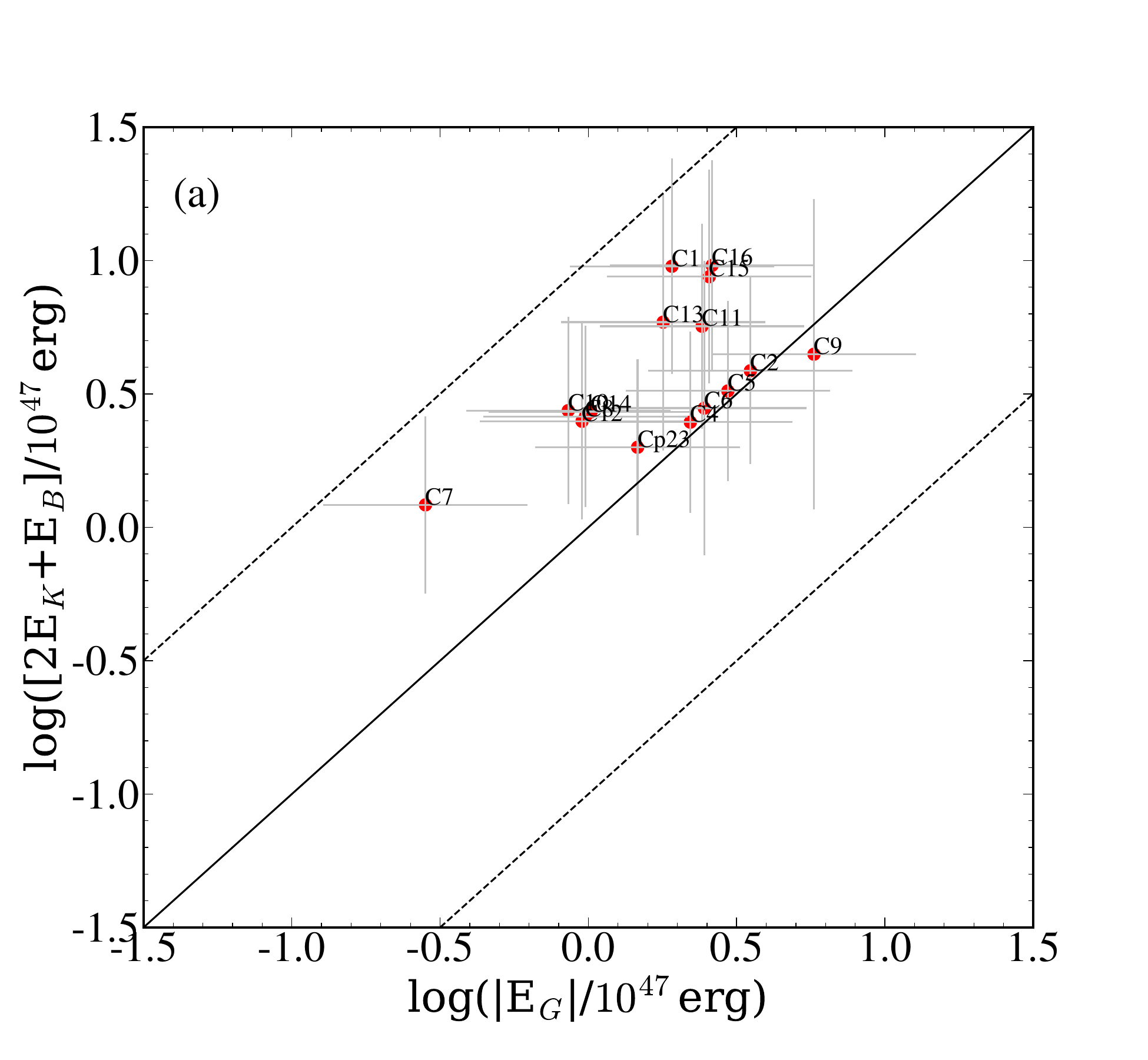}
\end{minipage}
\begin{minipage}{0.49\textwidth}
\includegraphics[width=\textwidth]{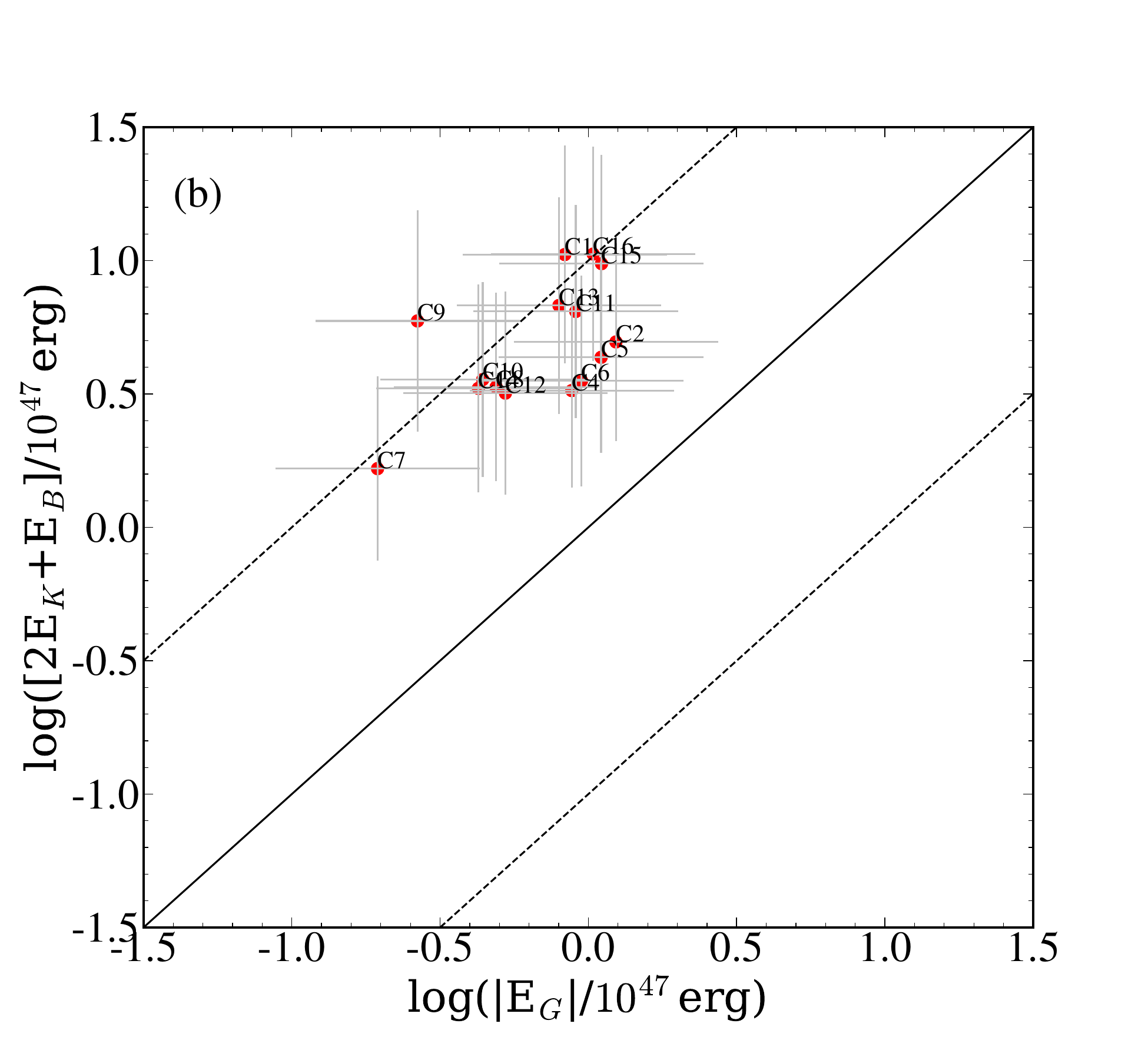}
\end{minipage}
\begin{minipage}{0.49\textwidth}
\includegraphics[width=\textwidth]{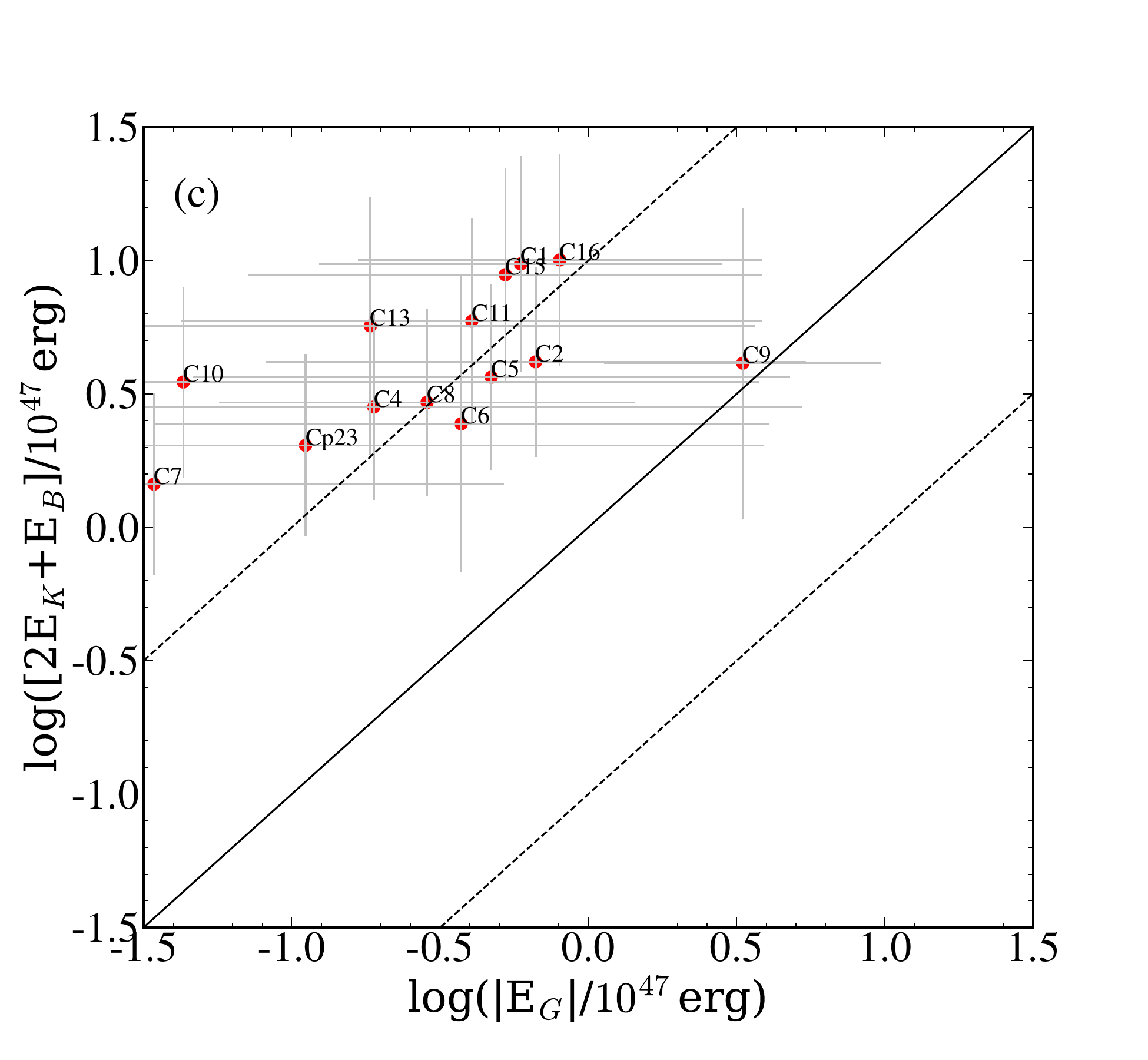}
\end{minipage}
\begin{minipage}{0.49\textwidth}
\includegraphics[width=\textwidth]{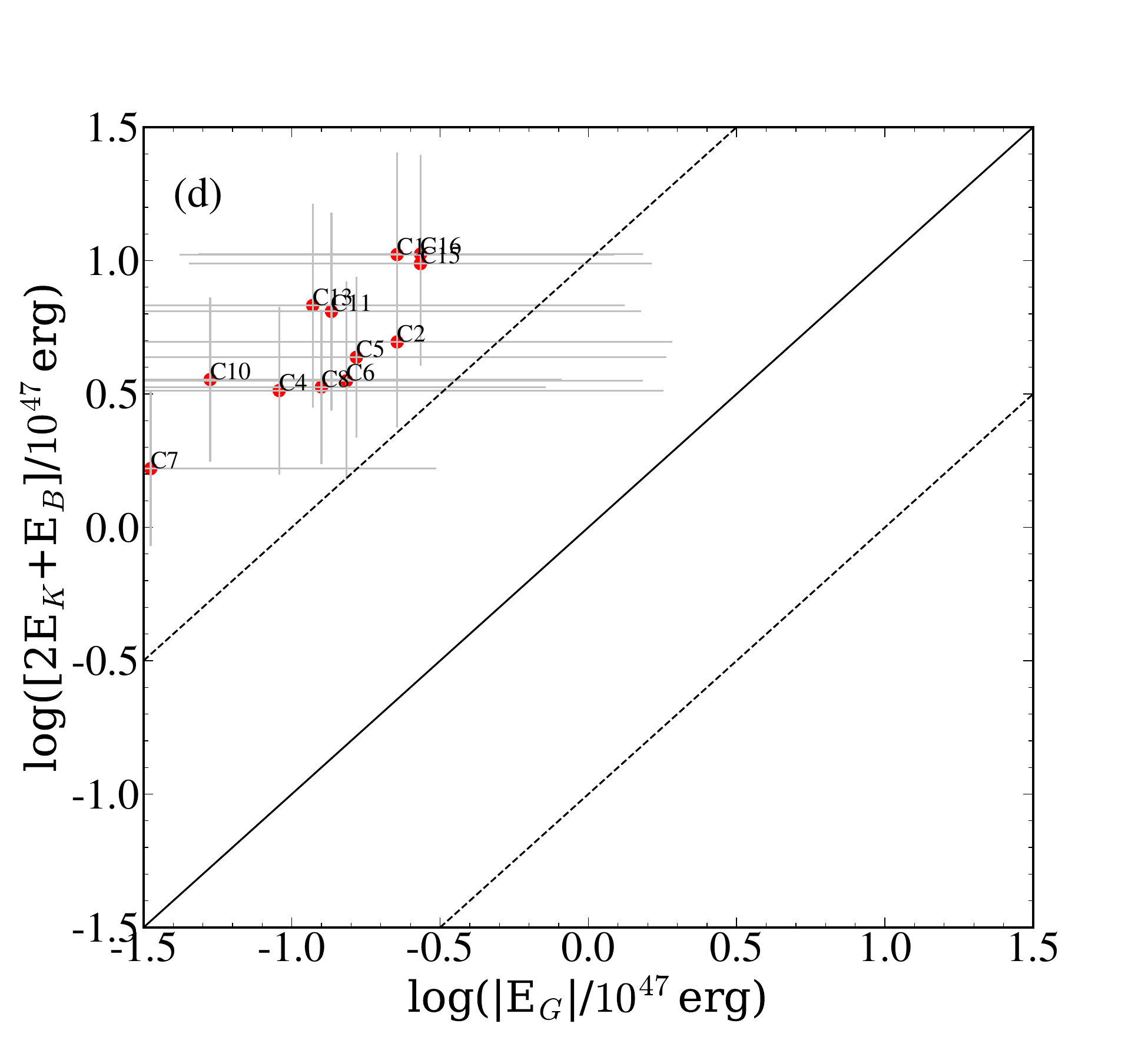}
\end{minipage}

\caption{{\it (a) Top Left:} As Fig.~\ref{figure:Kvir_vs_M}a, but now surface terms are not considered when evaluating $E_K$ and $E_B$.
{\it (b) Top Right:} As (a), but now with masses estimated from the MIREX $\Sigma$ map.
{\it (c) Bottom Left:} As (a), but now with background subtraction when evaluating source mass.
{\it (d) Bottom Right:} As (c), but now with masses estimated from the MIREX $\Sigma$ map.
}
\label{figure:Kvir_vs_M_nosurface}
\end{figure*}

To gauge the effects of the surface terms, Figure~\ref{figure:Kvir_vs_M_nosurface} shows the same quantities as Figure~\ref{figure:Kvir_vs_M}, but now with $E_K$ and $E_B$ evaluated without the influence of the surface terms. We see that the surface terms have a large impact the position of cores in this energy ratio diagram. Thus it is important to attempt to include them, even if their evaluation is somewhat uncertain.

\begin{figure*}[h!]
\begin{minipage}{0.49\textwidth}
\includegraphics[width=\textwidth]{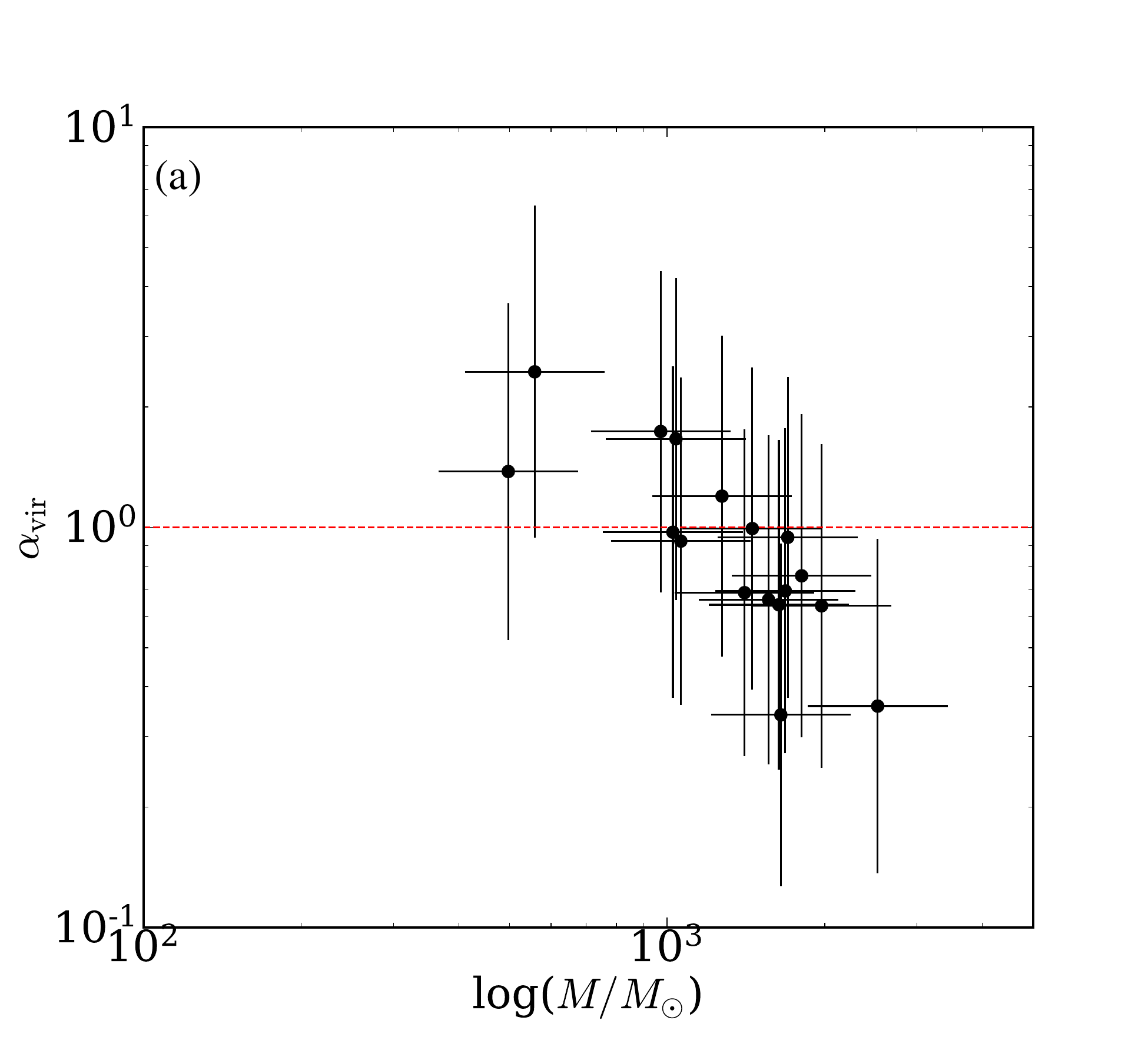}
\end{minipage}
\begin{minipage}{0.49\textwidth}
\includegraphics[width=\textwidth]{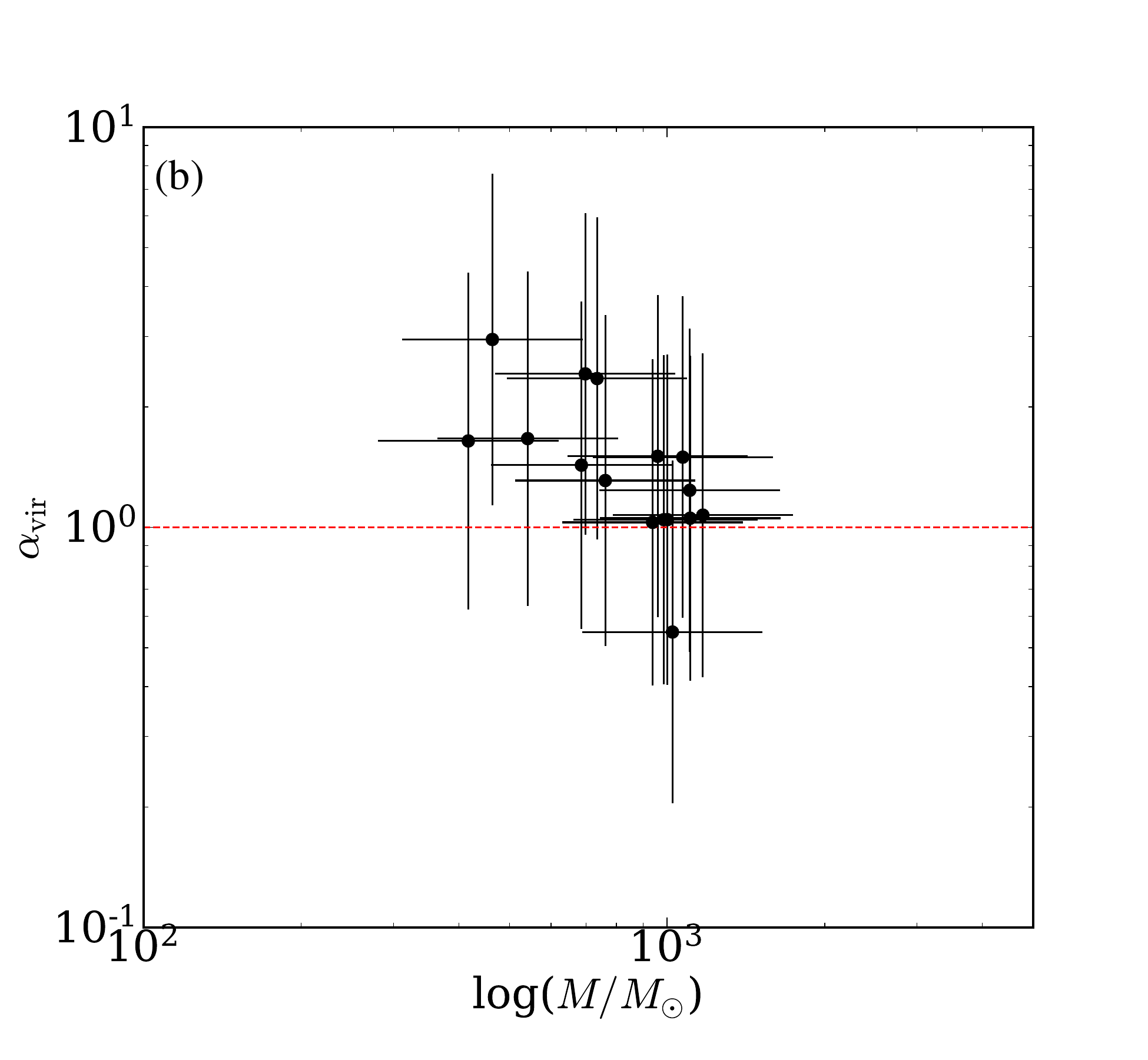}
\end{minipage}
\begin{minipage}{0.49\textwidth}
\includegraphics[width=\textwidth]{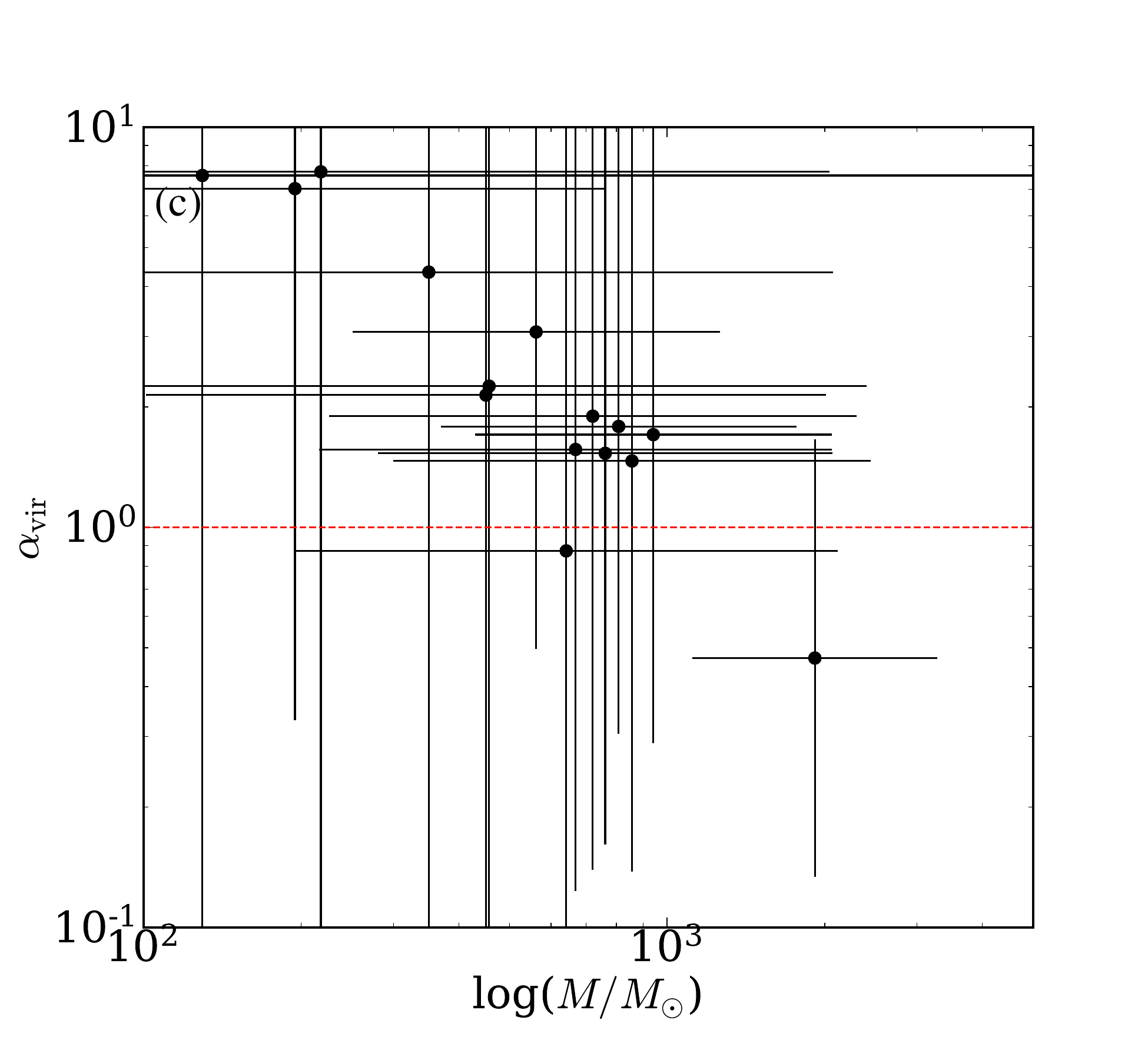}
\end{minipage}
\begin{minipage}{0.49\textwidth}
\includegraphics[width=\textwidth]{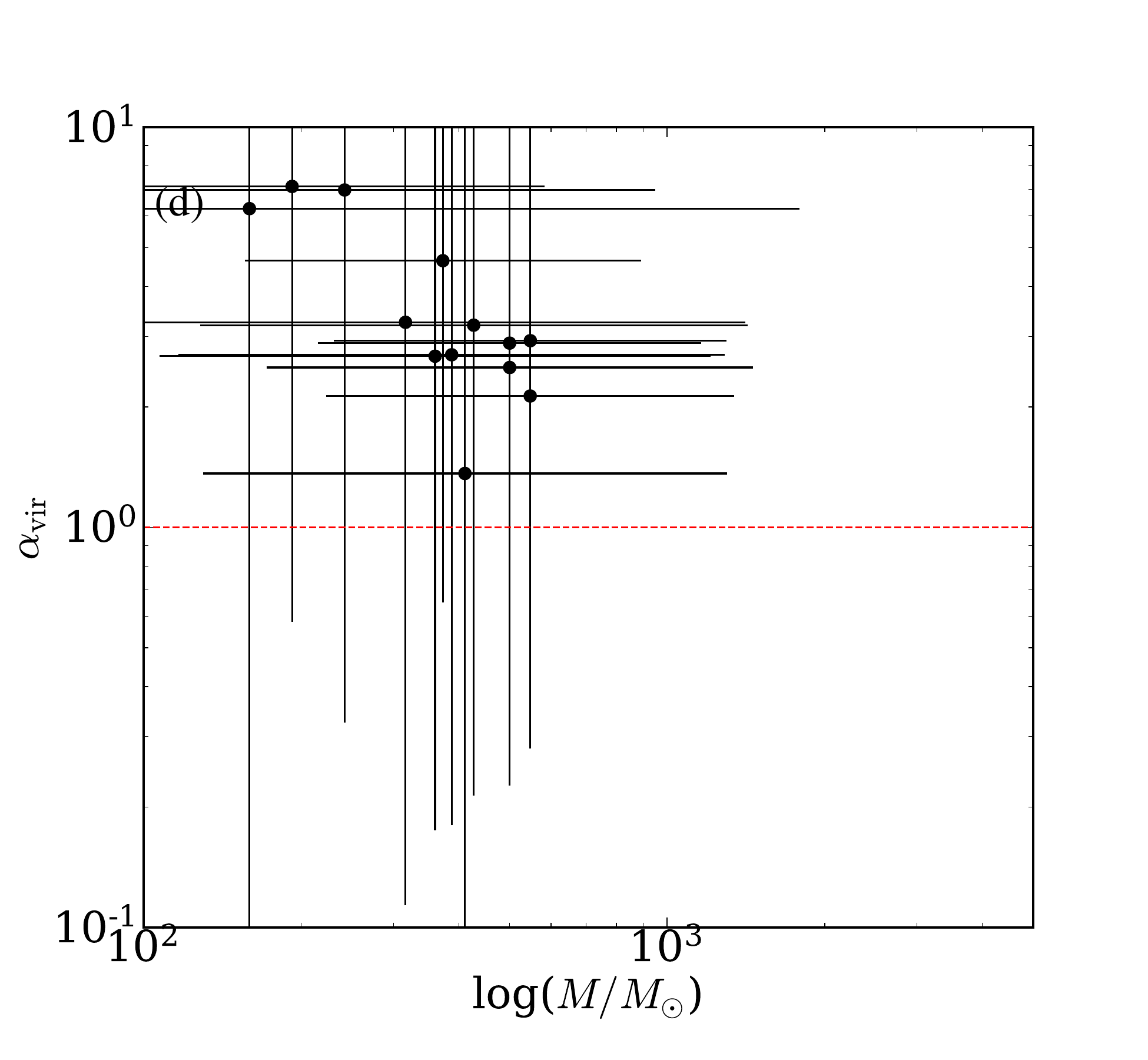}
\end{minipage}
\caption{{\it (a) Top Left:} Test of virial equilibrium of IRDC dense cores. Here masses are estimated from the FIR $\Sigma$ map and no background subtraction is done when evaluating source mass, velocity dispersion and $B$-field strength.
{\it (b) Top Right:} As (a), but now with measurements from the MIREX $\Sigma$ map.
{\it (c) Bottom Left:} As (a), but now with background subtraction when evaluating source mass.
{\it (d) Bottom Right:} As (b), but now with measurements from the MIREX $\Sigma$ map.
%
%
%
}
\label{fig:alphavir}
\end{figure*}

Traditional studies of cloud dynamics have focused mostly on measuring the virial parameter, which describes the ratio between kinetic and gravitational energies (BM92):
\begin{equation}
\alpha_{\rm vir} \equiv \frac{5 \sigma_v^2 R}{GM} =\frac{5\sigma_v^{2}}{G\Sigma \pi R} 
\end{equation} 
To place our sources in the context of such previous studies, here we also evaluate their virial parameters, which are shown in Fig.~\ref{fig:alphavir} and listed in Table~\ref{tab:core_summary2}. 

For the case of no background subtraction in estimating core mass, we find average and dispersion of $\alpha_{\rm vir} = 1.11\pm0.587$ for FIR derived masses and $1.65\pm0.666$ for MIREX derived masses. If we consider background subtracted masses, the corresponding average virial parameter values are $9.22\pm25.8$ and $2.89\pm7.89$ for FIR and MIREX traced cores, respectively. 

Next, we evaluate the Alfv\'en Mach number, $m_A$, of the cores, given by $m_A=\sqrt{3}\sigma_{v}/v_A = \sqrt{3} \sigma_v (B_{\rm tot}/\sqrt{4\pi \rho} )$. These values are also shown in Table~\ref{tab:core_summary2}. For the case of no background subtraction in estimating core mass, we find average and dispersion of $m_A = 0.438\pm0.130$ for FIR derived masses and $0.344\pm0.104$ for MIREX derived masses. If we consider background subtracted masses, the corresponding values are $0.277\pm0.100$ and $0.218\pm0.0756$ for FIR and MIREX traced cores, respectively.

Similarly, the mass-to-flux ratios are also presented in Table~\ref{tab:core_summary2}. For the case of no background subtraction in estimating core mass, we find average and dispersion of $\mu = 0.878\pm0.284$ for FIR derived masses and $0.539\pm0.155$ for MIREX derived masses. If we consider background subtracted masses, the corresponding mass to flux ratios are $0.380\pm0.231$ and $0.219\pm0.0820$ for FIR and MIREX traced cores, respectively.







Finally, we compare the observed quantities to the prediction of the Turbulent Core Model (TCM) of \citep{2003ApJ...585..850M}. The TCM describes the properties of virial equilibrium pressure-truncated singular polytropic spheres that are partially supported by large-scale $B-$fields and partially by turbulence. The expected 1D velocity dispersion of such a core in the fiducial case of Alfven Mach number of unity is given by \citep{2013ApJ...779...96T}:
\begin{equation}
\sigma_{\mathrm{c}, \mathrm{vir}} \rightarrow 1.09\left(\frac{M_c}{60 M_{\odot}}\right)^{1 / 4}\left(\frac{\Sigma_{\mathrm{cl}}}{1 \mathrm{~g} \mathrm{~cm}^{-2}}\right)^{1 / 4} \mathrm{~km} \mathrm{~s}^{-1}.
\end{equation}

\begin{figure*}[h!]
\begin{minipage}{0.49\textwidth}
\includegraphics[width=\textwidth]{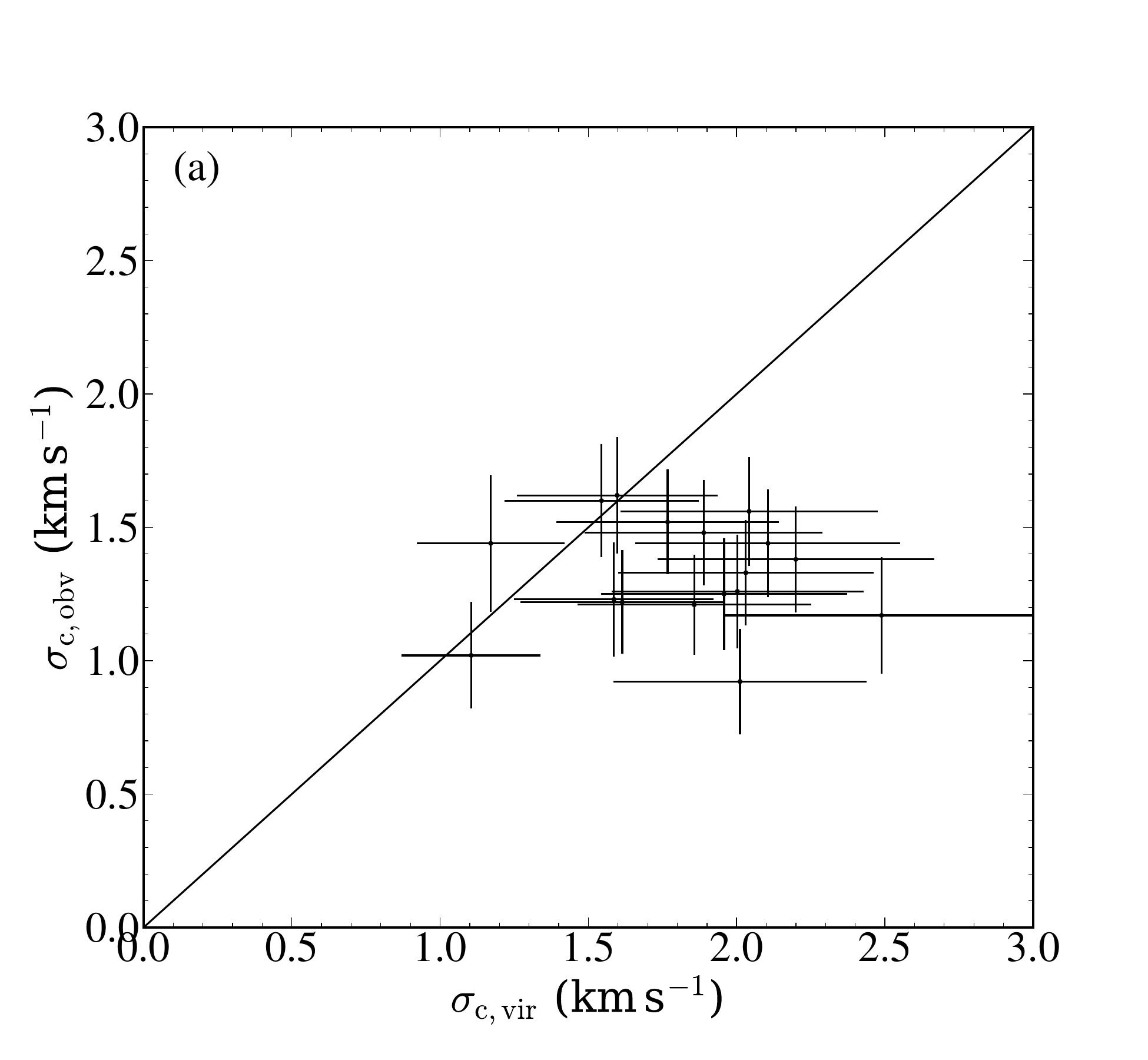}
\end{minipage}
\begin{minipage}{0.49\textwidth}
\includegraphics[width=\textwidth]{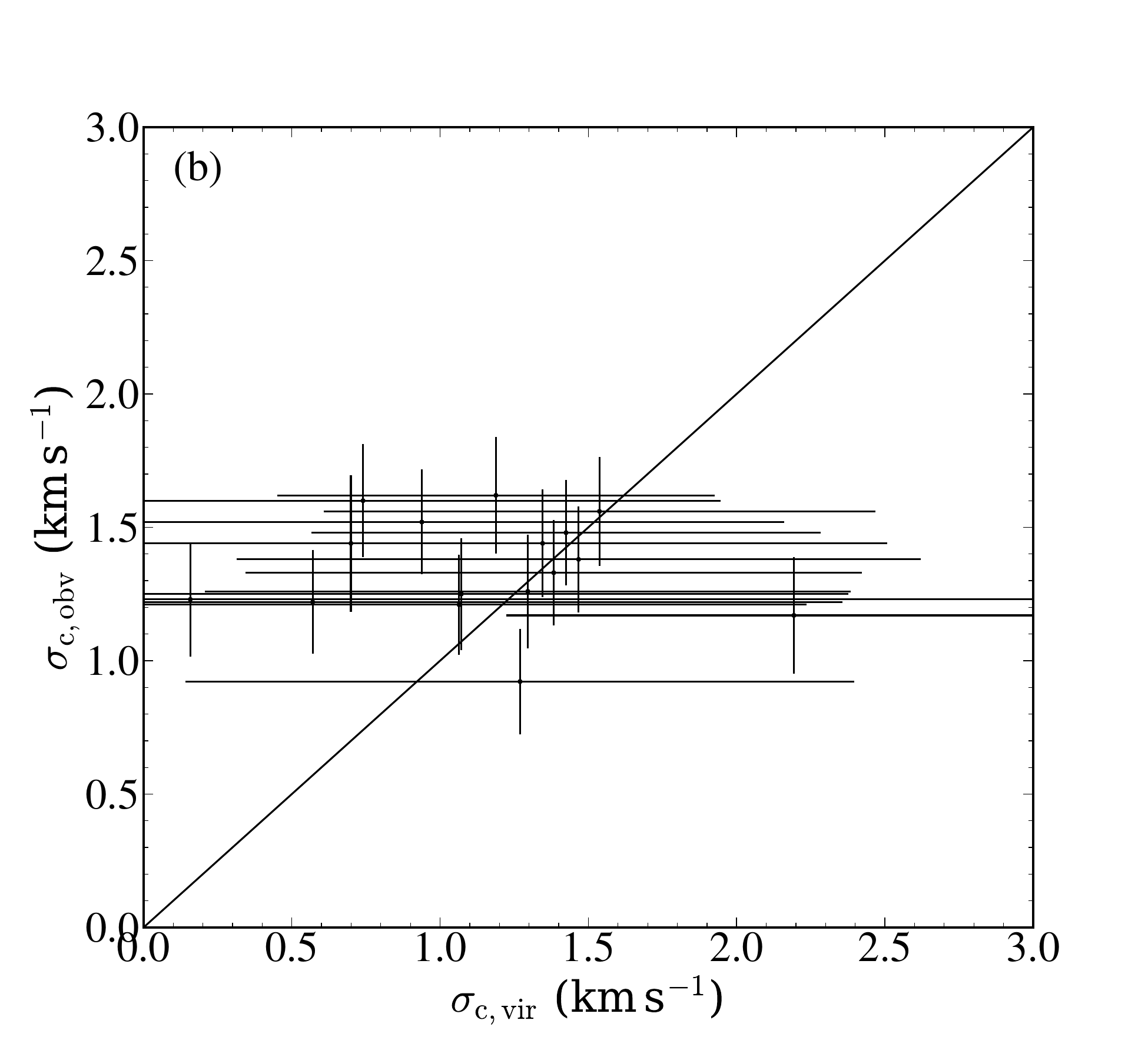}
\end{minipage}
\begin{minipage}{0.49\textwidth}
\includegraphics[width=\textwidth]{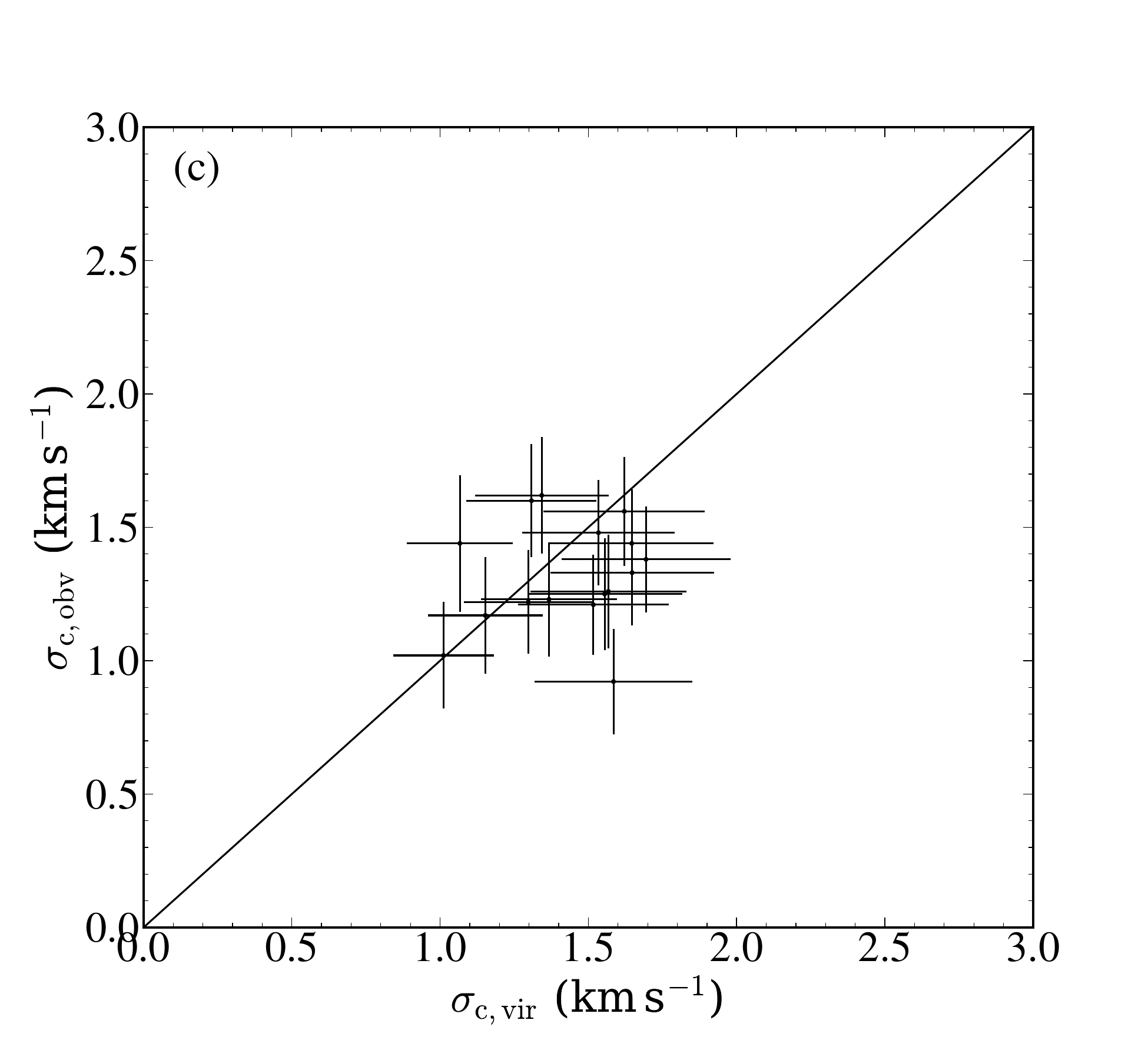}
\end{minipage}
\begin{minipage}{0.49\textwidth}
\includegraphics[width=\textwidth]{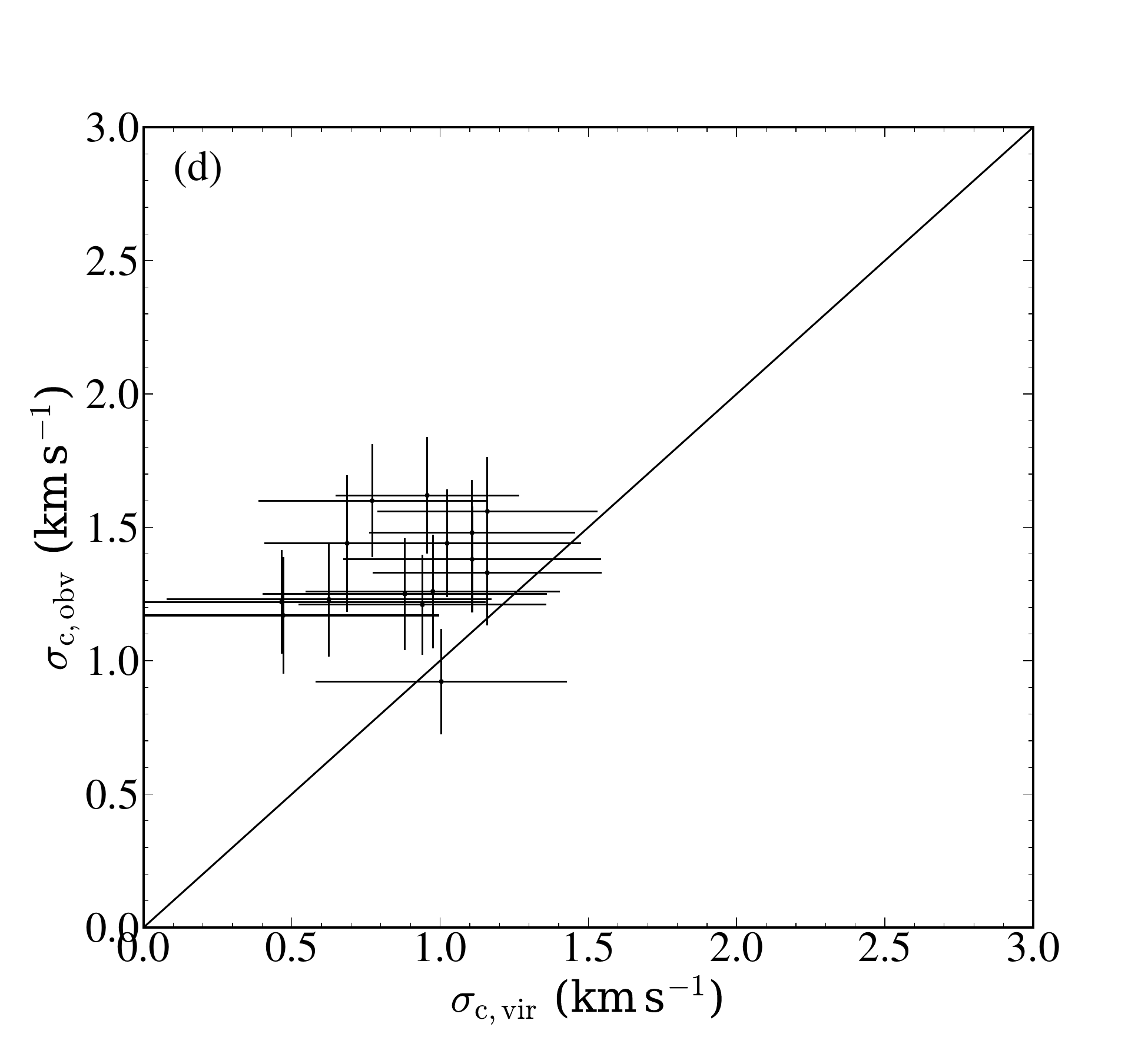}
\end{minipage}
\caption{{\it (a) Top Left:} Observed total 1-D velocity dispersion against predicted 1-D velocity dispersion from the TCM prediction based on the mass derived from the FIR mass surface density map. 
{\it (b) Top Right:} Similar to (a), but the predicted velocity dispersions are derived from the background-subtracted mass. 
{\it (c) Bottom Left:}Observed velocity dispersion against predicted 1-D velocity dispersions  from the TCM with masses are derived from the MIREX map.
{\it (d) Bottom Right:} Similar to (c), but the predicted velocity dispersions are derived from the background-subtracted mass. 
}
\label{figure:v_disp_compare}
\end{figure*}

In Figure~\ref{figure:v_disp_compare}, we compare the observed velocity dispersion of the cores to the expected velocity dispersion from a fiducial virialized core of given mass, $M_c$, and background clump mass surface density, $\Sigma_{\rm cl}$ (see also Table~\ref{tab:core_summary2}). For the FIR-based masses, 
we find that most observed velocity dispersions are smaller than predicted by about a factor of 0.8.
For masses based on the MIREX $\Sigma$ map, the observed velocity dispersion is closer to that predicted by the TCM.

\begin{figure*}[h!]
\begin{minipage}{0.49\textwidth}
\includegraphics[width=\textwidth]{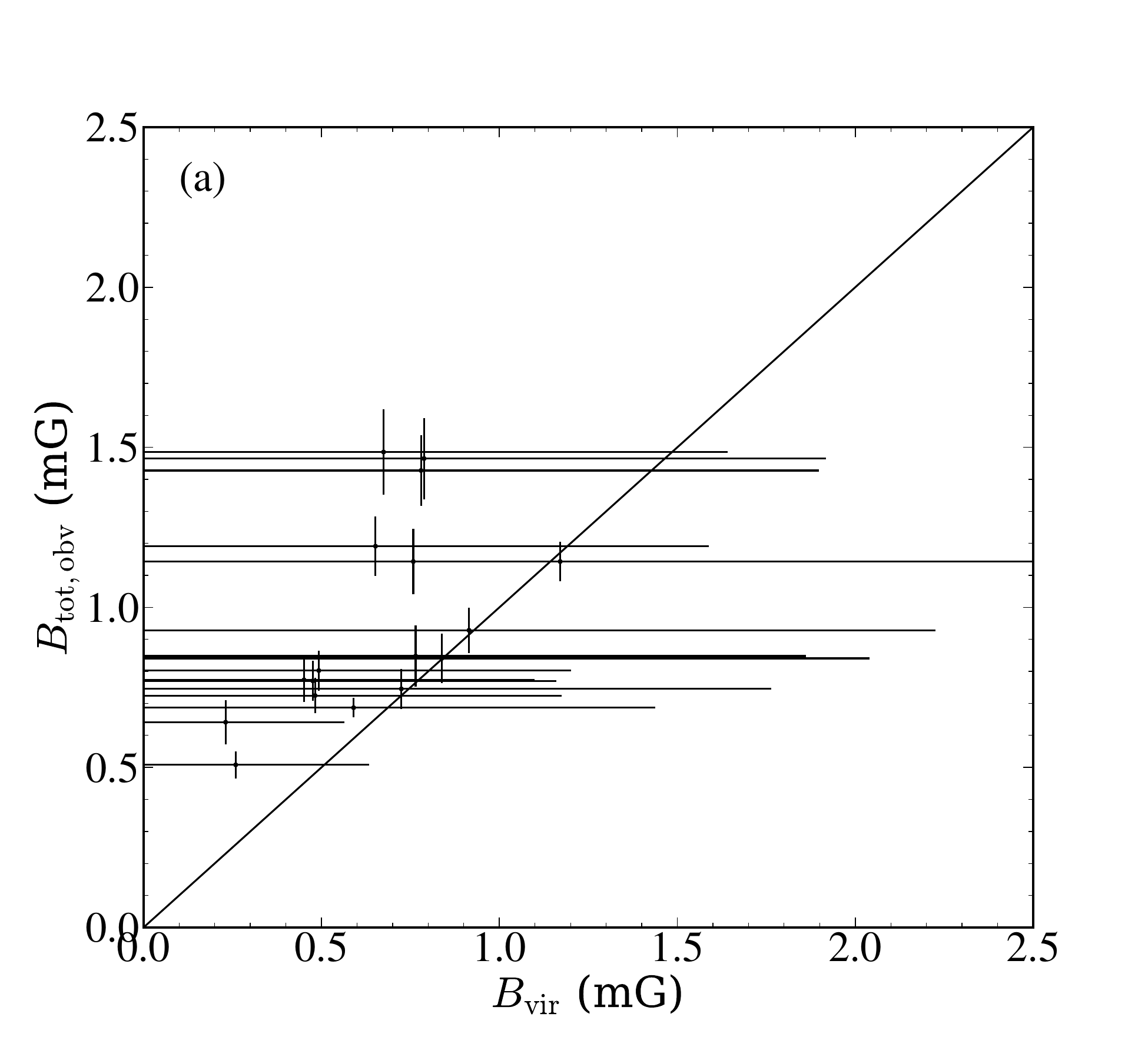}
\end{minipage}
\begin{minipage}{0.49\textwidth}
\includegraphics[width=\textwidth]{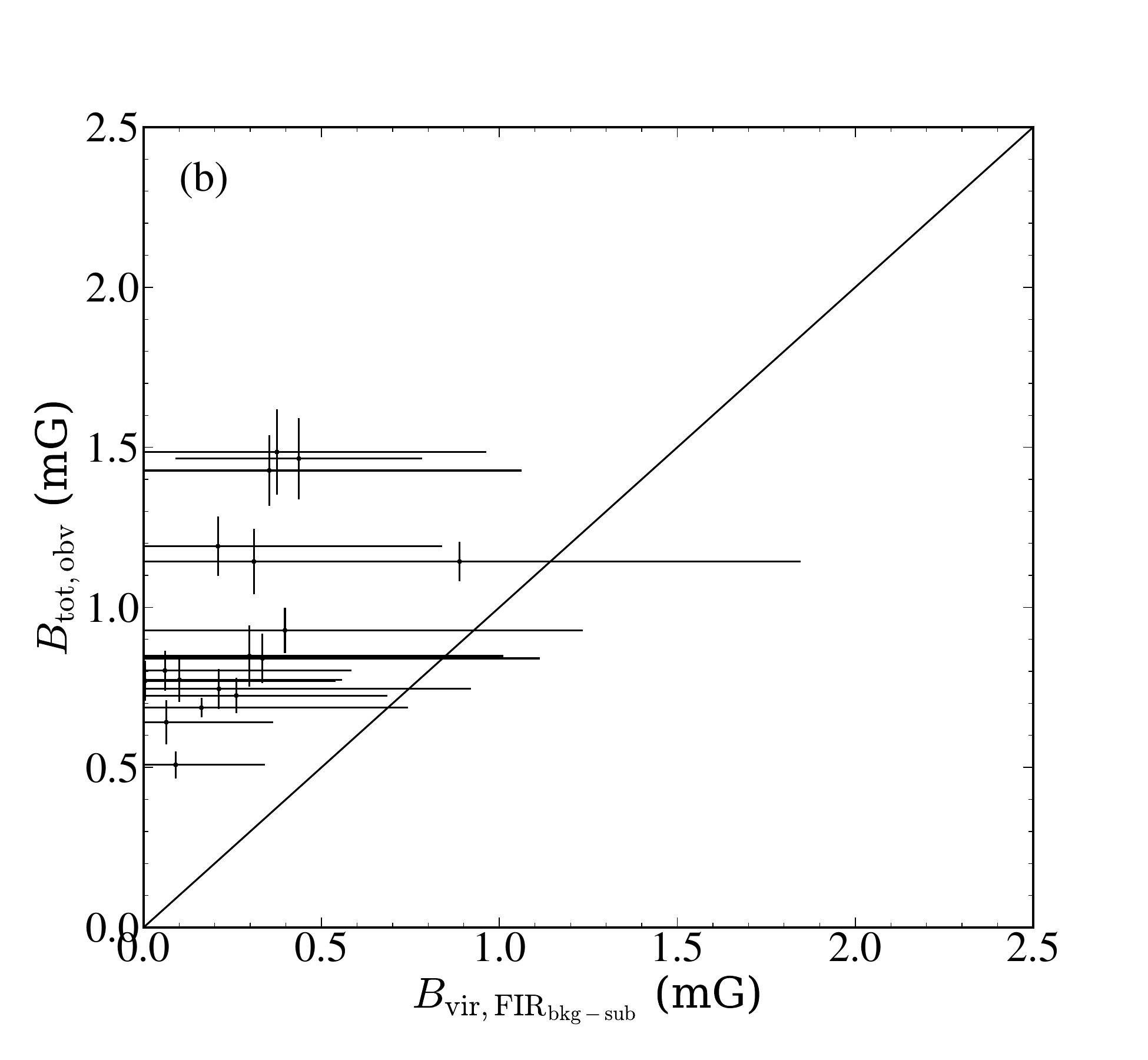}
\end{minipage}
\begin{minipage}{0.49\textwidth}
\includegraphics[width=\textwidth]{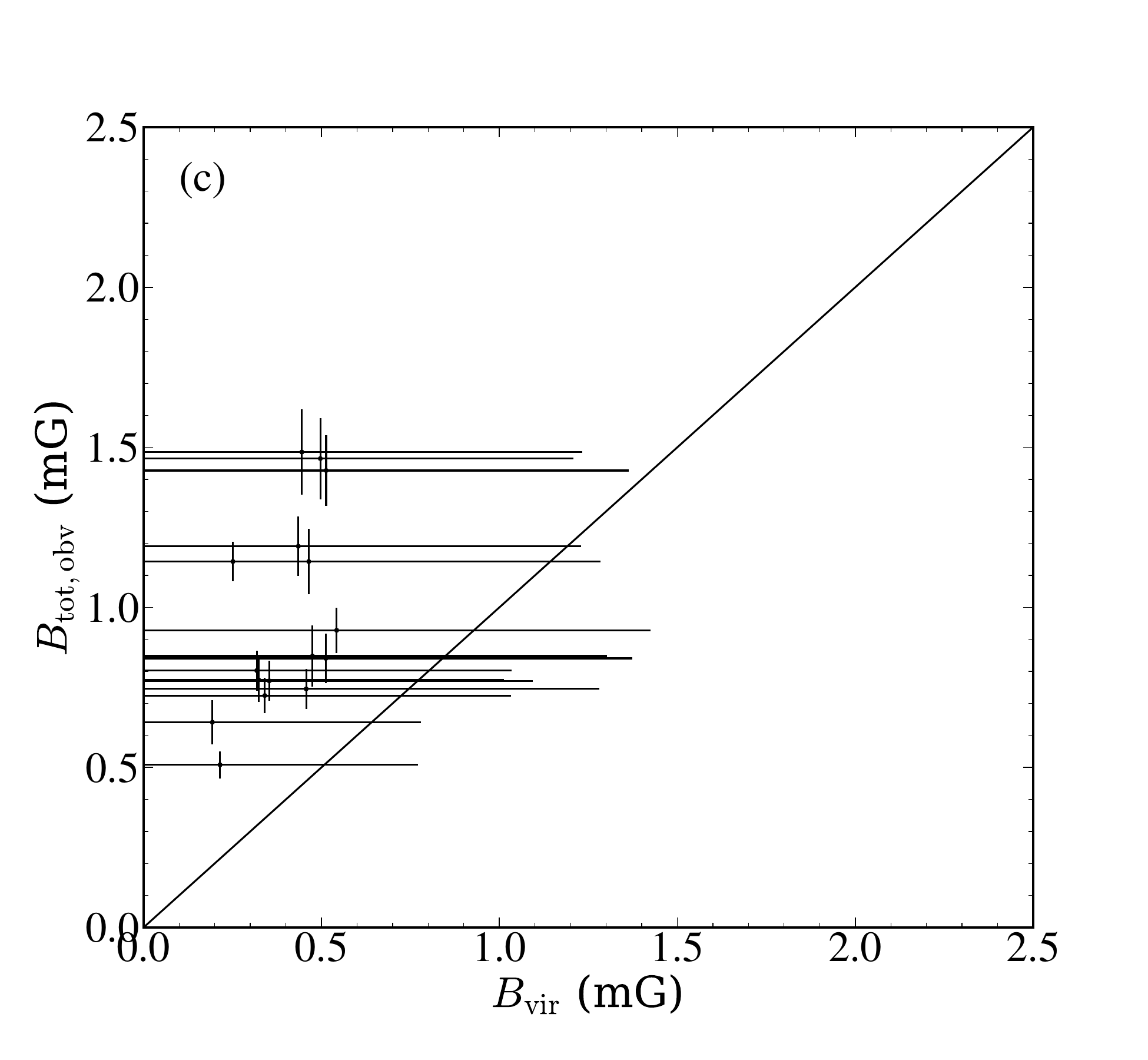}
\end{minipage}
\begin{minipage}{0.49\textwidth}
\includegraphics[width=\textwidth]{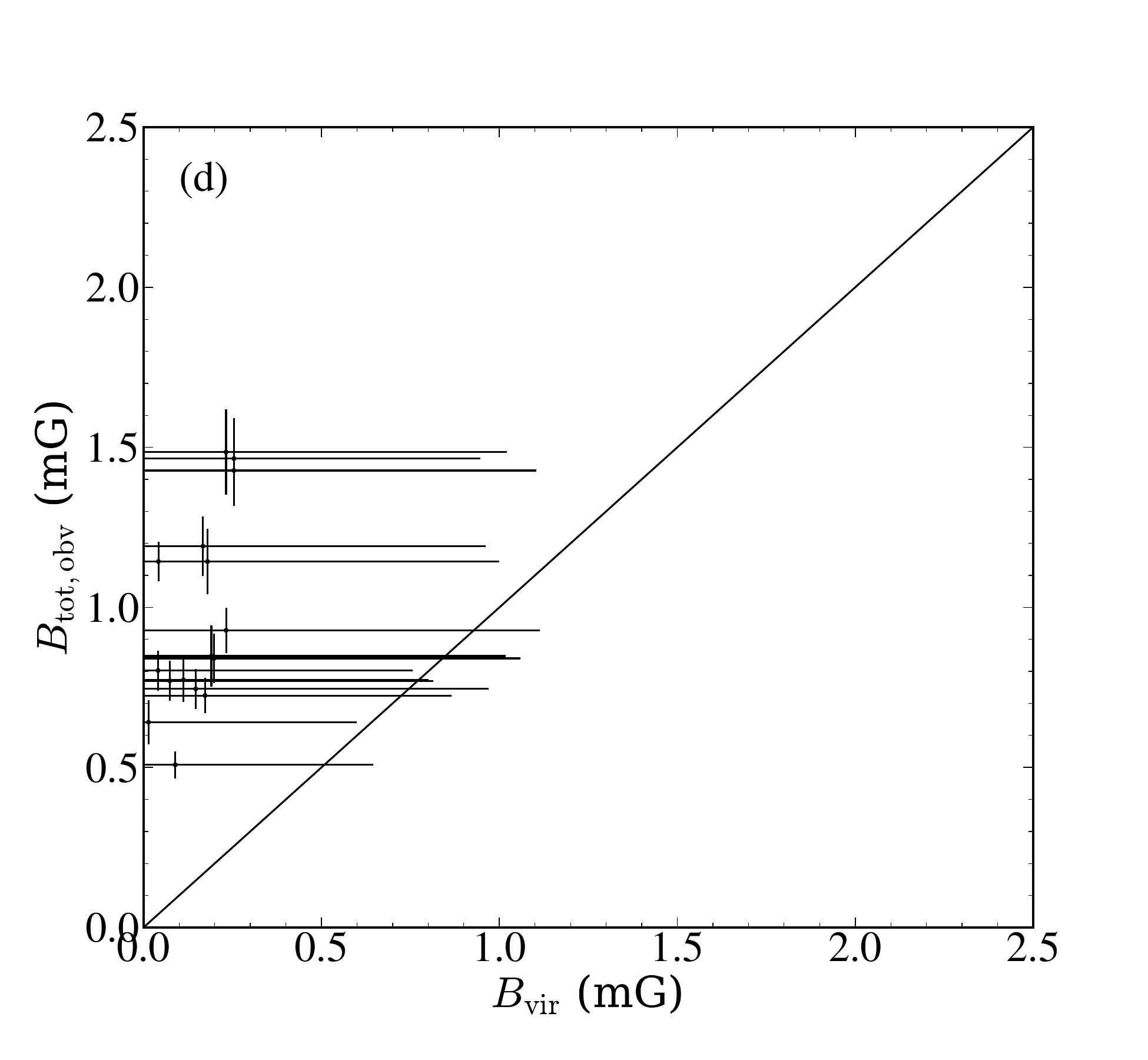}
\end{minipage}
\caption{{\it (a) Top Left:} Scatter plot of the observed total magnetic field strength ($B_{\rm tot}=\sqrt{3}B_x$) to the magnetic field strength assuming the core is at equilibrium based on mass derived from the FIR mass surface density map .
{\it (b) Top Right:} Similar to (a), but here the predicted virial magnetic field is based on the background subtracted mass
{\it (c) Bottom Left:} Scatter plot of the observed total magnetic field strength ($B_{\rm tot}=\sqrt{3}B_x$) to the magnetic field strength assuming the core is at equilibrium but based on mass derived from the MIREX map .
{\it (d) Bottom Right:} Similar to (c), but here the predicted virial magnetic field is based on the background subtracted mass
}
\label{figure:obv_B_vs_tcm_B}
\end{figure*}
\begin{figure*}[h!]
\begin{minipage}{0.49\textwidth}
\includegraphics[width=\textwidth]{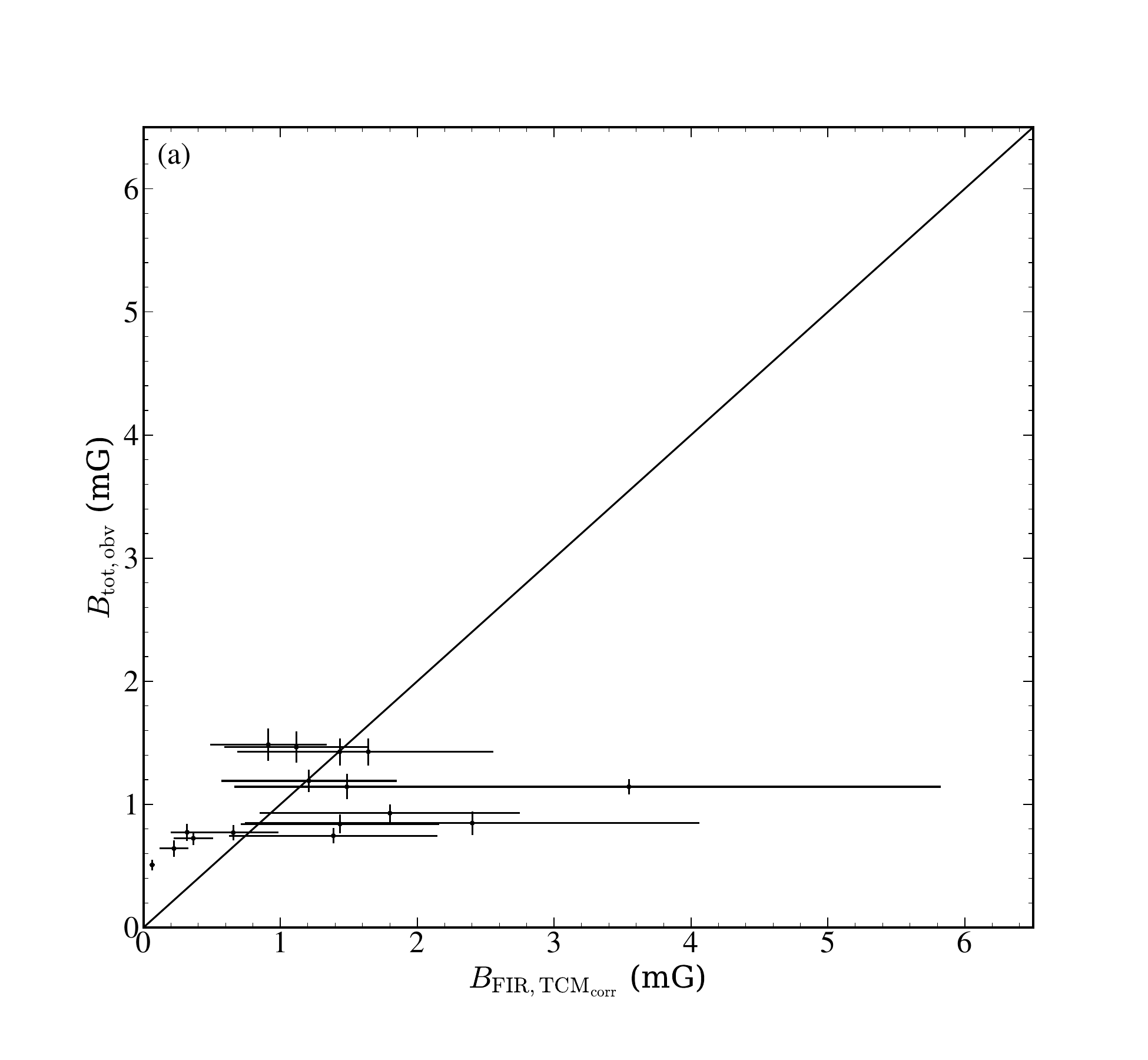}
\end{minipage}
\begin{minipage}{0.49\textwidth}
\includegraphics[width=\textwidth]{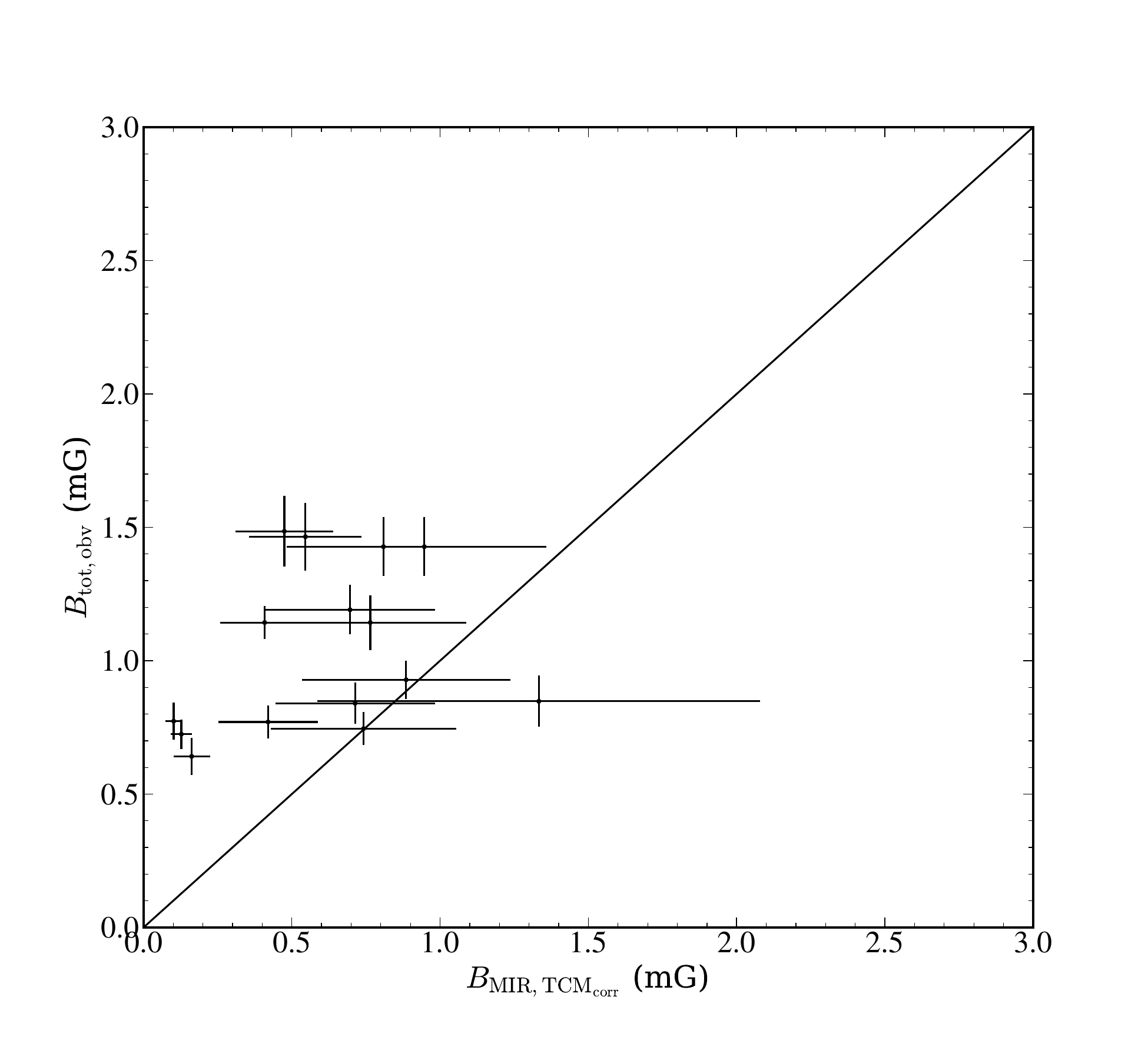}
\end{minipage}
\begin{minipage}{0.49\textwidth}
\includegraphics[width=\textwidth]{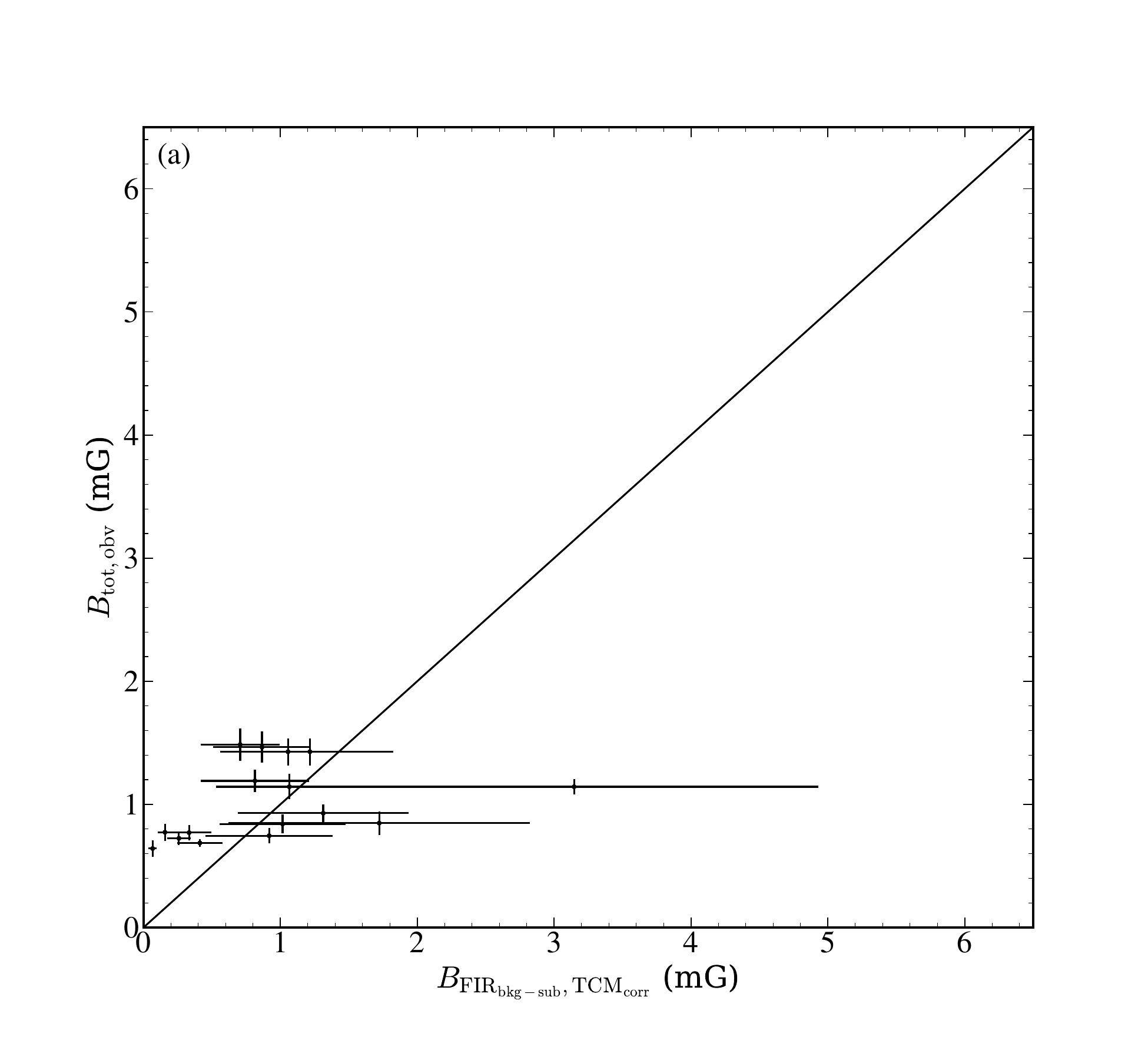}
\end{minipage}
\begin{minipage}{0.49\textwidth}
\includegraphics[width=\textwidth]{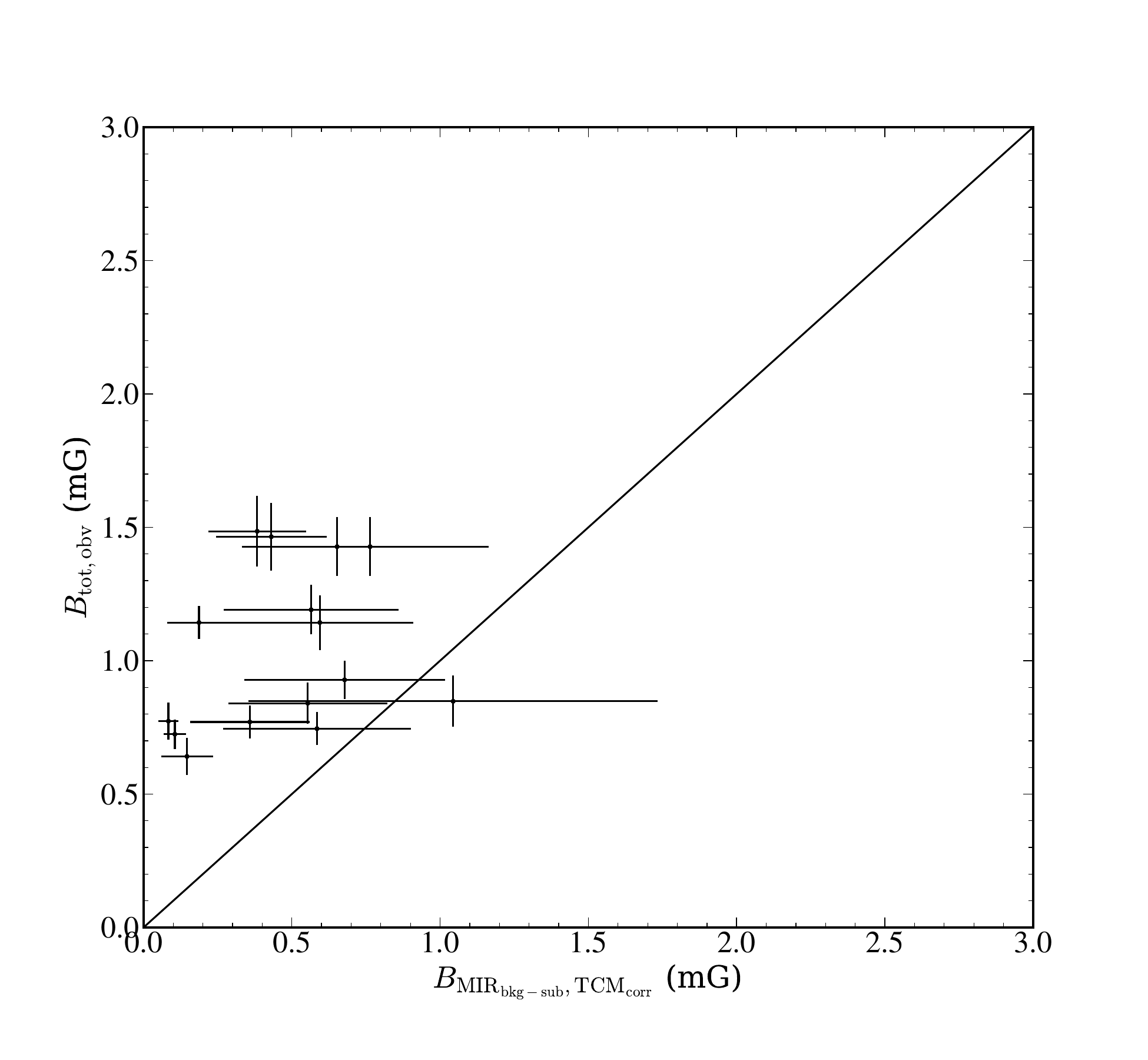}
\end{minipage}
\caption{{\it (a) Top Left:} Scatter plot of the observed total magnetic field strength ($B_{\rm tot}=\sqrt{3}B_x$) to the magnetic field strength assuming the core is at equilibrium with correction of higher magnetic field support \citep{2013ApJ...779...96T} based on mass derived from the FIR mass surface density map.
{\it (b) Top Right:} Similar to (a), but the predicted virial magnetic field is based on the background subtracted mass.
{\it (c) Bottom Left:} Scatter plot of the observed total magnetic field strength ($B_{\rm tot}=\sqrt{3}B_x$) to the magnetic field strength assuming the core is at equilibrium with correction of higher magnetic field support but based on mass derived from the MIREX map .
{\it (d) Bottom Right:} Similar to (c), but here the predicted virial magnetic field is based on the background subtracted mass
}
\label{figure:obv_B_vs_tcm_B_corr}
\end{figure*}

We also estimate the magnetic field at which a core is virialized as predicted by the TCM, i.e., assuming an Alfv\'en Mach of unity. 
We compare in Figure~\ref{figure:obv_B_vs_tcm_B} the predicted magnetic field strengths and the observed field strengths. Typically, the observed field strengths are stronger than predicted by factors of up to a few. For the predicted TCM field strengths to be larger would imply sub-Alfv\'enic conditions in the cores.
\citet{2013ApJ...779...96T} used the ratio of the observed velocity dispersion to the predicted velocity dispersion to estimate what magnetic field strength is needed to support a core in virial equilibrium.
In Figure~\ref{figure:obv_B_vs_tcm_B_corr}, we plot the corresponding corrected magnetic field strengths to the observed field strengths and find that with these corrections, the predicted field strengths are now in better agreement with those that are observed in the cores. This indicates that the observed magnetic field strengths are at a level to control support of the cores under sub-Alfv\'enic conditions.


\begin{table*}
\caption{Physical properties of IRDC G28.37+0.07 (Cloud C) and its dense core sample$^{\dagger}$}
\label{tab:core_summary}
\setlength\tabcolsep{1.9pt}
\begin{tabular}{|c|cccccc|ccc|}
\hline
 & $\Sigma$ & $M$ & ${n_{\rm H}}$& ${n_{\rm H,0}}$ & $\sigma_{v}$  &$B_{\rm tot}$& $E_G$ & $E_{K}$ & $E_{B}$\\
 & ($\rm g\,cm^{-2}$) & ($M_{\odot}$) & ($10^{4}$cm$^{-3}$)&($10^{4}$cm$^{-3}$) & (km/s) & (mG) & ($10^{47}\,$erg) & ($10^{47}$erg) & ($10^{47}$erg)\\
\hline
CloudC & 0.0868(-) & 78400(-) & 0.114(-) &(-)& 3.42 & 0.474(-) & 314(-) & 273(-) &5240(-) \\
 & 0.0601(-) & 54300(-) & 0.0787(-) & (-) &[-]& [-] & 151(-) & 189(-) &  \\
 \hline
Cp23 & 0.326(0.0941) & 1270(351) & 10.2 & 7.41&1.52 &0.687 & 1.47(0.111) & 0.0229(0.00632) & 0.124 \\
 &  &  & & &[1.50] & [0.665] &  &   & \\
 \hline
C1 & 0.372(0.217) & 1540(808) & 11.7 &5.12& 1.48 & 1.49 & 1.91(0.591) & 0.217(0.121) & 7.99 \\
 & 0.246(0.128) & 959(500) & 7.7 & 3.69& [1.30]& [0.558] & 0.833(0.226) & 0.143(0.0747) &  \\
\hline
C2 & 0.505(0.230) & 1970(857) & 15.8 &8.96& 1.38 & 0.928 & 3.52(0.664) & 0.273(0.119) & 1.43 \\
 & 0.299(0.128) & 1170(500) & 9.40&5.38 & [1.20]& [0.721] & 1.24(0.227) & 0.162(0.0693) &  \\
\hline
C3 & 0.127(-) & 497(-) & 4.00 &5.16& 1.02 & 0.641 & 0.224(-) & 0.224(-) & 0.332 \\
 & 0.107(-) & 417(-) & 3.35 & (-) & [1.20] &[0.576]& 0.157(-) & 0.129(-) &  \\
\hline
C4 & 0.400(0.123) & 1560(457) & 12.5 &8.87& 1.25 & 0.745 & 2.21(0.189) & 0.243(0.0712) & 0.193 \\
 & 0.252(0.081) & 985(316) & 7.92 & &[1.02] &[0.714]  & 0.879(0.0907) & 0.154(0.0493) &  \\
 \hline
C5 & 0.462(0.193) & 1810(721) & 14.5 &8.72& 1.44 & 0.841 & 2.95(0.47) & 0.341(0.136) & 1.13 \\
 & 0.283(0.109) & 1100(427) & 8.87 &5.44& [1.20] & [0.662] & 1.10(0.165) & 0.209(0.0807) &  \\
 \hline
C6 & 0.422(0.172) & 1650(642) & 13.2 &8.09& 0.922 & 0.848 & 2.46(0.373) & 0.418(-0.113) & 1.31 \\
 & 0.262(0.105) & 1020(411) & 8.22 &4.92& [1.20] & [0.639] & 0.948(0.153) & 0.260(-0.0723) &  \\
 \hline
C7 & 0.143(0.0522) & 559(195) & 4.49 &2.92& 1.44 & 0.508 & 0.282(0.0343) & 0.183(0.0638) & -0.322 \\
 & 0.119(0.0492) & 464(192) & 3.72 &2.18& [0.987] & [0.578]  & 0.195(0.0334) & 0.152(0.063) &  \\
 \hline
C8 & 0.266(0.151) & 1040(562) & 8.35&3.84 & 1.62 & 0.725 & 0.978 & 0.367(0.199) & 1.11 \\
 & 0.188(0.0955) & 735(373) & 5.90 &2.91& [1.20] & [0.512]  & 0.489(0.126) & 0.260(0.132) &  \\
 \hline
C9 & 0.646(0.514) & 2520(1910) & 20.3&4.89& 1.17 & 1.14 & 5.77(3.32) & 1.03(-0.520) & 3.98 \\
 & 0.139(0.0232) & 541(90.5) & 4.35&3.62 & [1.51] & [0.579]  & 0.265(0.00742) & 0.221(-0.0246) &  \\
 \hline
C10 & 0.249(0.0585) & 972(218) & 7.81&6.06& 1.6 & 0.774 & 0.855(0.0431) & 0.495(0.111) & 0.997 \\
 & 0.179(0.062) & 698(242) & 5.60 & 3.66& [0.924] & [0.602]  & 0.441(0.0531) & 0.355(0.123) &  \\
\hline
C11 & 0.418(0.179) & 1630(668) & 13.1&7.76 & 1.26 & 1.14 & 2.42(0.405) & 0.216(0.0883) & 3.94 \\
 & 0.256(0.0992) & 1000(388) & 8.04 &4.93& [1.07] & [0.605] & 0.907(0.136) & 0.132(0.0512) &  \\
\hline
C12 & 0.263(0.00266) & 1030(9.91) & 8.24 &8.16& 1.23 & 0.771 & 0.952(8.90e-05) & 0.0369(0.000356) & 0.474 \\
 & 0.195(0.0408) & 762(159) & 6.12 &4.84& [1.18] & [0.694]  & 0.525(0.023) & 0.0274(0.00573) &  \\
 \hline
C13 & 0.360(0.121) & 1410(451) & 11.3 &7.67& 1.21 & 1.19 & 1.79(0.184) & 0.614(-0.0446) & 3.89 \\
 & 0.240(0.0923) & 938(360) & 7.53 &4.64& [1.34] & [0.701] & 0.796(0.118) & 0.410(-0.0356) &  \\
\hline
C14 & 0.272(0.0347) & 1060(130) & 8.53 &7.49& 1.22 & 0.803 & 1.02(0.0152) & 0.746(0) & 0.656 \\
 & 0.176(0.0226) & 686(88.2) & 5.51 &4.80&[1.22] & [0.699] & 0.426(0.00705) & 0.291(0) &  \\
\hline
C15 & 0.430(0.204) & 1680(762) & 13.5 &7.39& 1.33 & 1.43 & 2.56(0.525) & 0.141(0.0637) & 6.66\\
 & 0.283(0.140) & 1110(548) & 8.89 &4.49& [1.22] & [0.677] & 1.11(0.271) & 0.0926(0.0458) &  \\
\hline
C16 & 0.435(0.252) & 1700(941) & 13.7 &6.09& 1.56 & 1.46 & 2.61(0.801) & 0.516(0.286) & 7.60 \\
 & 0.274(0.140) & 1070(548) & 8.60 &4.20 &[1.19] & [0.578]  & 1.04(0.272) & 0.325(0.166) &  \\
 \hline
 \end{tabular}
\small
\raggedright
\\
$^{\dagger}$An effective radius of 7.06~pc, i.e., 5.37$^\prime$, is used for the global IRDC, while a radius of 0.399~pc, i.e., 0.303$^\prime$, is used for the dense cores. For each source, the top row reports mass surface densities and masses based on the Herschel (FIR) mass surface density map, while the bottom row uses the MIREX map. Values in ``()'' are background subtracted estimates. When a background-subtracted mass or density is found to be negative, it is not evaluated and the symbol ``-'' is used. The 
values of $\sigma_v$ and $B_{\rm tot}$ in ``[]'' in the second rows are background values, i.e., $\sigma_{v,0}$ and $B_{\rm tot,0}$.
\end{table*}

\begin{table*}
\caption{Additional dynamical properties of IRDC G28.37+0.07 and its dense core sample$^{\dagger}$}
\label{tab:core_summary2}
\setlength\tabcolsep{1.9pt}
\begin{tabular}{|c|cccccc|}
\hline
 & $\alpha_{\rm vir}$ & $m_{\rm A}$ & $\mu$& ${\sigma_{\rm vir}}$ & $B_{\rm vir}$  &$B_{\rm vir,corr}$\\
 &  &  & &($\rm km\,s^{-1}$) &  (mG) & (mG) \\
\hline
CloudC & 1.93(-) & 0.230(-) & 0.430(-)&  3.56(-) & 0.116(-) & 0.124(-) \\
 & 2.79(-) & 0.19(-) & 0.298(-) & 2.79(-) &  0.0804(-) & 0.0595(-) \\
 \hline
Cp23 & 1.33(4.83) & 0.670(0.350) & 1.12(0.322) &  1.77(1.18) & 0.59(0.0616) & 0.736(0.0506) \\
 &  &&  &   &  &  \\
 \hline
C1 & 1.10(1.99) & 0.320(0.240) & 0.587(0.343) &  1.89(1.31) & 0.675(0.0508) & 0.999(0.218) \\
 & 1.67(3.21) & 0.260(0.190) &0.298(0.324)  & 1.53(1.08) & 0.445(0.232) & 0.537(-) \\
 \hline
C2 & 0.707(1.63) & 0.560(0.370) & 1.28(0.583) &  2.2(1.53) & 0.915(0.0741) & 1.95(0.48) \\
 & 1.19(2.79) & 0.430(0.280) & 0.388(0.256) &  1.69(1.19) & 0.543(0.232) & 0.973(0.141) \\
 \hline
C3 & 1.53(-) & 0.300(-) & 0.466(-) &  1.1(-) & 0.231(-) & 0.249(-) \\
 & 1.83(-) & 0.270(-) & 0.758(-) & 1.01(-) & 0.194(-) & 0.186(-) \\
 \hline
C4 & 0.733(2.5) & 0.560(0.300) & 1.26(0.388) &  1.96(1.31) & 0.725(0.074) & 1.5(0.266) \\
 & 1.16(3.62) & 0.440(0.250) & 0.393(0.305) & 1.56(1.06) & 0.457(0.147) & 0.814(0.0934) \\
 \hline
C5 & 0.841(2.11) & 0.610(0.390) & 1.29(0.540) &  2.11(1.46) & 0.838(0.075) & 1.56(0.343) \\
 & 1.38(3.56) & 0.480(0.300) & 0.792(0.291) &  1.65(1.15) & 0.512(0.198) & 0.791(0.0788) \\
 \hline
C6 & 0.378(0.97) & 0.370(0.230) & 1.17(0.477) &  2.01(1.39) & 0.765(0.0678) & 2.58(0.648) \\
 & 0.608(1.51) & 0.290(0.190) & 0.727(0.228) &  1.58(1.11) & 0.475(0.191) & 1.44(0.28) \\
 \hline
C7 & 2.72(7.8) & 0.560(0.330) & 0.662(0.242) & 1.17(0.803) & 0.259(0.03) & 0.106() \\
 & 3.28(7.9) & 0.510(0.330) & 0.551(0.310) &  1.07(0.749) & 0.215(0.0892) & nan(nan) \\
 \hline
C8 & 1.85(3.42) & 0.610(0.450) & 0.863(0.490) & 1.6(1.11) & 0.482(0.04) & 0.418(-) \\
 & 2.62(5.15) & 0.510(0.360) & 0.610(0.0479) &  1.34(0.949) & 0.341(0.173) & 0.188(-) \\
  \hline
C9 & 0.397(0.524) & 0.430(0.380) & 1.33(1.06) & 2.49(1.56) & 1.17(0.0499) & 3.81(1.03) \\
 & 1.85(11.1) & 0.200(0.0800) & 0.287(0.188) & 1.15(0.704) & 0.251(0.042) & 0.47(-) \\
  \hline
C10 & 1.93(8.6) & 0.540(0.260) & 0.757(0.178) &  1.54(0.994) & 0.451(0.0504) & 0.37() \\
 & 2.69(7.74) & 0.460(0.270) & 0.544(0.188) &  1.31(0.903) & 0.324(0.112) & 0.163() \\
 \hline
C11 & 0.712(1.74) & 0.380(0.240) & 0.862(0.369)&  2.00(1.39) & 0.758(0.0679) & 1.61(0.38) \\

 & 1.16(3) & 0.290(0.180) & 0.528(0.205) &  1.57(1.09) & 0.464(0.18) & 0.839(0.117) \\
 \hline
C12 & 1.08(112) & 0.430(0.0400) & 0.802(0.00812) &  1.59(0.496) & 0.476(0.0667) & 0.718(nan) \\
 & 1.45(6.95) & 0.370(0.170) & 0.595(0.124) & 1.37(0.872) & 0.353(0.0739) & 0.468(nan) \\
  \hline
C13 & 0.763(2.38) & 0.320(0.180) & 0.712(0.239) &  1.86(1.26) & 0.652(0.0639) & 1.31(0.249) \\
 & 1.14(2.98) & 0.260(0.160) &  0.474(0.182)& 1.52(1.06) & 0.435(0.167) & 0.764(0.112) \\
 \hline
C14 & 1.03(8.42) & 0.420(0.150) & 0.797(0.192) &  1.61(0.922) & 0.493(0.0661) & 0.777(0.0118) \\

 & 1.59(12.4) & 0.340(0.120) &0.516(0.0662)  &  1.3(0.751) & 0.318(0.0409) & 0.411(nan) \\
 \hline
C15 & 0.771(1.7) & 0.320(0.220) & 0.707(0.336)&  2.03(1.42) & 0.78(0.0618) & 1.55(0.383) \\
 & 1.17(2.37) & 0.26(0.180) & 0.466(0.230) &  1.65(1.17) & 0.513(0.254) & 0.888(0.153) \\
  \hline
C16 & 1.05(1.89) & 0.370(0.280) & 0.701(0.406) &  2.04(1.42) & 0.788(0.0579) & 1.22(0.279) \\
 & 1.66(3.25) & 0.290(0.210) & 0.442(0.226) &  1.62(1.15) & 0.497(0.254) & 0.617(nan)\\
 \hline
 \end{tabular}
\small
\raggedright
\\
$^{\dagger}$An effective radius of 7.06~pc, i.e., 5.37$^\prime$, is used for the global IRDC, while a radius of 0.399~pc, i.e., 0.303$^\prime$, is used for the dense cores. For each source, the table reports: virial parameter ($\alpha_{\rm vir}$); Alfv\'en Mach number ($m_A$); mass-to-flux ratio ($\mu$); fiducial TCM predicted virialized 1D velocity dispersion ($\sigma_{\rm vir}$); fiducial TCM predicted core $B-$field strength ($B_{\rm vir}$); and TCM $B-$field strength needed to achieve virial equilibrium give observed velocity dispersion ($B_{\rm vir,corr}$). The top row for each source reports quantities
based on masses and densities measured from the Herschel (FIR) mass surface density map, while the bottom row uses the MIREX map. Values in ``()'' are background subtracted estimates. When a background-subtracted mass or density is found to be negative, it is not evaluated and the symbol ``-'' is used. Background subtraction is not evaluated for the global properties of Cloud C.
\end{table*}

\section{Connection of $B-$field morphology to massive star formation}\label{sec:fragmentation}

\subsection{Magnetic field orientation and mass surface density gradient}

\begin{figure*}
\centering
\begin{minipage}{0.32\textwidth}
\includegraphics[width=\columnwidth]{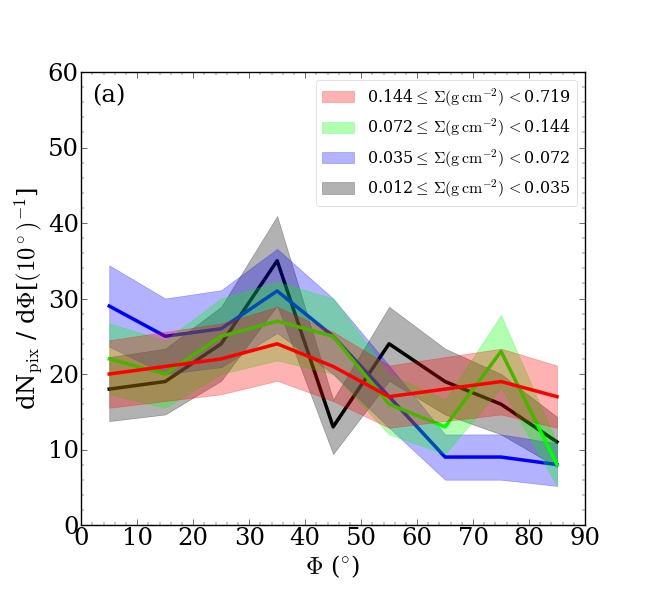}
\end{minipage}
\begin{minipage}{0.32\textwidth}
\includegraphics[width=\textwidth]{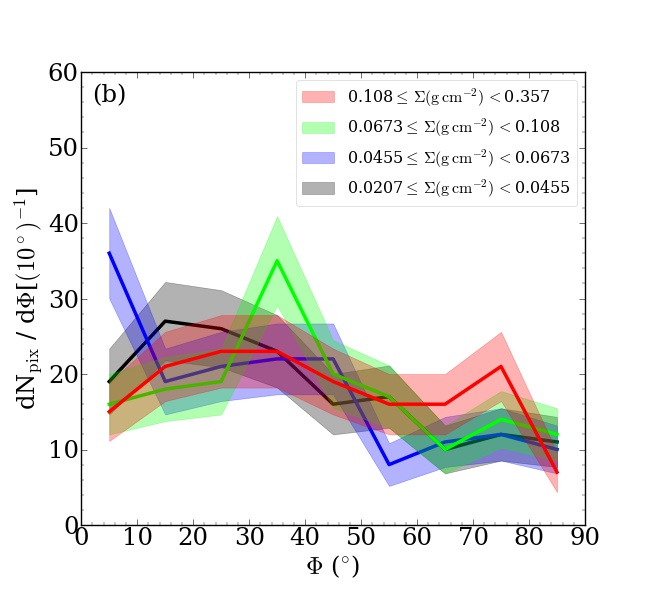}
\end{minipage}
\begin{minipage}{0.32\textwidth}
\includegraphics[width=\textwidth]{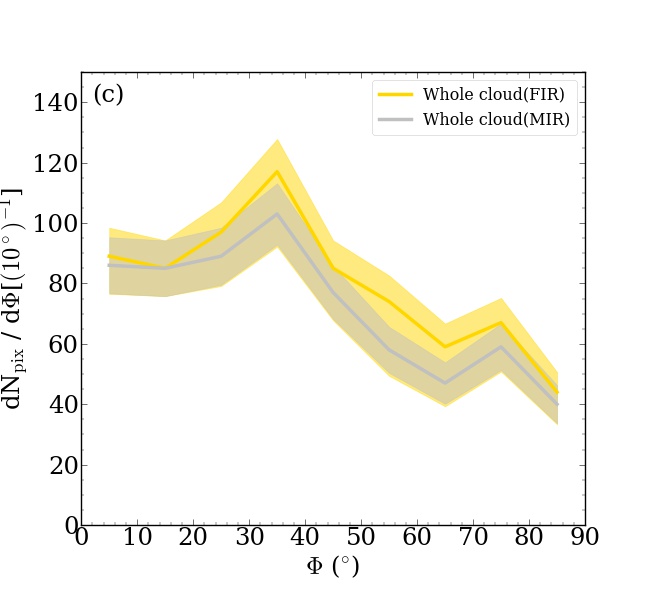}
\end{minipage}
\caption{{\it (a) Left:} HROs constructed from the FIR mass surface density ($\Sigma$) map with a constant bin-width of 10 degrees (solid lines) for four quartiles of the $\Sigma$ distribution (each with $N_{\rm pix}=179$), as labelled. Shaded regions show $\pm 1 \sigma$, where $\sigma=\sqrt{N_{\rm pix}}$ is the Poisson error of the count in a given bin. {\it (b) Middle:} As (a), but now for $\Sigma$ derived from the MIREX map.
{\it (c) Right:} HROs for the entire pixel sets from the FIR mass surface density ($\Sigma$) map (gold) and from the MIREX map (silver).
}
\label{fig:HRO_same_pixel}
\end{figure*}

Here, we examine potential connections between magnetic field morphology and the formation of dense gas structures in the IRDC. First we examine the distribution of angles, $\Phi$, between direction of plane of sky magnetic field orientation and gradient in the mass surface density.
We carry out a histogram of relative orientations (HRO) analysis following the method described in \citep{Soler2013}. 
We examine cases with the Herschel (FIR)-derived $\Sigma$ map and the MIREX $\Sigma$ map, each regridded to the $18.^{\prime\prime}2$ pixel scale of the HAWC+ image.
We compute the gradient by convolving with a Gaussian kernel with a FWHM of 3 pixels (i.e., $55.^{\prime\prime}6$). The HROs are presented in Figure~\ref{fig:HRO_same_pixel}. We set the ranges of $\Sigma$ so that each HRO set contains 25\% of the pixels. 

The results of Figure~\ref{fig:HRO_same_pixel} generally show a greater number of pixels having smaller values of $\Phi$, i.e., indicating that magnetic field direction tends to align with gradient in $\Sigma$ (and thus also volume density). However, the distributions are quite broad, with significant numbers of pixels at large values of $\Phi$. There is no clear trend of the HROs varying systematically with $\Sigma$.

We evaluate the significance of any overall parallel/perpendicular alignment by evaluating the projected Rayleigh statistic (PRS) \citep[e.g.,][]{2018MNRAS.474.1018J,2019ApJ...878..110F}.
The PRS is used to test for the existence of a preference for perpendicular or parallel alignment and has the following equation
\begin{equation}
Z_x=\frac{\sum_i^N \cos \theta_i}{\sqrt{N / 2}}
\end{equation}
where $N$ is the number of relative orientations between the two sets of angles.
The corresponding uncertainty in the PRS is given by \citep{2018MNRAS.474.1018J}:
\begin{equation}
\sigma_{Z_x}=\sqrt{\frac{2 \sum_i^{N}\left(\cos \theta_i\right)^2-\left(Z_x\right)^2}{N}}.
\end{equation}
We apply the PRS and present the $Z_x$ values in Table~\ref{tab:PRS_kuiper}. A value of $Z_x > 0$ indicates a preferred parallel alignment, while $Z_x < 0$ indicates a preferred perpendicular alignment. While the values are all $>0$, given the uncertainties, there is no strong evidence for preferred parallel alignment in any of the individual quartiles. Table~\ref{tab:PRS_kuiper} also reports the value for the PRS of the whole pixel sample.

The PRS can infer if there is a uni-modal distribution but cannot reflect on the existence of any multi-modal, non-uniform distributions, e.g., a bimodal distribution. Thus, next we perform a Kuiper test, which is closely related to the Kolmogorov–Smirnov (KS) test. The Kuiper test compares the distribution to the cumulative density function of a certain distribution, which in this case is a uniform distribution with the same number of measurements within 0 to 90 degrees, i.e., no preferred direction. 
We use the python function {\it kuiper\_two} in {\it astropy} to carry out the Kuiper two samples test. The output $p$-value reflects whether the sample distribution
differs from that of a uniform distribution. We also summarize the Kuiper test results in Table~\ref{tab:PRS_kuiper}. We see that certain quartiles, corresponding to the ones with the largest PRS values, have small ($\lesssim 0.05$) values of $p_{\rm Kuiper}$, indicating distributions that are inconsistent with a uniform, i.e., random, distribution. In addition, the result for the entire pixel set shows a significant result that the distribution is inconsistent with that of a uniform distribution.

\begin{table}
    \centering
    \begin{tabular}{|c|cc|}
    \hline
        Stat test & $Z_x$ &$p_{\rm Kuiper}$\\
        \hline
        \multicolumn{3}{|c|}{FIR}\\
        \hline
         Whole cloud&$5.13\pm4.86$&0.000571 \\
         $0.144\leq \Sigma\,({\rm g\,cm}^{-2})<0.719$& $0.902\pm 3.62$ &0.164 \\
         $0.072\leq \Sigma\,({\rm g\,cm}^{-2})<0.144$& $2.27\pm3.44$ & 0.260\\
         $0.035\leq \Sigma\,({\rm g\,cm}^{-2})<0.072$& $5.21\pm3.16$ &0.0001 \\
         $0.012\leq \Sigma\,({\rm g\,cm}^{-2})<0.035$& $1.89\pm3.41$ & 0.208\\
         \hline
         \multicolumn{3}{|c|}{MIR}\\
        \hline
          Whole cloud&$6.05\pm4.76$&0.000138 \\
         $0.108\leq \Sigma\,({\rm g\,cm}^{-2})<0.375$& $1.65\pm 3.46$ &0.693 \\
         $0.0673\leq \Sigma\,({\rm g\,cm}^{-2})<0.108$& $0.322\pm3.45$ & 0.011\\
         $0.0455\leq \Sigma\,({\rm g\,cm}^{-2})<0.0673$& $5.21\pm3.16$ &0.005 \\
         $0.0207\leq \Sigma\,({\rm g\,cm}^{-2})<0.0455$& $2.33\pm3.41$ &0.238\\
         \hline
    \end{tabular}
    \caption{Results of PRS and Kuiper tests toward the HROs constructed from the FIR and MIREX $\Sigma$ maps.}
    \label{tab:PRS_kuiper}
\end{table}

\subsection{Magnetic fields and protostellar outflows}\label{sec:B_and_outflow}


Here, we investigate if there is any correlation between magnetic field orientation and protostellar outflow axis orientation. We use the sample of 64 averaged blue- and red-shifted CO(2-1) plane-of-sky outflow axis position angles, $\theta_{\rm outflow}$, presented by \citet{2019ApJ...874..104K}. These authors already noted that their sample tends to align perpendicularly to the main IRDC filament.

\begin{figure*}[]
\centering
\includegraphics[width=\textwidth]{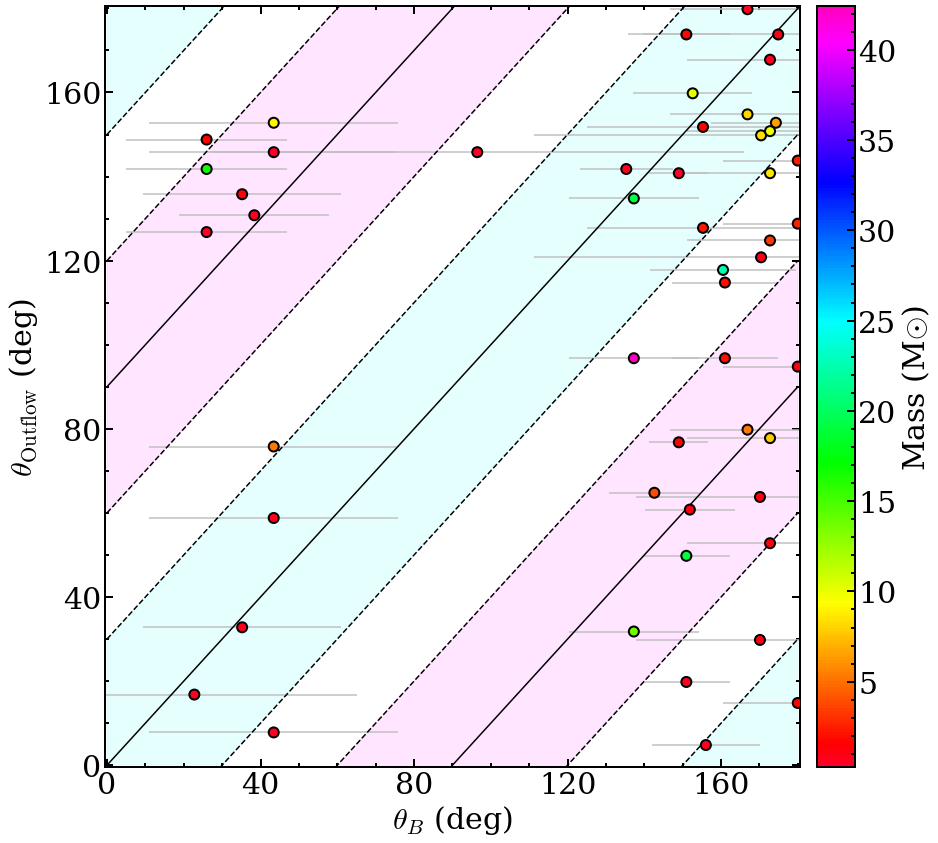}
\caption{Scatter plot of outflow axis position angle \citep{2019ApJ...874..104K} versus local magnetic field orientation. Protostellar core masses are evaluated from ALMA mm continuum data (see text).
} 
\label{fig:cloudC-outflow_B_scatter}
\end{figure*}

Figure~\ref{fig:cloudC-outflow_B_scatter} shows a scatter plot of $\theta_{\rm outflow}$ versus $\theta_B$. Note, that multiple outflows can be located within a SOFIA beam, i.e., sharing the same $18.^{\prime\prime}2$-size pixel in the $B-$field map. In this case, we assume that all the outflows share that same field orientation. Indeed, the distribution of outflows is clustered so that 49 of the outflows are found to be launched from within just 7 of the pixels of the $B-$field map. Examining Figure~\ref{fig:cloudC-outflow_B_scatter}, we notice that there seems to be a tentative trend that the relative orientations between outflows and local magnetic fields are either parallel or perpendicular. However, there are also significant numbers of sources that have no preferential orientations. We also color code the outflows with masses that are measured base on the ALMA continuum image \citep{2019ApJ...874..104K}.

\begin{figure}[]
\centering
\includegraphics[width=\columnwidth]{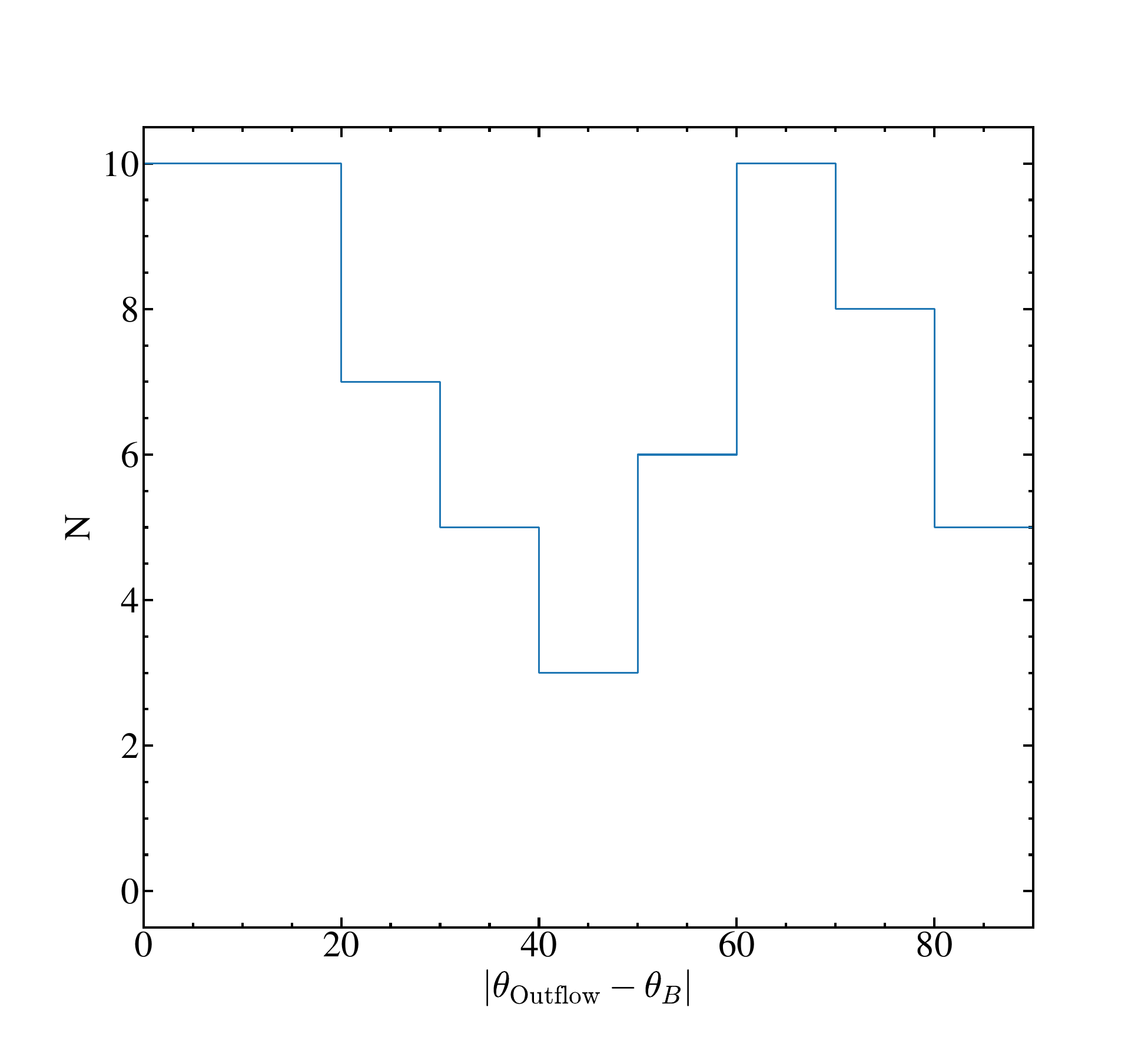}
\caption{Histogram of angle difference between protostellar outflow and local magnetic field orientations. 
} 
\label{fig:cloudC-outflow_B}
\end{figure}

In Figure~\ref{fig:cloudC-outflow_B} we present a histogram of the relative orientations between the average outflow orientations and the magnetic field orientations. The apparent bimodality is also apparent in this histogram. However, when we apply the PRS and Kuiper tests to the distribution of relative orientations, the results are $Z_x = 1.81\pm2.59$ and $p=0.65$. Thus the data do not indicate any significant evidence for a preferential alignment direction or a distribution that is distinct from a random, uniform distribution.

\section{Discussion \& Conclusions}
\label{sec:conclusions}

In this paper we have presented the first global, relatively high-resolution ($18^{\prime\prime}$) magnetic field strength map of the massive IRDC G28.37+0.07 utilizing SOFIA-HAWC+ $214 \:{\rm \mu m}$ polarized dust emission data, combined with a velocity dispersion map obtained with GBT-Argus C$^{18}$O(1-0) observations and mass surface density and associated inferred volume density maps derived from Herschel FIR emission and Spitzer MIR extinction maps.
We have assessed the role of magnetic fields on the star formation in of G28.37+0.07 based on an evaluation of the $B$-$\Sigma$ and $B$-$\rho$ relations in the maps and more directly by measuring the energy balances between magnetic fields, turbulence and gravity in the global cloud and a sample of dense cores. We have also examined the morphological connection of the magnetic field orientation with density (i.e., $\Sigma$) gradients and protostellar outflow axis orientations. Here, we discuss the implications of these results for a more general understanding of the role of magnetic fields in massive star formation.


The relative importance of magnetic fields compared to gravity may influence the observed $B$-$\Sigma$ and $B$-$\rho$ relations of the cloud.
Gravity-dominated isotropic contraction is expected to lead to a $B$-$\rho$ index of about 2/3, while our results imply a shallower index of about $0.5\pm 0.05$. Such a shallow value may be 
consistent with dynamically important magnetic field regulated collapse via ambipolar diffusion \citep[e.g.,][]{2023ApJ...946L..46C}.


More directly, the mass to flux ratio maps indicate that magnetic fields are dominant (or at least significant) in large regions of the IRDC. This is especially true for values of $B$-fields estimated by the DCF method. However, even for the lower $B-$field values of the r-DCF method of \citet{2021A&A...647A.186S}, we find that major parts of the IRDC, including the eastern end of its dense filamentary spine are magnetically sub-critical ($\mu<1)$. An assessment of the global energetics finds that magnetic energy is much larger than gravitational and turbulent energies. We note here that a caveat of the DCF and r-DCF methods is that they ignore the influence of self-gravity in bending $B-$field orientations. However, such effects would lead us to systematically underestimate true $B-$field strengths, so our conclusions about the dynamical importance of the fields are quite conservative.


Zooming into sub-regions of the cloud, while there are some variations in the mass to flux ratio,
overall, we see that for most cores, magnetic fields play a dynamically important role compared to turbulence. A detailed study of virial balance indicates that cores may be close to virial equilibrium but with relatively strong $B-$fields that imply sub-Alfvénic conditions. Such $B$-fields with $\sim 1$~mG strengths are also likely to be important for setting the mass scale of the cores at a level of $\sim 100\:M_\odot$ and by preventing fragmentation to much smaller masses, such $B-$field strengths may be an essential requirement for the formation of massive stars \citep[e.g.,][]{2012ApJ...754....5B,2013ApJ...779...96T}. 

From the morphological point of view, we find strong evidence for a preferential alignment of $B-$field orientation with density gradient across the full range of $\Sigma$. This is consistent with previous results that found dense filaments have orientations perpendicular to $B-$field orientation \citep[e.g.,][]{Soler2013}. It appears that most of the range of $\Sigma$ that is probed by our maps is in this regime. Again, the implication is that magnetic fields are helping to shape the collapse of gas into structures that then tend to align perpendicularly to the field. 

Once protostellar sources form and drive outflows, one may also ask if the direction of the outflows is influenced by the large scale $B-$field. Here we do not find strong evidence for such an effect, even though a preferred orientation of the outflow axes has been claimed \citep{2019ApJ...874..104K} and such effects have been noted for mainly low-mass protostellar outflows \citep{2022ApJ...941...81X}.

A summary of our main conclusions is as follows:

\begin{itemize}

\item We derived the $B-$field strengths maps for G28.37+0.07 with both the DCF and r-DCF \citep{2021A&A...647A.186S} methods, with the latter considered to be most accurate, given the sub- to trans-Alfvénic conditions in the IRDC. The $B-$field strengths in the r-DCF map range from about 0.03 to 1~mG.


\item From the r-DCF map we derive a $B$-$N_{\rm H}$ relation of the form $B_{\rm 1D} = 0.15 N_{\rm H,22}^{0.38}\:$mG and a $B$-$n_{\rm H}$ relation of the form $B_{\rm 1D} = 0.19 n_{\rm H,4}^{0.51}\:$mG. The latter index is relatively shallow compared to some previous estimates, e.g., the classic relation of \citet{2012ARA&A..50...29C} with a power law index of 0.65. This difference may reflect the particular environment of IRDC G28.37+0.07 or be the result of different methodologies in the estimation of $B$-fields and density.


\item Magnetic fields appear to be playing a dynamically important role on both the global and local, dense core scales of the IRDC. Conditions in the dense cores are consistent with virial equilibrium, although there are significant uncertainties, especially associated with estimation of mass and the effect of clump envelope background subtraction. The most massive protostellar source in the IRDC (Cp23) is seen to be the most gravity-dominated source, while the highly deuterated massive pre-stellar core candidate (C1) is the most magnetically-dominated source.


\item A morphological study of magnetic field orientation with gradients in the mass surface density map finds strong evidence for a preferentially alignment of the $B-$fields with the density gradients, again indicating the fields play a role in regulating the collapse. However, the is no evidence that the field is able to control directions of protostellar outflows.


\end{itemize}
   
The overall findings of the study demonstrate that magnetic fields play an important role in regulating the formation processes of massive stars by setting up the initial conditions in the massive IRDC G28.37+0.07. Such strong magnetic field may leave traces in anisotropic gas kinematics. A follow-up work towards this IRDC (Law et al., in prep.) aims to investigate this aspect.

\begin{acknowledgments}
C.-Y.L. acknowledges support from an ESO Ph.D. student fellowship. J.C.T. acknowledges ERC project MSTAR, NSF grant AST-200967, and USRA-SOFIA grant 09-0104.
\end{acknowledgments}

%

\vspace{5mm}
\facilities{GBT, SOFIA}
\software{Matplotlib \citep{2007CSE.....9...90H}, APLpy \citep{2012ascl.soft08017R}, Astropy \citep{2013A&A...558A..33A,2018AJ....156..123A}, Astrodendro \citep{2019ascl.soft07016R}, Numpy \citep{harris2020array}, Reproject \citep{2020ascl.soft11023R}, and Spectral-Cube \citep{2019zndo...2573901G}.}
          

\section*{References}
\bibliography{HAWC_ARGUS.bib}
\bibliographystyle{aasjournal}
\appendix
\section{Stokes Q, U, p', and SNR maps}\label{sec:AppA}
\restartappendixnumbering

\begin{figure*}[]
\begin{minipage}{0.49\textwidth}
\includegraphics[width=\textwidth]{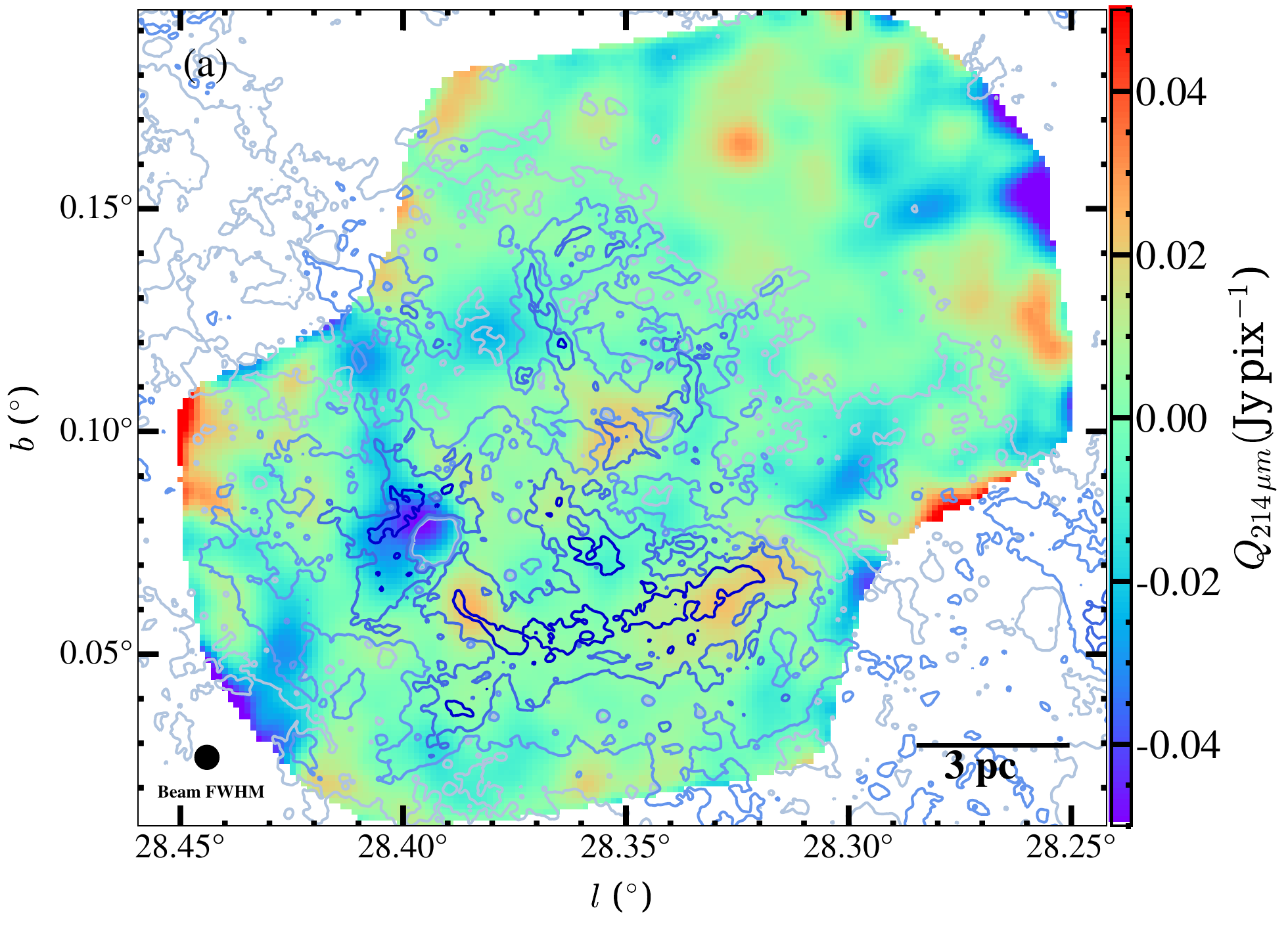}
\end{minipage}
\begin{minipage}{0.49\textwidth}
\includegraphics[width=\textwidth]{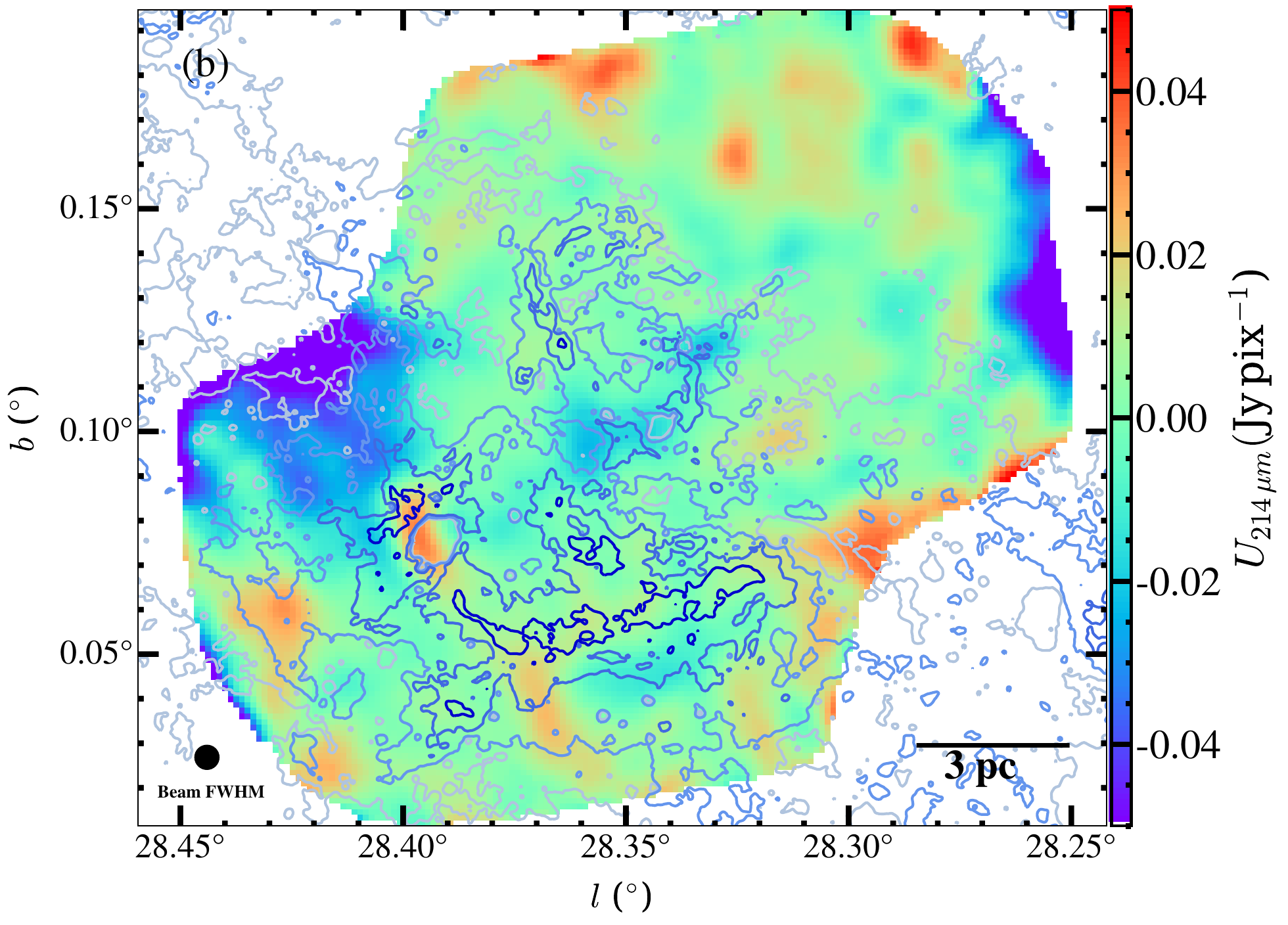}
\end{minipage}
\begin{minipage}{0.49\textwidth}
\includegraphics[width=\textwidth]{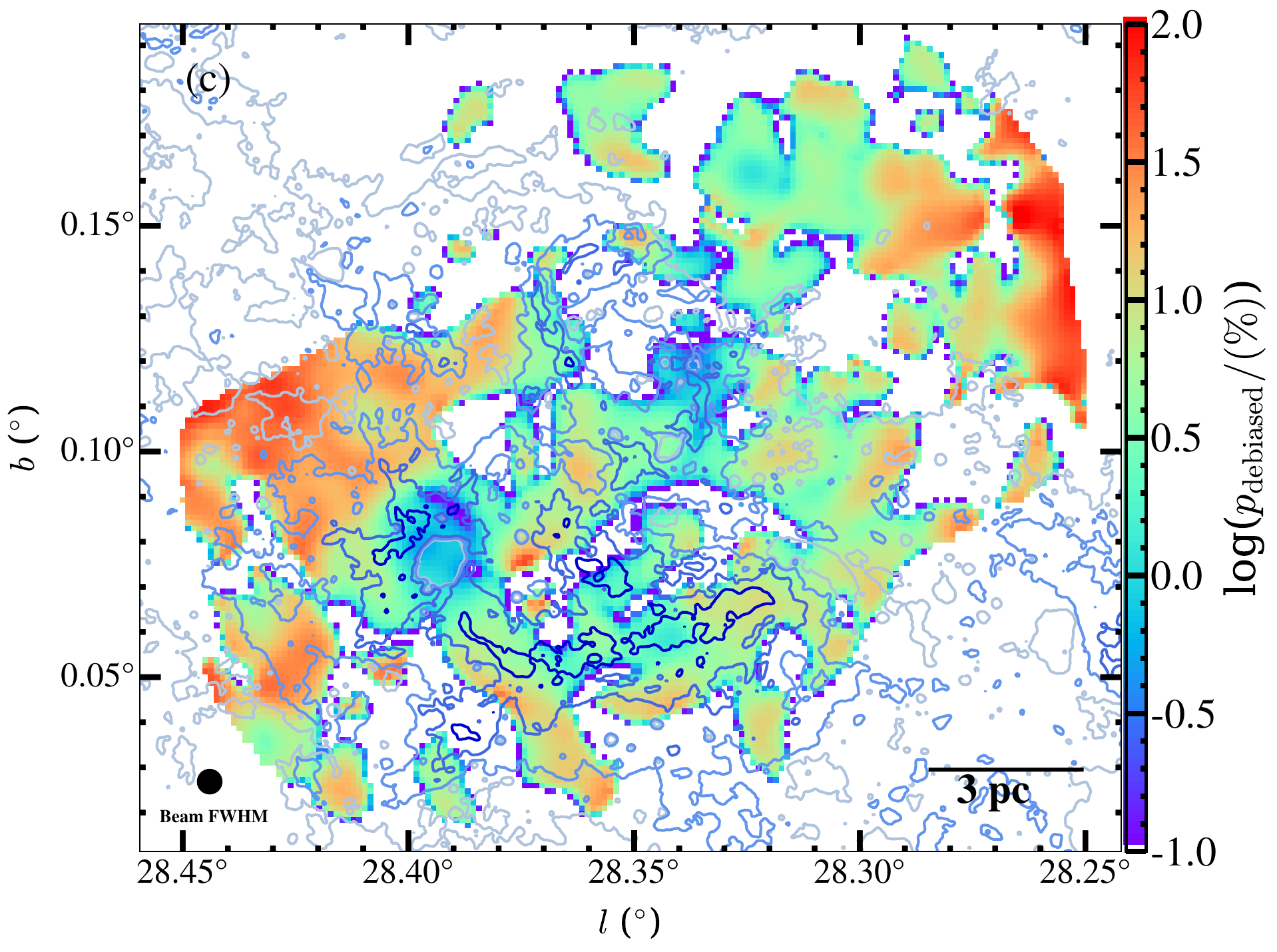}
\end{minipage}
\begin{minipage}{0.49\textwidth}
\includegraphics[width=\textwidth]{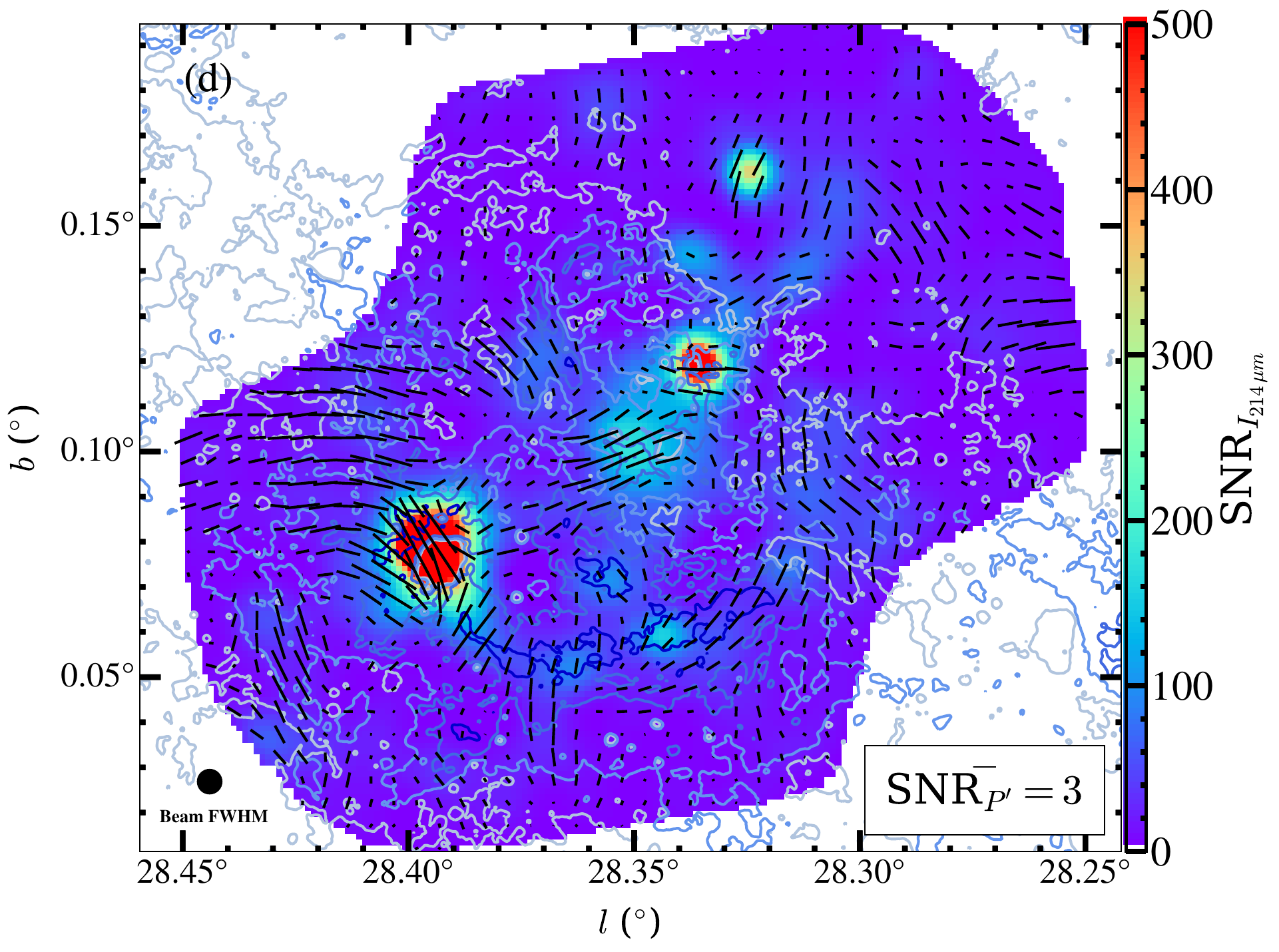}
\end{minipage}

\caption{{\it (a) Top Left:} Stokes $Q$ map of G28.37+0.07, with other contours following format of Fig.~2.
{\it (b) Top Right:} As (a), but now showing the Stokes $U$ map. 
{\it (c) Bottom Left:} As (a), but now showing the debiased polarization direction map of G28.37+0.07 in logarithmic scale.
{\it (d) Bottom Right:} As (a), but now showing SNR in Stokes $I$, overlaid with the vectors in which the lengths are scaled with the SNR in polarization fraction. 
%
}
\label{figure:appendex_figure_0}

\end{figure*}

Figure~\ref{figure:appendex_figure_0} presents maps of Stokes $Q$, $U$, $p^\prime$, and SNR.

\section{Effects on number of pixels in measuring dispersion in polarization angles}\label{sec:AppC}
\restartappendixnumbering
We have presented maps of angle dispersion and the corresponding magnetic field strength maps. We select the smallest window size of 4 pixels (equivalent to two SOFIA-HAWC+ beams). However, the low number statistics in the number of vectors in computing the field strengths via (refined) DCF methods may underestimate the angle dispersion, thus overestimating the field strength. 
We present the mean polarisation angle and the dispersion in the polarization angle based on a 3x3 window centred at each pixel. This analysis shows similar results as the map using the 2$\times$2 window, but with a slightly higher level of angle dispersion, given it is averaging over a larger region. We notice that the magnetic field strength is thus on average slightly lower.
Thus, we conclude that there is no significant impact on our main findings by the use of a $2\times2$ window function compared to a $3\times3$ window function.

\begin{figure*}[h!]
\begin{minipage}{0.49\textwidth}
\includegraphics[width=\textwidth]{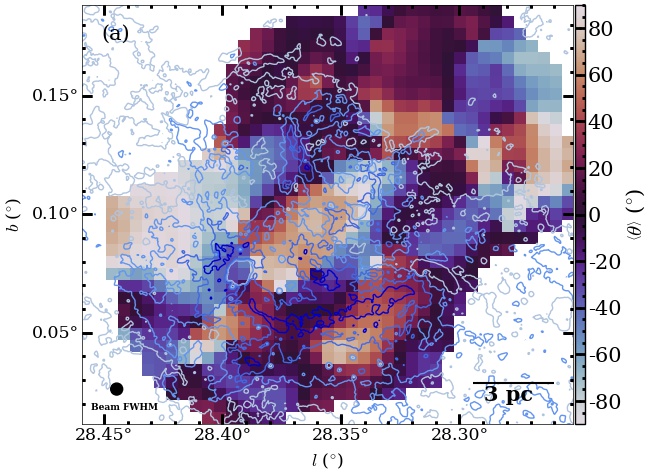}
\end{minipage}
\begin{minipage}{0.49\textwidth}
\includegraphics[width=\textwidth]{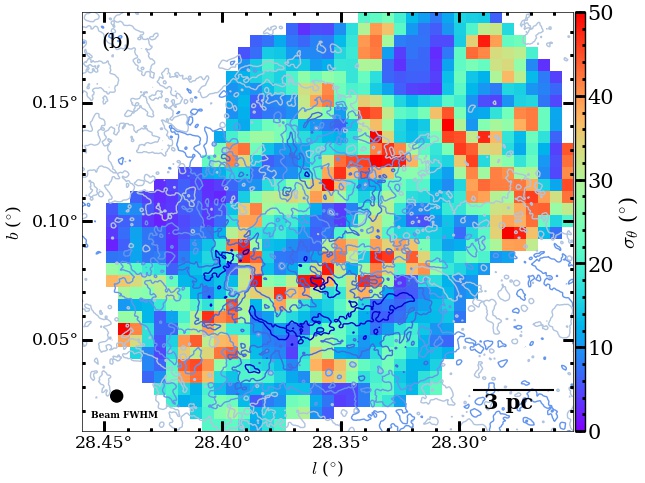}
\end{minipage}
\begin{minipage}{0.49\textwidth}
\includegraphics[width=\textwidth]{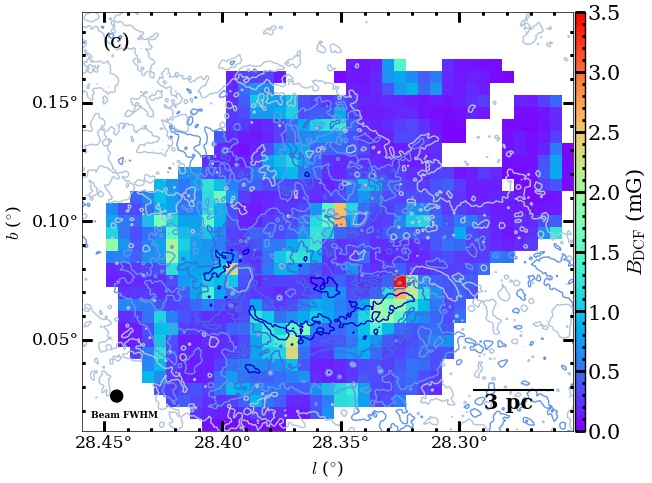}
\end{minipage}
\begin{minipage}{0.49\textwidth}
\includegraphics[width=\textwidth]{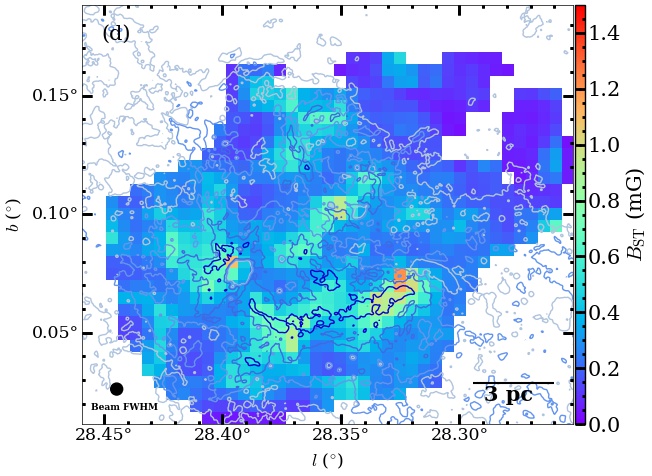}
\end{minipage}
\caption{{\it (a) Top Left:} Mean polarization angle map constructed based on a 3-by-3 window function.
{\it (b) Top Right:} The corresponding dispersion map of the magnetic field position angles. 
{\it (c) Bottom Left:} The corresponding magnetic field strength map based on the DCF method.
{\it (d) Bottom Left:} The corresponding magnetic field strength map based on the r-DCF method.
}
\label{figure:summary_B_measure_3b3}
\end{figure*}
%
%

\clearpage
\restartappendixnumbering



\end{document}